\documentclass[letterpaper,twocolumn,10pt]{article}

\usepackage[dvipsnames]{xcolor}
\usepackage{usenix-2020-09}

\usepackage{tikz}
\usepackage{amsmath}

\usepackage{setspace}
\usepackage{amsmath,amsfonts}
\usepackage{times}
\usepackage{makecell}
\usepackage[T1]{fontenc}
\usepackage[utf8]{inputenc}
\usepackage[figurename=Fig.]{caption}
\usepackage{svg}
\usepackage{ dsfont }
\usepackage{verbatim}
\usepackage[ruled,vlined]{algorithm2e}
\usepackage{diagbox}
\usepackage{subfigure}
\usepackage{graphicx}
\usepackage{url}
\usepackage{float}
\usepackage{multirow}
\usepackage{url}

\usepackage[normalem]{ulem}
\usepackage{setspace}
\usepackage[noend]{algpseudocode}
\usepackage[normalem]{ulem}
\usepackage{amssymb}
\usepackage{xparse}
\usepackage{mathtools}

\usepackage[thmmarks,  thref, amsmath]{ntheorem}
\theoremseparator{.}

\newtheorem{theorem}{Theorem}

\theoremstyle{nonumberplain}
\theorempostwork{\setcounter{case}{0}}
\theoremheaderfont{\scshape}
\theorembodyfont{\upshape}
\theoremsymbol{\ensuremath{\blacksquare}}
\newtheorem{proof}{Proof}

\usepackage[T1]{fontenc}    
\usepackage{authblk}
\usepackage{amsfonts}       
\usepackage{nicefrac}       
\usepackage{microtype}      

\usepackage{times}
\usepackage{soul}

\usepackage{subfigure}
\usepackage{amsmath}
\DeclareMathOperator*{\argmax}{arg\,max} 

\usepackage{enumitem}
\setlist{nolistsep}

\usepackage{atbegshi}
\usepackage{filecontents}
\DeclarePairedDelimiter{\norm}{\lVert}{\rVert} 
\usepackage{enumitem}

\usepackage{atbegshi}
\usepackage{filecontents}

\title{PredCoin: Defense against Query-based Hard-label Attack}
\newcommand{\name}{\textsf{PredCoin}}

\author[1]{Junfeng Guo}
\author[1]{Yaswanth Yadlapalli}
\author[2]{Thiele Lothar}
\author[3]{Ang Li}
\author[1]{Cong Liu}
\affil[1]{$UT~Dallas$}
\affil[2]{$ETH~Zurich$}
\affil[3]{$Google~Deep~Mind$}
\date{July 2020}

\begin{document}

\maketitle

\begin{abstract}

Many adversarial attacks and defenses have recently been proposed for Deep Neural Networks (DNNs). While most of them are in the white-box setting, which is impractical, a new class of query-based hard-label (QBHL) black-box attacks pose a significant threat to real-world applications (e.g., Google Cloud, Tencent API). Till now, there has been no generalizable and practical approach proposed to defend against such attacks.

This paper proposes and evaluates \name{}, a practical and generalizable method for providing robustness against QBHL attacks. \name{} poisons the gradient estimation step, an essential component of most QBHL attacks. \name{} successfully identifies gradient estimation queries crafted by an attacker and introduces uncertainty to the output. Extensive experiments show that \name{} successfully defends against four state-of-the-art QBHL attacks across various settings and tasks while preserving the target model's overall accuracy. 
\name{} is also shown to be robust and effective against several defense-aware attacks, which may have full knowledge regarding the internal mechanisms of \name{}.

\end{abstract}

\section{Introduction}






In recent years, Deep Neural Networks (DNNs) have been pervasively applied to many domains, such as facial recognition~\cite{deep_face,deep_id}, autonomous driving~\cite{deep_auto2,auto_deep,deepbillboard,deeptest}, natural language processing~\cite{nlp,nlp_2,nlp_3,nlp_4}, binary code analysis~\cite{binary,binary_2,binary_3,binary_4}, bug detection~\cite{bug,bug_2,bug_3,bug_4}, etc.  However, a series of vulnerabilities have been discovered by researchers from both academia and industries~\cite{phsGAN,adv_gan,adv,face_adv,guo2020practical,nattack,trojan,cw,fgsm,pgd}, which prevent DNN deployment in safety-critical applications (e.g., autonomous driving~\cite{deepbillboard,deeptest,phsGAN,deep_auto2,bojarski2017explaining}, health robotics~\cite{robot,robot_2}). Among the existing vulnerabilities, generating adversarial examples ~\cite{cw,adv,adv_gan,fgsm,pgd,phsGAN,nattack} is one of the most critical issues, attracting attention from academia and industry~\cite{cw,adv_gan,adv_train,e_adv_train,trades,region-based}.

Adversarial attacks are broadly categorized into white-box attacks~\cite{cw,fgsm,pgd} and black-box attacks~\cite{nattack,DBLP:journals/corr/abs-1909-10773,hsja,ql,nattack,sfa,ba,pa}. White-box attacks assume that the attacker has full knowledge of the target model (i.e., architecture and parameters), enabling the attacker to craft adversarial samples through optimization methods with known target model's weights. In contrast, the black-box setting assumes the attacker is restricted only to send queries and observe their predictions from the target model~\cite{pa,nattack,ql,sfa,ba,DBLP:journals/corr/abs-1909-10773,hsja}. Depending on the information obtained from the target model's prediction, black-box attacks can be further divided into soft-label~\cite{nattack,ql} and hard-label attacks~\cite{pa,DBLP:journals/corr/abs-1909-10773,hsja,sfa}. Soft-label attacks ~\cite{nattack,ql} assume the attacker can obtain the probability distribution across classes for each given input, while hard-label attacks ~\cite{pa,DBLP:journals/corr/abs-1909-10773,hsja,sfa} assume  that the attacker only gets the final decision. We focus on hard-label attacks in this paper, which are more realistic in practice and may pose a  significant  threat to real-world applications due to the minimal knowledge requirements.

Till now, only two types of hard-label attacks have been proposed. 
The first type utilizes surrogate models~\cite{pa} but was proven to be impractical by recent studies \cite{chen,e_adv_train}. The second type, Query-Based  Hard-Label attack (shortened as QBHL attack throughout the paper), utilizes outputs from a set of carefully crafted queries. QBHL attacks have been empirically shown (see related works in Sec. \ref{sec:background}) to be effective in many real-world applications, such as Google Cloud~\cite{google_cloud} and Tencent Image Classification API~\cite{tencent}.

Compared to the rapid development of QBHL attack techniques, the defensive methods against QBHL attacks remain underexplored. There are very few works on defense against QBHL attacks in the literature. 
Two existing defenses~\cite{stateful,blacklight} against black-box QBHL attacks exhibit  inherent shortcomings with several  impractical assumptions, restricting  their applicability (see related works in Sec. \ref{sec:background}).
On the other hand, most defenses against white-box QBHL attacks can be bypassed by the latest QBHL black-box attack ~\cite{sfa}.
Empirically evidenced, the most robust existing defense methods against QBHL are the optimization-based approaches (e.g., adversarial training~\cite{adv_train} and TRADES~\cite{trades}). However, these approaches are not scalable in practice due to extensive re-training overheads, limiting their applicability. 

This paper proposes \name{}, a practical and generalizable framework for providing robustness against QBHL black-box attacks by exploring the properties of the gradient estimation step, which is an essential step within the most successful QBHL attacks. 
The key idea behind  \name{} is to identify when an attacker sends adversarial queries to estimate the gradient of the decision boundary in the target model. When an input is identified as an adversarial query,  \name{} introduces non-determinism in the output, ensuring that the underlying gradient is not inferable. Supported by a set of analytical reasoning (detailed in Sec.~\ref{sec:analytical}), \name{} can preserve the overall accuracy and provide robustness against QBHL attacks for the target model.

We evaluate \name{} on 4 state-of-the-art QBHL attacks: Boundary Attack (BA)~\cite{ba}, Sign-OPT Attack (Sign-OPT) ~\cite{DBLP:journals/corr/abs-1909-10773}, HopSkipJumpAttack (HSJA) ~\cite{hsja} and Sign-based Flip Attack (SFA) ~\cite{sfa}. For each attack, we evaluate the robustness of \name{} under four tasks: MNIST, CIFAR-10, GTSRB, and ImageNet in both $\ell_{2}$ and $\ell_{\infty}$ settings. Experimental results show that \name{} is effective against the state-of-the-art QBHL attacks, even under excessive query budgets.
Importantly, \name{} is shown to be robust against defense-aware attacks, even considering the worst-case scenario where an attacker knows the detailed internal mechanism of \name{}. 
Moreover, \name{} is shown to be superior compared to existing  defensive methods against QBHL attacks. 
Additionally, compared to optimization-based methods against black-box QBHL attacks, \name{} is shown to be more scalable as it incurs much lower training time and accuracy loss. 

In summary, our work contributes in the following ways:

\begin{itemize}
    
    \item  We propose a practical and generalizable defensive framework against black-box query-based hard-label  attacks via invalidating the essential gradient estimation step. \name{} is a certifiable defense, fundamentally supported by analytical reasoning on this invalidation process.
    
    
    \item We perform extensive experiments under various tasks and settings to evaluate \name{} against four state-of-the-art QBHL attacks. Results show \name{} significantly improves the robustness of the model against QBHL attacks in both $\ell_2$ and $\ell_{\infty}$ settings, even assuming an excessive query budget (50K queries).


    \item \name{} is shown to be robust and effective against defense-aware attacks where the attacker may know \name{}'s internal mechanisms. Also interestingly, \name{} can be combined with state-of-art white-box defensive methods to further enhance the robustness against query-based hard-label attacks. 
\end{itemize}

\section{Background}
\label{sec:background}
In the context of single-label classification, the inference of DNNs can be formulated as $y=\mathcal{F}(x;\theta)$. Specifically, a DNN  $\mathcal{F}(x;\theta)$ : $\mathbb{R}^d \to \mathbb{R}^{m}$ maps input $x \in [0,1]^d$ to $y \in \{ y \in [0,1 ]^{m}\mid\sum_{i=1}^{m}y_i=1\}$ through a series of computations. Each $y_i=\mathcal{F}_i(x;\theta)$ represents the predictive confidence score for $x$ belonging to label $i$ in the label set $[m] : \{ 1,...,m \}$. Typically, a DNN-based classifier $\mathcal{C}$ will assign input $x$ to the class $c$, i.e., $\mathcal{C}(x)=c:\{ c\in[m]\mid c=\argmax_i \mathcal{F}_i(x;\theta) \}$. A DNN model may either give the user $y\in[0,1]^{m}$ (soft-label) or a class $ c \in [m]$ (hard-label) for each input $x$. In this paper, we focus on the hard-label scenario.\\

\noindent\textbf{Definition of Query-Based Hard-Label Attack.} In the context of adversarial machine learning~\cite{cw,adv,hsja,sfa,adv_gan}, evasion attacks refer to the task of producing an adversarial example $x_t$ given an input $x^*$ with its correct label $c^*$. However, according to the goal of the attacker, evasion attacks can be further categorized into \textit{targeted} and \textit{untargeted} attacks.  The goal of an untargeted attack
is to craft $x_t$ which would be misclassified as any $c\in [m] \setminus {c^*}$ by the target model, whereas the goal of a targeted attack is to craft $x_t$, which would be misclassified as certain pre-specified $c_t \in [m] \setminus{c^*}$. Consistent with prior defensive methods, we focus on defending the untargeted attack since the untargeted attack is harder for the defender~\cite{adv_train,adv_2,cw}. 

We adopt two most common conditions an adversarial sample $x_t$ has to satisfy in the context of an untargeted attack: 
\begin{itemize}\label{condition}
\item $\mathcal{S}_{x^*}(x_t)\geq0$, where function $\mathcal{S}_{x^*}$ : $\mathbb{R}^{d}\to\mathbb{R}$ is defined as:  
\begin{equation}
\mathcal{S}_{x^{*}} (x_t) = \max\limits_{c \neq c^{*}} \mathcal{F}_{c}(x_t;\theta)- \mathcal{F}_{c^{*}}(x_t;\theta)~;\\
\end{equation}

\item $\ell_p=\norm{x_t-x^*}_p \leq \epsilon$ for $p \in \{ 2,\infty \} $. $\boldsymbol{\epsilon}$ is an arbitrarily small constant which ensures $x_{t}$ is visually-indistinguishable to $x^*$. 
\end{itemize}
To indicate success of $x_t$ against input $x^*$ a new Boolean-valued function is defined $\phi_{x^*} : [0, 1]^d \to \{-1, 1\} $
\begin{equation}
\label{eq:phi_def}
\phi_{x^*}(x_t) = \text{sign}(S_{x^*}(x_t)) = \left\{
             \begin{array}{lr}
                1,& \text{if} \; \mathcal{S}_{x^*}(x_t) > 0, \\
            
                 -1,& \text{otherwise.}
             
             \end{array}
\right.
\end{equation}

Evasion attacks are categorized into white-box and black-box attacks, based on the attacker's knowledge specifications. A white-box attacker has access to the target model $\mathcal{C}$ as well as its learned parameters $\theta$ ~\cite{cw,pgd,fgsm}. 
As a fundamental step to explore the vulnerabilities within DNNs, numerous white-box attacks have been proposed~\cite{cw,fgsm,pgd,phsGAN}.
Meanwhile, a black-box attacker is blind to the target model's structure and parameters. The attacker can only obtain the target model's output for a given input. Moreover, in the hard-label setting, the attacker can only get the final predicted label $c \in [m]$.

\noindent\textbf{Related Works.}  Prior works generate adversarial samples by observing the target model's outputs for a set of cleverly crafted queries ~\cite{pa,sfa,hsja,DBLP:journals/corr/abs-1909-10773,ba}. Papernot et al.~\cite{pa} present a practical black-box attack (PA), which crafts transferable adversarial samples by constructing (using queries) a surrogate model whose decision boundaries are close to the target model. However, Chen et al. ~\cite{chen} report PA to be ineffective and query-inefficient for large-scale datasets (CIFAR-10, ImageNet). Li et al.~\cite{ql} present an approach that can effectively estimate the target model's gradient in the soft-label setting but is ineffective and query-inefficient in the hard-label setting. More recent works ~\cite{ba,DBLP:journals/corr/abs-1909-10773,sfa,hsja} propose methods to improve  the  query-efficiency for hard-label attacks. Brendel et al.~\cite{ba} proposed the decision boundary attack (BA), which first initializes the adversarial image in the target class. 
Iteratively, the distance between the adversarial image and the original input is reduced by sampling a perturbation from a Gaussian or uniform distribution. BA returns the adversarial image, which is misclassified by the target model.
Cheng et al.~\cite{DBLP:journals/corr/abs-1909-10773} introduced the Sign-Opt attack, which finds a direction vector pointing towards the nearest decision boundary points from $x^*$. Chen et al. ~\cite{hsja} proposed HopSkipJumpAttack (HSJA), which improves upon the decision-boundary attack by introducing a novel gradient direction estimation, achieving the best performing hard-label attack in both $\ell_2$ and $\ell_\infty$ settings with a query budget of 5K. Chen et al. ~\cite{sfa} propose the Sign Flip attack (SFA), which utilizes an evolutionary algorithm~\cite{evolutionary} to improve query-efficiency.
In this work, we focus on these four state-of-art Query-based Hard-label attacks (QBHL): BA~\cite{ba}, Sign-OPT~\cite{DBLP:journals/corr/abs-1909-10773}, HSJA~\cite{hsja}, SFA~\cite{sfa}.    

Unfortunately, there is no generalizable and practical defense approach against hard-label black-box attacks. 
Recent related works~\cite{blacklight,stateful} focus on detecting black-box attacks by storing all the queries sent by an attack and classifying a user as malicious when similar queries are seen. 
These methods have two serious disadvantages. On the one hand, there is no evidence showing that an average user will not ask such queries. On the other hand, they fail to consider scenarios where an attacker could own many user/zombie accounts.
HSJA~\cite{hsja} proposed a potential defense method to assign an "Unknown" class for low confidence inputs, which is effective against targeted attacks but fails in the case of untargeted attacks. Finally, defensive methods for white-box attacks are ineffective against some query-based hard-label attacks~\cite{sfa,hsja}. Other effective defensive methods have several practical issues ~\cite{adv_train,e_adv_train}.
Our work is the first generalizable and practical approach that provides robustness against black-box query-based attacks to the best of our knowledge.

\section{Adversarial Model}
\subsection{Attack Model}
We considered a consistent model with the recent set of state-of-the-art hard-label attacks~\cite{hsja}, where the attacker aims to generate an adversarial sample $x_{t}$ from a given input image $x^*$. The attacker can only obtain the hard-label prediction on a limited ($\leq 20$K) number of inputs (queries) from the target model $\mathcal{C}$.

\subsection{Defense Model}

\noindent\textbf{Defender Goal.}
For each given DNN model $\mathcal{C}$, the defender aims to design a mechanism that prevents the attacker from crafting adversarial samples under the attack model while having a negligible effect on the performance (accuracy and computational cost) of $\mathcal{C}$.


\noindent\textbf{Defender Capability.} 
We assume the defender can only access the current query to achieve its goal. Furthermore, the defender has access to a validation set of images and their corresponding labels.

\section{Analytical Reasoning Supporting \name{}}
\label{sec:analytical}


This section describes our analysis of the gradient estimation method, which is standard for all state-of-the-art QBHL attacks~\cite{hsja,sfa,DBLP:journals/corr/abs-1909-10773,ba}. In addition, we prove why \name{} could provide certificated robustness against attacks such as HSJA~\cite{hsja}, Boundary Attack~\cite{ba}, and Sign-opt Attack~\cite{DBLP:journals/corr/abs-1909-10773} while preserving the overall accuracy of the target model. 


\subsection{Analysis of QBHL Gradient Estimation}
\label{sec:analysis}

\begin{figure*}[!t]
    \centering
    \includegraphics[width=1\textwidth]{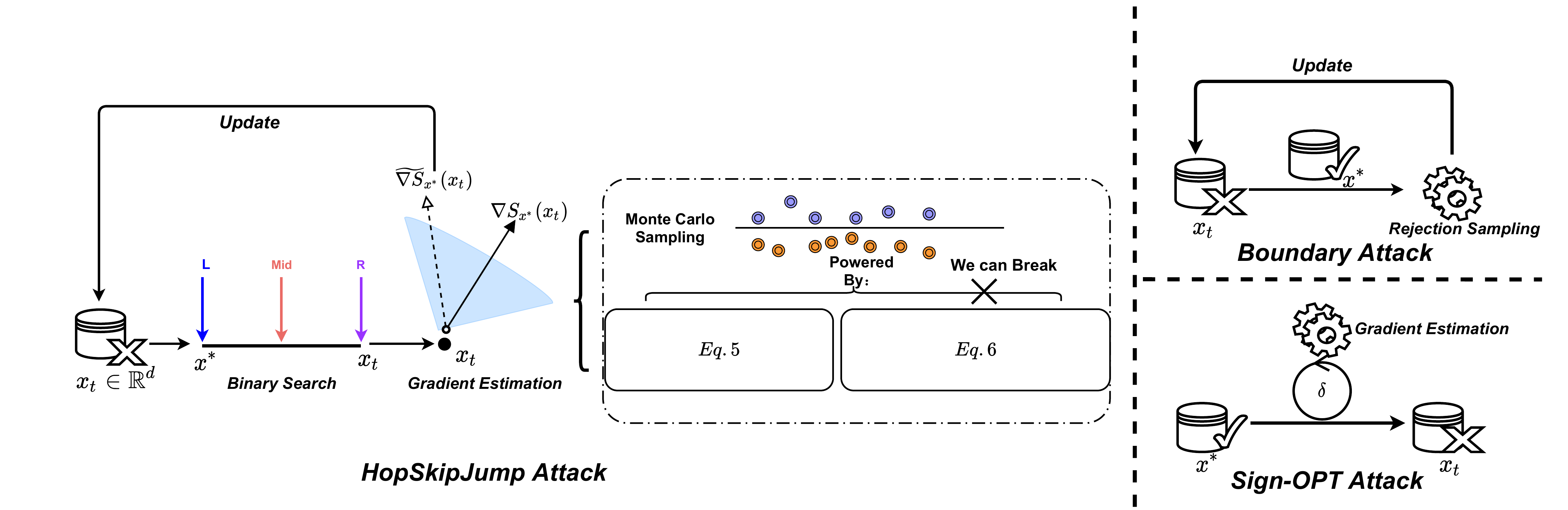}
    \caption{An Overview Illustration of Boundary Attack, Sign-opt, and HSJA.}
    \label{fig:overview}
\end{figure*}

\noindent\textbf{Overview of state-of-the-art QBHL attacks.} 
Fig. \ref{fig:overview} illustrates several QBHL techniques \cite{ba,hsja,DBLP:journals/corr/abs-1909-10773}, which contain various procedures (e.g., binary search, rejection sampling, gradient estimation) to ensure their effectiveness. Take the example of HSJA, which starts with a binary search to force the adversarial sample $x_t$ to approach a decision boundary  and then estimates the gradient $\frac{\nabla \mathcal{S}_{x^*}(x_{t})}{||\nabla \mathcal{S}_{x^*}(x_{t})||}$  through Monte Carlo sampling. Other QBHL techniques (BA, Sign-OPT) exhibit workflows that contain similar gradient estimations, which is essential for their efficacy. \footnote{Note that the rejection sampling step in the Boundary Attack (BA) which seeks the minimal $\delta$ for $x_{t}$, has the same goal as gradient estimation.} Intuitively, if we could invalidate the gradient estimation step, then the most successful QBHL techniques are mitigated.

As for HSJA, through theoretical analysis, we find that Monte Carlo sampling requires two critical conditions to be true for accurately estimating the gradient. As shown in the dashed box in Fig. \ref{fig:overview}, breaking one of the two conditions hinders the gradient estimation step.

\noindent\textbf{Analysis of Gradient Estimation.} 
Without loss of generality, we analyze the gradient estimation step used in HSJA as it is a state-of-the-art generalized QBHL technique useful in both $\ell_2$ and $\ell_\infty$ settings.
Technically, HSJA is incremental upon existing attacks~\cite{ba}, with an improved gradient estimation that utilizes binary information at the nearest decision boundary~\cite{hsja}. Intuitively, if \name{} could defend against HSJA by invalidating its gradient estimation, it would likely be resilient to other QBHL methods.
As shown in Fig. \ref{fig:overview}, HSJA's gradient estimation step aims to determine the direction of $\nabla S_{x^*}(x_{t})$ through Monte Carlo sampling:
\begin{equation}
    \label{eq:point_of_defense}
    \widetilde{\nabla S}_{x^*}(x_{t}, \delta) = \frac{1}{B}\sum_{b=1}^{B} \phi_{x^{*}}(x_{t}+\delta u_b)u_b,
\end{equation}
where $\{ u_b \}_{b=1}^{B}$ are i.i.d draws from a uniform distribution over a d-dimensional sphere and $\delta$ is a small positive parameter.


Theorem 2 from the HSJA paper \cite{hsja} proves that if $x_t$ is a boundary point of the decision function, then the following holds:

\begin{equation}
\label{eq:to_break}
 \lim_{\delta \to 0} cos\angle(\mathbb{E}[ \widetilde{\nabla S}_{x^*}(x_{t}, \delta)] ,\nabla \mathcal{S}_{x^*}(x_{t}))=1~.
\end{equation}

\noindent The proof for Eq. \ref{eq:to_break} is given in Appendix \ref{sec:theorem2} for reference. 
We analyze the steps in this proof to figure out a feasible defense methodology that could disrupt the computation of Eq. \ref{eq:point_of_defense}. Broadly, the gradient estimation works because $\mathbb{E}[|\beta_1|]$ satisfies the following equations:

\begin{equation}\label{eq:bounce11}
    ||\widetilde{\nabla S}_{x^*}(x_{t}, \delta)-\mathbb{E}[|\beta_1|v_1]||\leq 3q~,
\end{equation}
\begin{equation} \label{eq:bridge1}
    \mathbb{E} [|\beta_1|v_1]= \frac{1}{B} \sum_{b=1}^{B} |\beta_1^b|\frac{\nabla \mathcal{S}_{x}(x_{t})}{||\nabla \mathcal{S}_{x}(x_{t})||}=  \frac{\nabla \mathcal{S}_{x}(x_{t})}{||\nabla \mathcal{S}_{x}(x_{t})||}\mathbb{E}[|\beta_1|]~.
\end{equation}
\noindent where $\beta_1, v_1$, and $q$ are as defined in Appendix \ref{sec:theorem2}. Combining Eq. \ref{eq:bounce11} and Eq. \ref{eq:bridge1}, we have a bound on the accuracy of gradient estimate as:

\begin{equation}\label{eq:final1}
    cos\angle(\widetilde{\nabla S}_{x^*}(x_{t}, \delta) ,\nabla \mathcal{S}_{x}(x_{t})) \geq 1-\frac{1}{2}(\frac{3q}{\mathbb{E} [|\beta_1|]})^2
\end{equation}

Through further exploring the distribution of $q$, Eq.~\ref{eq:bounce11} can be bounded to 1, which is proved in Theorem~\ref{theorem:1} given in the Appendix. Thus, to ensure that the accuracy of gradient estimation is no longer bounded, we can disrupt Eq. \ref{eq:bounce11} and/or Eq. \ref{eq:bridge1}. We find it is much easier to manipulate Eq. \ref{eq:bridge1}, details of which are in the next section.


\subsection{Breaking the Estimate $\widetilde{\nabla\mathcal{S}}_{x^*}(x_{t})$}
\label{sec:break}

As previously stated,  \name{} transforms the model's behavior when the attacker tries to do gradient estimation (i.e., compute Eq. \ref{eq:point_of_defense}). We observe that  when a gradient estimation query $x_{t}+\delta u_b$ is sent to the model if we return $\overline{\mathcal{\phi}_{x^*}}(x_{t}+\delta u_b)$, which is defined as:

\begin{equation}\label{eq:manipulate}
    \overline{\phi}_{x^*}(x_{t}+\delta u_b)=\left\{
             \begin{array}{lr}
                \phi_{x^*}(x_{t}+\delta u_b) & with~ 0.5 ~ probability\\ 
             
                 -\phi_{x^*}(x_{t}+\delta u_b) & with ~0.5~ probability
             
             \end{array}
\right.
\end{equation}
\noindent instead of $\mathcal{\phi}_{x^*}(x_{t}+\delta)$ to the attacker, then the accuracy of gradient estimation drops significantly.  Plugging in $\overline{\mathcal{S}}_{x^*}$ (corresponding to $\overline{\phi}_{x^*}$) in place of $\mathcal{S}_{x^*}$ into Eq. \ref{eq:bridge1}, we have:

\begin{align}
    \mathbb{E} [| \beta_1 | v_1 ]  &= \frac{1}{B} \sum_{b=1}^{B} |\beta^{b}_1|\frac{\nabla\overline{ \mathcal{S}}_{x^*}(x_{t})}{||\nabla \overline{\mathcal{S}}_{x^*}(x_{t})||} \to 0 \label{eq:fi}.  
\end{align} 

\noindent Here $ \nabla \overline{\mathcal{S}}_{x^*}(x_t) $ denotes a random variable that reflects the  non-deterministic behavior introduced by  Eq. \ref{eq:manipulate}. Intuitively, and through empirical observations, since $\overline{\phi}_{x^*}$ flips with a probability of 50\%, the sum of $ \nabla \overline{\mathcal{S}}_{x^*}(x_t) $ over randomly sampled queries should approach zero.  Hence, by applying our defense approach, $\mathbb{E}[ |\beta_{1}|v_1]$ becomes irrelevant to the actual $\nabla \mathcal{S}_{x^*}(x_{t})$  and $\mathbb{E}[| \beta_1 |] \to 0$.  This ensures that the accuracy of the gradient estimate is no longer bounded (Eq. \ref{eq:final1}). Hence the computation of Eq. \ref{eq:to_break} results in inaccurate gradients estimates ( i.e., $cos \angle (\widetilde{\nabla \overline{ S}_{x^*}}(x_{t}) ,\nabla \overline{\mathcal{S}}_{x^*}(x_{t})) \geq -1$).

Replacement of  $\phi_{x^*}$ with $\overline{\phi}_{x^*}$  also affects the accuracy of the target model. Hence, the mechanism for detecting these queries need to be very accurate. To address this issue, we use a second neural network $\mathcal{F_Q}$ as a classifier to help distinguish between regular inputs ($x^{*}$) and queries for gradient estimation ($x_{t}+\delta u_b$ or $x_t$). From our evaluations (see Sec. \ref{sec:build}), we know that such an $\mathcal{F_Q}$ can distinguish between normal inputs and adversarial samples with the accuracy of $s \; (\geq 97\%)$. The expected accuracy loss ($\bigtriangleup _{acc}$) of the target model can be bounded as follows:

\begin{equation}
     \mathbb{E} [\bigtriangleup_{acc} ]\ \leq \frac{1-s}{2},
\end{equation}

\noindent As $s \to 100\%$, \name{} will preserve the target model's accuracy.

\subsection{Effects of \name{} on Other Components}
In the case of HSJA, \name{} also affects the steps other than gradient estimation, namely the binary searches and the geometric progression, since they also involve sending queries to the target model. While those methods ensure HSJA's efficacy, they are not as essential as gradient estimation, as evidenced by its prevalence in other QBHL techniques. 

SFA~\cite{sfa} utilizes an evolutionary algorithm~\cite{evolutionary} to improve its query-efficiency and is very effective in the $\ell_{\infty}$ setting. Still, our approach provides significant robustness against SFA, as shown in Sec. \ref{sec:exp}.

\section{Design of \name{}}

\begin{figure*}[!t]

    \includegraphics[width=1\textwidth]{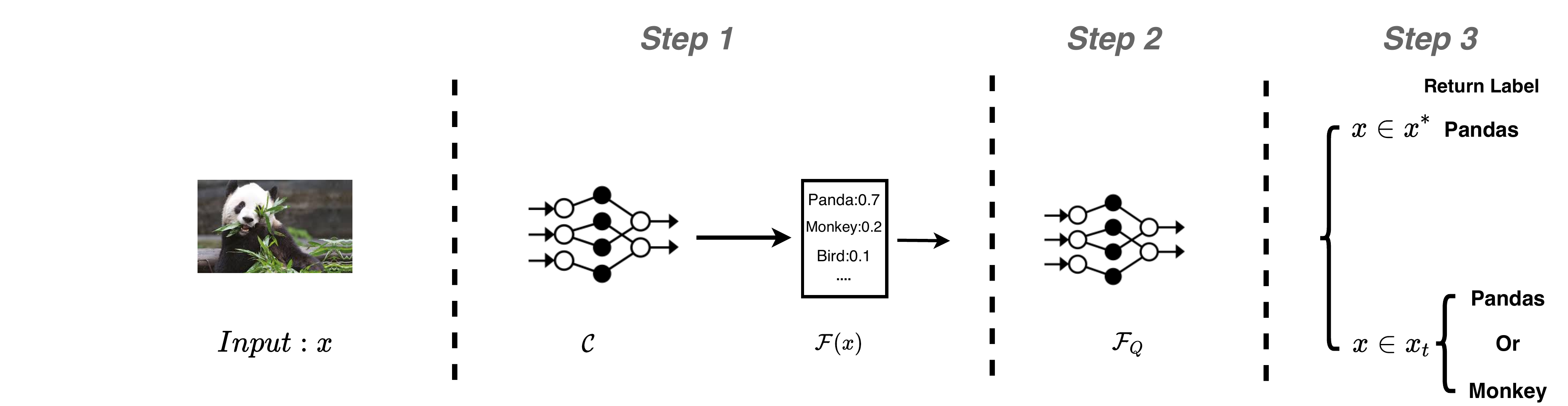}
    \caption{The Architecture of \name{} }
    \label{fig:arch}
\end{figure*}

\subsection{\name{} Overview}
\label{sec:overview}

In this section, we describe the design of \name{}. 
\name{} is an inference phase defensive method consisting of two DNN models: $\mathcal{F_Q}$, which determines if an input ($x$) is adversarial ($x_t$), and $\mathcal{C}$, which is the same as the target model. The architecture of \name{} is illustrated in Fig. \ref{fig:arch}. For each input $x$, $\mathcal{C}$ is used to calculate $\mathcal{F}(x;\theta)$. Next, $\mathcal{F_Q}$ uses $\mathcal{F}(x;\theta)$ to determine if $x$ is adversarial. Finally, if $x$ is adversarial with a 50\% probability, we replace $\mathcal{C}(x)$ with the label having the second-largest confidence score within $\mathcal{F}(x;\theta)$, else we predict the label of $x$ as $\mathcal{C}(x)$. Note that all the above operations can be done in parallel on a GPU for batches of inputs without I/O between devices, ensuring low inference computational cost. $\mathcal{F_Q}$, as supported by our analysis in Sec. \ref{sec:break}, needs to very accurate. In the next section, we describe how to create such an effective $\mathcal{F_Q}$.  

\subsection{Creating $\mathcal{F_{Q}}$}
\label{sec:build}

In the following, we investigate two options to construct the detection network $\mathcal{F_Q}$. As we will see, Options 2 is superior due to its detection capabilities and independence of $\mathcal{C}$.

\vspace{2mm}
\noindent\textbf{Option 1: Taking $x$ as input.}
A straightforward way to build $\mathcal{F_{Q}}$, similar to current methods for detecting adversarial samples~\cite{carlini2017adversarial},  is to take $x$ as input and output a binary variable that determines if it is adversarial. Here, $\mathcal{F_Q}$ manifests as a DNN with a similar structure as $\mathcal{C}$ except for the final layer.

We first create $\mathcal{D}^{'}_{tr}:\{ \mathcal{G}(x)~|~x \in \mathcal{D}_{tr} \}$  and  $\mathcal{D}^{'}_{te}:\{ \mathcal{G}(x)~|~x \in \mathcal{D}_{te} \}$ for the training phase, where $\mathcal{G}(x)$ represents the Monte Carlo sampling procedure in HSJA. The Monte Carlo sampling takes a regular input $x^*$ and returns a randomly selected adversarial sample $x_t$. $\mathcal{D}_{tr}$ and $\mathcal{D}_{te}$ are the training and testing datasets for $\mathcal{C}$.
We label $\mathcal{D}_{tr}$, $\mathcal{D}_{te}$ as $x^*$, and $\mathcal{D}^{'}_{tr}$, $\mathcal{D}^{'}_{te}$ as $x_t$,
and train $\mathcal{F_{Q}}$ with $\mathcal{D}_{tr} \cup \mathcal{D}^{'}_{tr} $.
During the validation procedure, we observe that $\mathcal{F_{Q}}$ has low accuracy on $\mathcal{D}_{te} \cup \mathcal{D}^{'}_{te}$ ($ \leq 71\%$) with ResNet50~\cite{resnet} on the CIFAR-10 dataset; which might be due to extreme similarities between $\mathcal{G}(x)$ and $x$, especially under an untargeted attack scenario, which may lead to overfitting. 
However, similar to related works \cite{grosse2016adversarial,gong2017adversarial}, this approach requires a large amount of training data that belongs to the same distribution as $\mathcal{D}_{tr}$, thus limiting its real-world applicability.




\begin{table}[!t]
\centering
\begin{tabular}{ c c c  } 
\hline
Layer Type & Model Size\\
\hline
Full Connected~(Input Layer)& the size of $\mathcal{F}(x;\theta)$\\
Fully Connected + ReLU&64\\
Fully Connected + ReLU&64\\
Fully Connected + ReLU&32\\
Softmax~(Output Layer)&2\\
\hline
\end{tabular}
\caption{DNN Structure of $\mathcal{F}^i_\mathcal{Q}$}
\label{table:F_Q}

\end{table}

\noindent\textbf{Option 2: Taking $\mathcal{F}(x;\theta)$ as input.} To address the challenge of distinguishing $x_t$ from $x^*$, we infer additional information from the given input. Membership Inference Attack (MIA) \cite{shokri2017membership} successfully identifies the distribution of an input based on its prediction vector $\mathcal{F}(x;\theta)$ from the target model $\mathcal{C}$. Motivated by this approach, we build $\mathcal{F'_Q}$ through  shadow models $\mathcal{F}_{Q}^{i}$($i\in[1,m]$)$:\mathbb{R}^{m}\to\mathbb{R}^{2}$ for each label $i$ of $\mathcal{C}$, which takes $\mathcal{F}(x;\theta)$ as input and gives a score vector $(y_{x_t}^{1},y_{x_t}^{2})$, where $y_{x_t}^{1}$ and $y_{x_t}^{2}$ represent the confidence of $x$ predicted as $x_t$ and $x^*$ respectively, such that $y_{x_t}^{1}+y_{x_t}^{2}=1$ and  $y_{x_t}^{1}$, $y_{x_t}^{2}$ $\in (0,1)$. The DNN structure of $\mathcal{F}_{Q}^{i}$ is shown in Table \ref{table:F_Q}. We define:
$$\mathcal{F'_Q}(\mathcal{F}(x;\theta))= \mathcal{F}_{Q}^{i} (\mathcal{F}(x;\theta))~\text{where}~ i = \mathcal{C}(x)$$

To validate our approach, we follow the MIA training configurations and define the following four sets:
   $$\mathcal{F}_{tr}^{'i}:\{ \mathcal{F}(x;\theta)~|~x \in  \mathcal{D}^{'}_{tr} \wedge i=\mathcal{C}(x) \}~,$$
   $$\mathcal{F}_{tr}^i:\{ \mathcal{F}(x;\theta)~|~x \in  \mathcal{D}_{tr} \wedge i=\mathcal{C}(x) \}~,$$
   $$\mathcal{F}_{te}^{'i}:\{ \mathcal{F}(x;\theta)~|~x \in  \mathcal{D}^{'}_{te}  \wedge i=\mathcal{C}(x) \}~,$$
   $$\mathcal{F}_{te}^i: \{ \mathcal{F}(x;\theta)~|~x \in  \mathcal{D}_{te}  \wedge i=\mathcal{C}(x) \}~,$$
where $\mathcal{D}^{'}_{tr}$ and $\mathcal{D}^{'}_{te}$ are obtained in the same way as in option 1. In an iteration, we train and validate each $\mathcal{F}_{Q}^i \in \mathcal{F'_Q}$ with $\mathcal{F}_{tr}^{'i} \cup \mathcal{F}_{tr}^i $ and $\mathcal{F}_{te}^{'i} \cup \mathcal{F}_{te}^i $, respectively. We label $\{\mathcal{F}(x;\theta)~|~x\in x^*\}$ and $\{\mathcal{F}(x;\theta)~|~x\in x_t\}$ as $x^*$ and $x_t$, respectively. We use two metrics to evaluate the performance of the MIA-based approach:
\begin{itemize}
    \item \textbf{False positive rate (FP).} Number of times $x^*$ is incorrectly identified as $x_t$ by $\mathcal{F'_Q}$, which influences $\bigtriangleup _{acc}$.
    \item \textbf{False negative rate (FN).} Number of times $x_t$ is misclassified as $x^*$, which influences the defense efficacy.
\end{itemize}
We observe each $\mathcal{F'_Q}$ achieves $\leq 2.5\%$ FP and FN on ResNet for CIFAR-10 dataset with an overall accuracy $>97.7\%$. Empirically, taking only $0.05$ times the original size of $\mathcal{D}_{tr}$ for $\mathcal{C}$ as training data is sufficient to achieve the above results.

However, the computation of the MIA-based approach $(\mathcal{F'_Q})$ grows proportional to $m$ at the training time. To reduce the computation cost, instead of using shadow models like $\mathcal{F}_{Q}^i$, which depend upon $\mathcal{C}(x)$, we propose to directly build a single DNN, $\mathcal{F_Q}$  which takes a modified version of $\mathcal{F}(x;\theta)$ as the input, and outputs the combined scores $(y_Q^{1},y_Q^{2})$ directly. $\mathcal{F_Q}$'s design and computation cost are independent of $\mathcal{C}$. From our experiments, this new $\mathcal{F_Q}$ achieves performance similar to $\mathcal{F}_{Q}^i$.

By observation, we find in most cases the sum of the largest three $\mathcal{F}_{i}(x;\theta)$ is close to 1. So we craft $\overline{\mathcal{F}}(x;\theta)$ by selecting the largest three  $\mathcal{F}_{i}(x;\theta)$ and re-arranging them in the descending order to replace $\mathcal{F}(x;\theta)$ as the input for $\mathcal{F_Q}$. Thus, we have $\mathcal{F}_{Q}:\mathbb{R}^{3}\to\mathbb{R}^{2}$, with $\mathcal{F_Q}(\overline{\mathcal{F}}(x ; \theta)) = (y^1_Q, y^2_Q)$. We train and test the proposed $\mathcal{F_Q}$ with $\overline{\mathcal{F}_{tr}}: \{ \overline{\mathcal{F}}(x;\theta)~|~x \in \mathcal{D}_{tr} \} \cup \overline{\mathcal{F}^{'}_{tr}}: \{ \overline{\mathcal{F}}(x;\theta)~|~x \in \mathcal{D}_{tr}^{'} \}$  and $\overline{\mathcal{F}_{te}}: \{ \overline{\mathcal{F}}(x;\theta)~|~x \in \mathcal{D}_{te} \} \cup \overline{\mathcal{F}^{'}_{te}}: \{ \overline{\mathcal{F}}(x;\theta)~|~x \in \mathcal{D}_{te}^{'} \}$ to validate our approach. The structure of $\mathcal{F_Q}$, is the same as $\mathcal{F}_{Q}^{i}$ (shown in Table \ref{table:F_Q}), except for the input layer size, which becomes 3.
\begin{table}[!t]
\centering
\begin{tabular}{|c|c|c|c|}
\hline
Attack Method & FP($\%$)   & FN($\%$)  & Accuracy($\%$) \\ \hline \hline
BA            & 2.16 & 8.82 &94.51\\ \hline
Sign-OPT      & 2.16 & 2.42 &97.71\\ \hline
HSJA          & 2.16 & 2.41 &97.71\\ \hline
SFA           & 2.16 & 2.47 &97.69\\ \hline
\end{tabular}
\caption{\textbf{False Positive}, \textbf{False Negative} rates of $\mathcal{F_Q}$ and \textbf{Accuracy} of \name{} under various attack techniques.}
\label{table:fn}
\end{table}

 Finally, We test the improved version of $\mathcal{F_Q}$ on validation data generated from  other attack techniques for CIFAR-10 task, as shown in Table. \ref{table:fn}. $\mathcal{F_Q}$ performs similarly for Sign-Opt attack, SFA, and HSJA, while having higher FN for BA ($\approx 9\%$). 
 However, the gradient estimation within BA is less effective than the other three methods (revealed by its query-inefficiency), which counteracts the high FN rate. In Sec \ref{sec:exp}, we show that our proposed $\mathcal{F_Q}$ provides significant robustness against all four attack techniques. Note that, since we use the same base images to evaluate each QBHL attack, the FP rate under each attack shall be the same (i.e., 2.16). Moreover, to better balance the trade-off between FP and FN rates for various tasks, we introduce a threshold value $\gamma$ for $y_Q^{1}$. Specifically, if $y_Q^{1} \geq \gamma$, $x$ will be predicted as $x_t$. In Sec. \ref{sec:trade-off}, we further investigate the trade-off between  $\bigtriangleup_{acc}$ and the robustness of \name{} by adjusting  $\gamma$.

To summarize, the improved version of $\mathcal{F_Q}$ described in Option 2 has two advantages: (\textit{i}) it is rather effective in distinguishing $x_t$ and $x$, and (\textit{ii}) it yields low and model-agnostic computation costs both at the training and inference phases.

\begin{table*}[h!]
\centering
\scalebox{0.8}{
 \begin{tabular}{||c |c |c |c |c |c|c||} 
 \hline
 Task & Datasets & $\#$ of Labels & Input Size & \makecell{$\#$ of Training\\  Images} & Model Architecture & Accuracy\\ [0.5ex] 
 \hline\hline
 \makecell{Hand-written Digit\\ Recoginition}& MNIST & 10 & 28x28x1 & 60,000& 3 Conv + 2 Dense&$99.72\%$\\ 
 \hline
\makecell{Traffic Sign\\ Recoginition}& GTSRB & 43 & 32x32x3 & 35,288& 5 Conv + 3 Dense&$98.01\%$\\
\hline
\makecell{Object Classification}& CIFAR-10 & 10 & 32x32x3 & 50,000& ResNet.50&$92.41\%$\\
\hline
\makecell{Object Classification}& ImageNet & 1,000 & 224*224*3 & 1,281,167& \makecell{ResNet.50}&$73\%$\\
\hline
 \end{tabular}}
  \caption{Detailed dataset information and model architecture for each task.}
 \label{table:tasks} 
\end{table*}


\begin{figure}[!t]
    \centering
    \scalebox{0.8}{
    
    \includegraphics[width=0.6\textwidth]{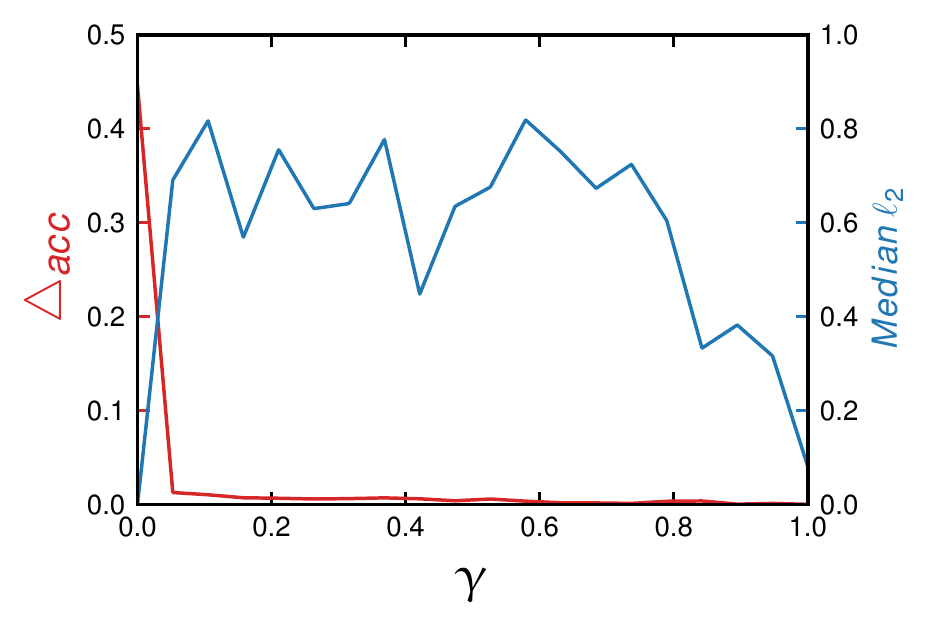}}
    \caption{Behavior of $\bigtriangleup_{acc}$ and Median $~\ell_2$ distance with $\gamma$.}
    \label{fig:trade_off}
\end{figure}

\begin{algorithm}[h]
    \DontPrintSemicolon
    \LinesNumbered
    \SetKwInOut{Input}{input}
    \SetKwInOut{Output}{output}
    \SetKwComment{Comment}{\#}{}

    \Input{Target Model $\mathcal{C}$ AND Model $\mathcal{F_Q}$}
    \Comment{Validation Data is used to calculate $\bigtriangleup_{acc}$}
    \Input{Validation Data}

    Set \textbf{$\gamma_{low}$}=0;
    
    Set \textbf{$\gamma_{high}$}=1;

    \While{|\textbf{$\gamma_{high}$}-\textbf{$\gamma_{low}$}|$\geq 0.01$}{
        Set $\gamma \leftarrow \frac{\gamma_{high}+\gamma_{low}}{2}$
        
        \If{$\bigtriangleup_{acc} \leq 10\%$}
        {
        
        Set $\gamma_{low} \leftarrow \gamma$
        
        }
        \Else
        {
        Set $\gamma_{high} \leftarrow \gamma$
        }
            }
    \Return{$\gamma$}

    \caption{\textbf{Binary Search For Searching An Appropriate $\gamma$}}
    \label{alg:binary_search}
\end{algorithm}

\begin{table*}[!t]
\centering
\scalebox{0.8}{

\begin{tabular}{|c|l|c|l|c|c|c|c|c|c|c|c|}
\hline
\multicolumn{2}{|c|}{\multirow{3}{*}{$p$}} & \multicolumn{2}{c|}{\multirow{3}{*}{Task}} & \multicolumn{8}{c|}{Model Queries}                                      \\ \cline{5-12} 
\multicolumn{2}{|c|}{}                          & \multicolumn{2}{c|}{}                      & \multicolumn{4}{c|}{30K}           & \multicolumn{4}{c|}{50K}           \\ \cline{5-12} 
\multicolumn{2}{|c|}{} &
  \multicolumn{2}{c|}{} &
  BA &
  \begin{tabular}[c]{@{}c@{}}Sign-\\ OPT\end{tabular} &
  SFA &
  HSJA &
  BA &
  \begin{tabular}[c]{@{}c@{}}Sign-\\ OPT\end{tabular} &
  SFA &
  HSJA \\ \hline\hline
\multicolumn{2}{|c|}{\multirow{4}{*}{$\ell_{2}$}}       & \multicolumn{2}{c|}{MNIST}                 &\textbf{4.995}(1.780)  & \textbf{8.066}(1.341)  & \textbf{4.123}(2.231) & \textbf{2.851}(1.510) &\textbf{4.170}(1.603)  & \textbf{7.256}(1.339)  & \textbf{3.161}(2.29)  & \textbf{2.434}(1.510) \\ \cline{3-12} 
\multicolumn{2}{|c|}{} &
  \multicolumn{2}{c|}{CIFAR-10} &
  \textbf{1.673}(0.1695) &
  \textbf{1.525}(0.106) &
  \textbf{2.423}(0.142) &
  \textbf{0.564}(0.098) &
   \textbf{1.119}(0.132)&
  \textbf{1.523}(0.102) &
  \textbf{2.398}(0.134) &
  \textbf{0.431}(0.094) \\ \cline{3-12} 
\multicolumn{2}{|c|}{}                          & \multicolumn{2}{c|}{GTSRB}                 &  \textbf{3.743}(0.667)&\textbf{4.966}(0.494)  &  \textbf{5.585}(0.743)            & \textbf{1.854}(0.462)             & \textbf{2.571}(0.624) &   \textbf{4.714}(0.374)&              \textbf{5.383}(0.701)&  \textbf{1.549}(0.4511)            \\ \cline{3-12} 
\multicolumn{2}{|c|}{}                          & \multicolumn{2}{c|}{ImageNet}              & \textbf{11.298}(2.716) &  \textbf{9.762}(1.134) &            \textbf{38.602}(5.111)  &\textbf{8.425}(1.019)             &  \textbf{10.180}(2.709)& \textbf{9.201}(0.912)  &   \textbf{39.048}(2.295)           &      \textbf{7.214}(0.842)        \\ \hline\hline
\multicolumn{2}{|c|}{\multirow{4}{*}{$\ell_{\infty}$}}       & \multicolumn{2}{c|}{MNIST}                 &  \textbf{0.578}(0.37)& - & \textbf{0.550}(0.251) & \textbf{0.544}(0.193) & \textbf{ 0.565}(0.351)& - & \textbf{0.554}(0.251) & \textbf{0.513}(0.186) \\ \cline{3-12} 
\multicolumn{2}{|c|}{}                          & \multicolumn{2}{c|}{CIFAR-10}              &\textbf{0.082}(0.018)  & - & \textbf{0.056}(0.003) & \textbf{0.065}(0.006) & \textbf{0.074}(0.0166) & - & \textbf{0.055}(0.003) & \textbf{0.043}(0.006) \\ \cline{3-12} 
\multicolumn{2}{|c|}{}                          & \multicolumn{2}{c|}{GTSRB}                 &  \textbf{0.310}(0.080)& - & \textbf{0.166}(0.018)&            \textbf{ 0.191 }(0.035)&  \textbf{0.258}(0.072)& - &  \textbf{0.166}(0.017)            &\textbf{0.169}(0.031)              \\ \cline{3-12} 
\multicolumn{2}{|c|}{}                          & \multicolumn{2}{c|}{ImageNet}           &  \textbf{0.196}(0.030)   & - &    \textbf{0.127}(0.003)          &        \textbf{0.137}(0.007)      &  \textbf{0.166}(0.030)  & - &   \textbf{0.126}(0.0027)        &\textbf{0.122}(0.006)               \\ \hline
\end{tabular}}
\vspace{2mm}
\caption{Median $\ell_p$ distances under various queries budgets. The bold values are results with our approach incorporated and the values inside the brackets are results for the standalone target model.}
\vspace{3mm}
 \label{table:median_distance} 
\end{table*}

\section{Experiments \& Evaluation}
\label{sec:exp}
This section evaluates the overall performance of \name{} against four existing Hard-label attacks on five prevalent tasks. Then, we compare the efficacy of \name{} against several existing defense or detection mechanisms. We perform all experiments using a system with an Intel(R) Xeon(R) CPU E5-2650 CPU and an  NVIDIA 2080 Ti GPU and report results as averages over ten runs. 
\subsection{Experimental Setup}

\noindent\textbf{Experimental Datasets.}
We evaluate \name{} against existing attacks with four prevalent tasks, including Hand-written
Digit Recognition (\textit{MNIST}),  Object classification (\textit{CIFAR-10}), Traffic sign recognition (\textit{GTSRB}),  and Object classification (\textit{ImageNet}).

We describe each task and its corresponding datasets below, with a summary in Table \ref{table:tasks}. We include the detailed model architectures and corresponding training configurations for each task in Tables: \ref{table:MNIST1}, \ref{table:MNIST2}, and \ref{table:training_configuration} in the Appendix due to space constraints.
\begin{itemize}
    \item Hand-written Digit Recognition (\textit{MNIST}). Recognize handwritten gray-scale images of 10 digits (0-9)~\cite{deng2012mnist}. The dataset contains 60,000 training samples and 10,000 testing samples. This task's test model is a plain convolution network consisting of 3 convolution layers and 2 fully connected layers.  

    \item Object Classification (\textit{CIFAR-10}). Classify various images into 10 different classes (e.g., plane, truck). The dataset contains 50,000 training images and 10,000 testing images. We implement the state-of-art model ResNet50 and crop every image to 32x32x3 for this task.
    
    \item Traffic Sign Recognition (\textit{GTSRB}). Recognize 43 German traffic signs in color images. The dataset contains 35,288 training images and 12,600 testing images. Every image is cropped to 32x32x3, and an 8-layer convolution network is used as the test model.
    
    \item Object Classification (\textit{ImageNet}). ImageNet is a large-scale image classification task, which contains 1,281,167 training inputs across 1,000 classes and 50,000 test inputs. For the ImageNet task, we implement ResNet50 with the input size as 224x224x3 to evaluate our approach.
\end{itemize}

\begin{figure*}[!t]
\centering
\subfigure[MNIST(HSJA)\label{fig:mnist_hsja}]{\includegraphics[width=0.24\textwidth]{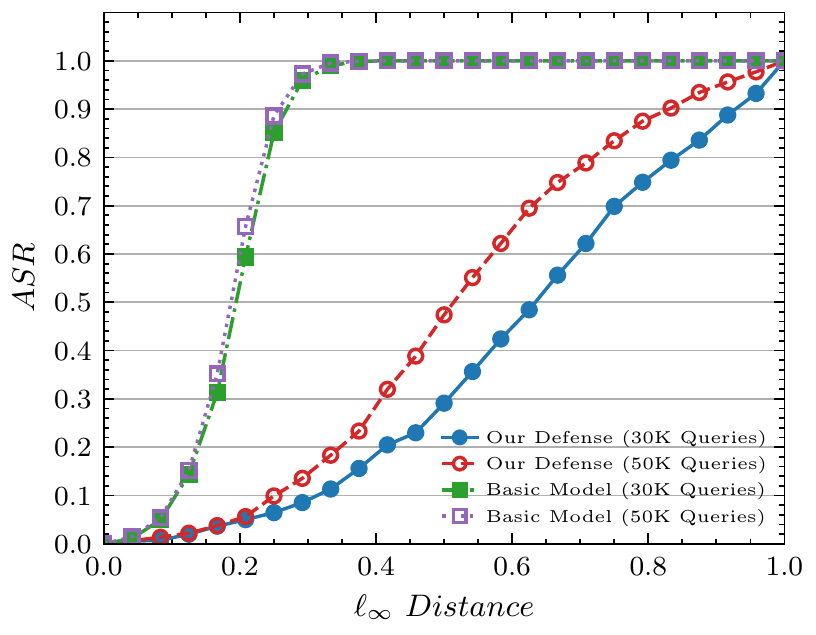}}
\subfigure[MNIST(HSJA)\label{fig:mnist_ba_li}]{\includegraphics[width=0.24\textwidth]{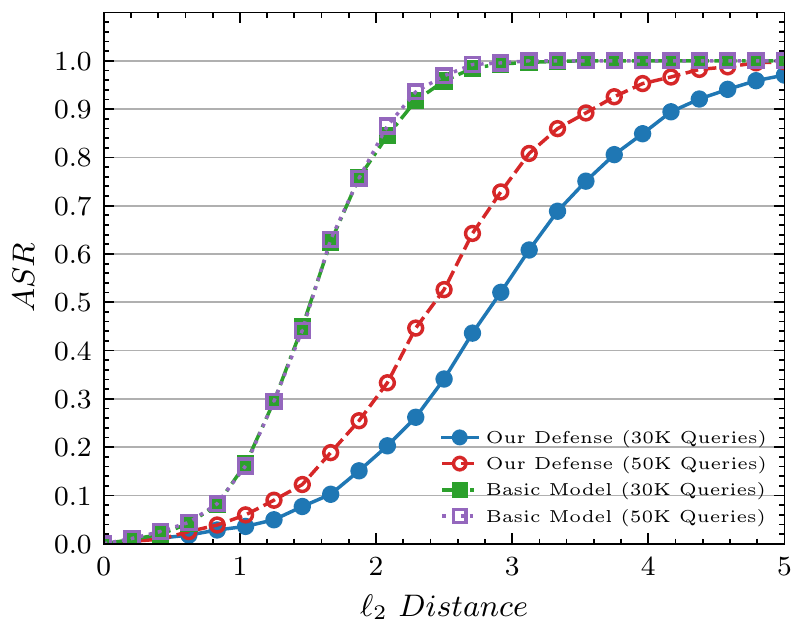}}
\subfigure[CIFAR-10(HSJA)\label{fig:cifar_ba}]{\includegraphics[width=0.24\textwidth]{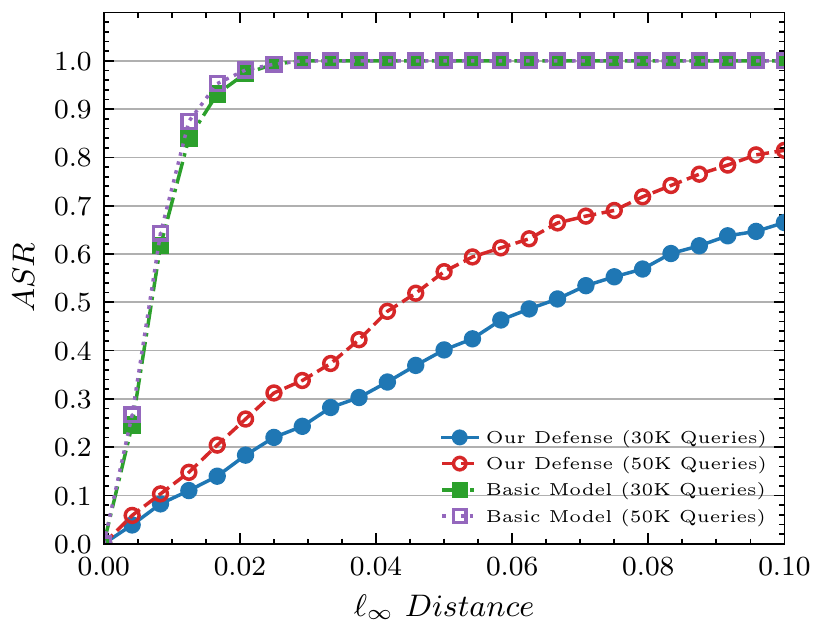}}
\subfigure[CIFAR-10(HSJA)\label{fig:cifar_ba_li}]{\includegraphics[width=0.24\textwidth]{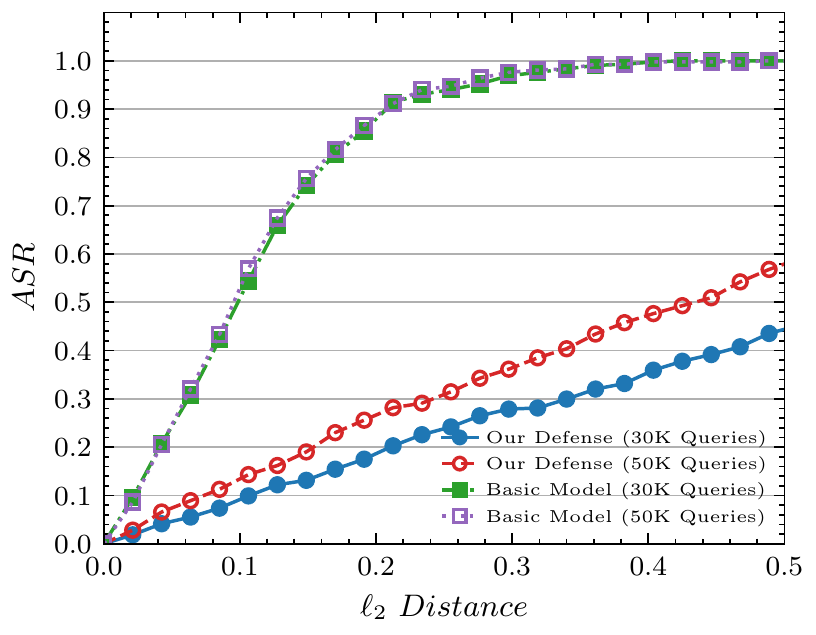}}
\subfigure[ImageNet(HSJA)\label{fig:imagenet_hsja}]{\includegraphics[width=0.24\textwidth]{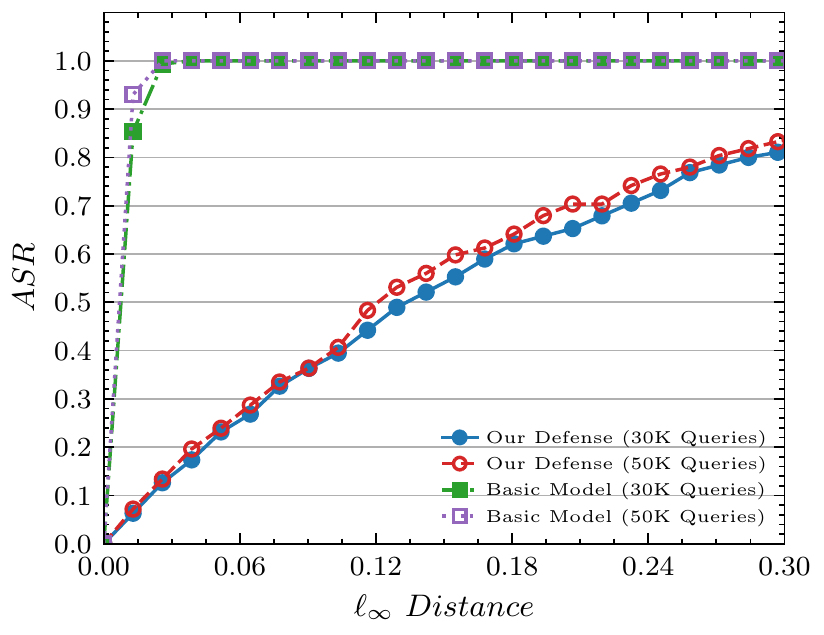}}
\subfigure[ImageNet(HSJA)\label{fig:imagenet_hsja_l2}]{\includegraphics[width=0.24\textwidth]{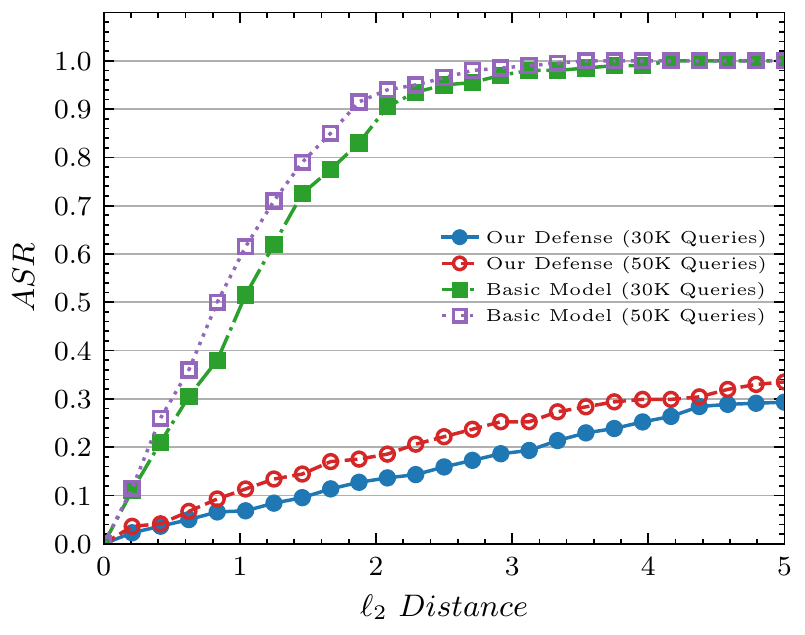}}
\subfigure[MNIST(SFA)\label{fig:mnist_hsja}]{\includegraphics[width=0.24\textwidth]{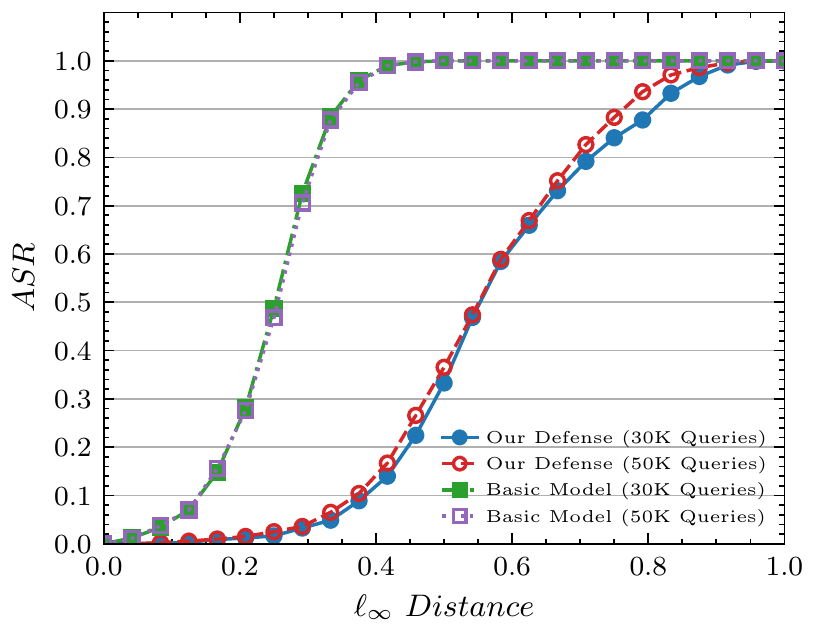}}
\subfigure[MNIST(SFA)\label{fig:mnist_ba_li}]{\includegraphics[width=0.24\textwidth]{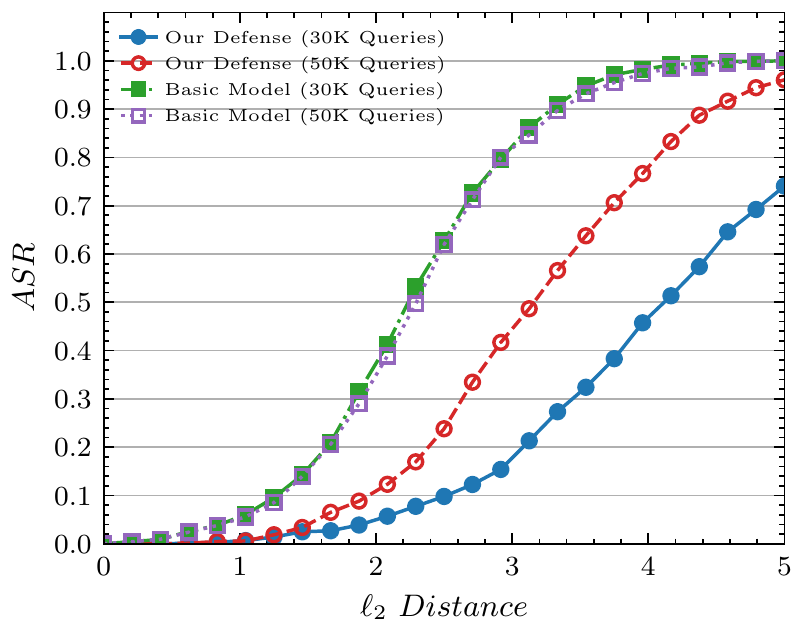}}
\subfigure[CIFAR-10(SFA)\label{fig:cifar_ba}]{\includegraphics[width=0.24\textwidth]{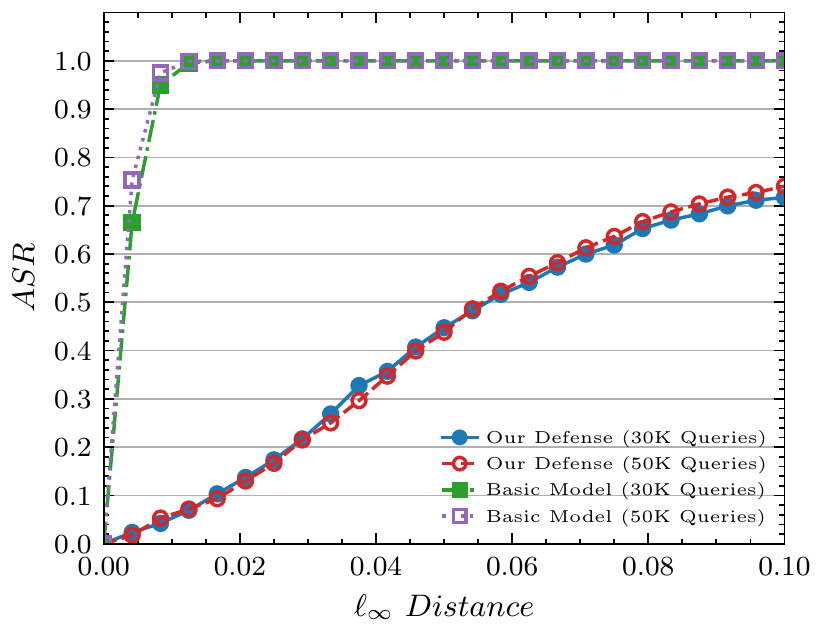}}
\subfigure[CIFAR-10(SFA)\label{fig:cifar_ba_li}]{\includegraphics[width=0.24\textwidth]{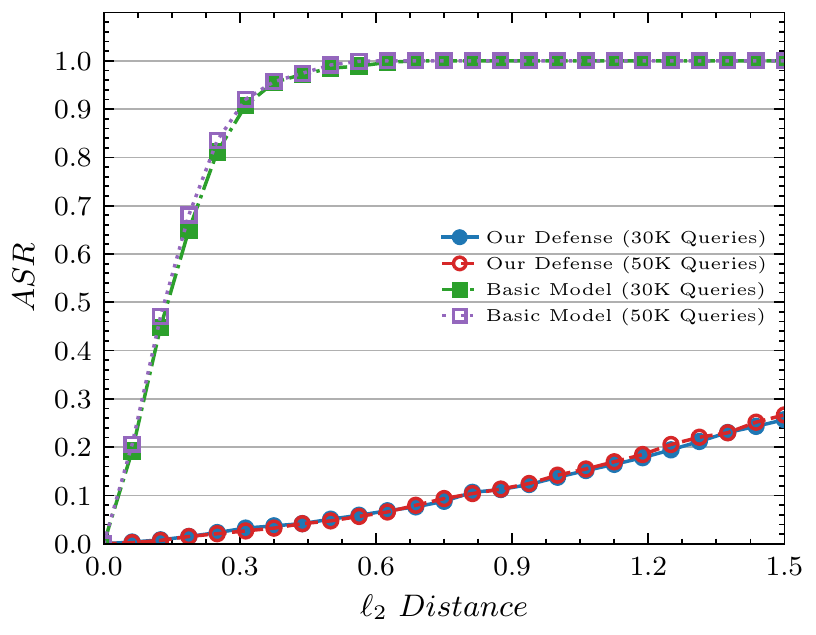}}
\subfigure[ImageNet(SFA)\label{fig:imagenet_hsja}]{\includegraphics[width=0.24\textwidth]{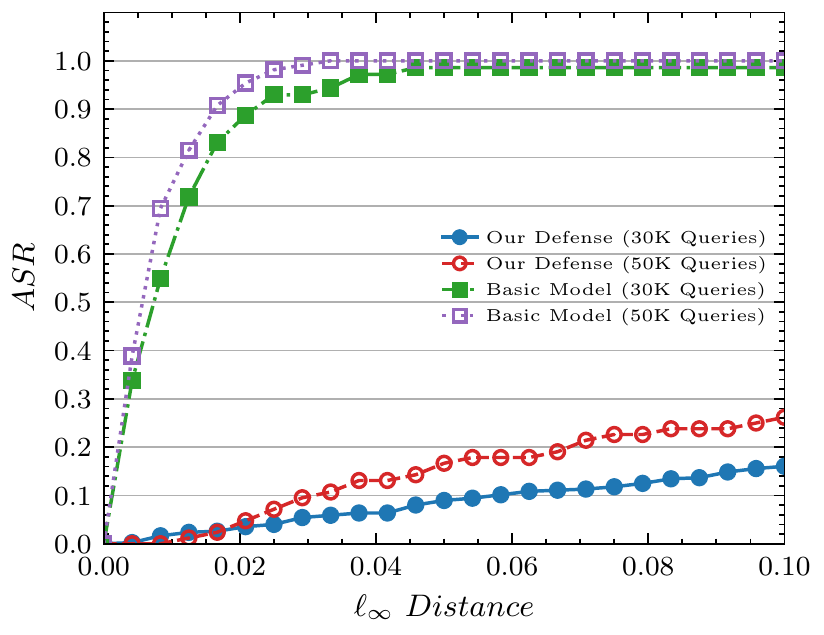}}
\subfigure[ImageNet(SFA)\label{fig:imagenet_hsja_l2}]{\includegraphics[width=0.24\textwidth]{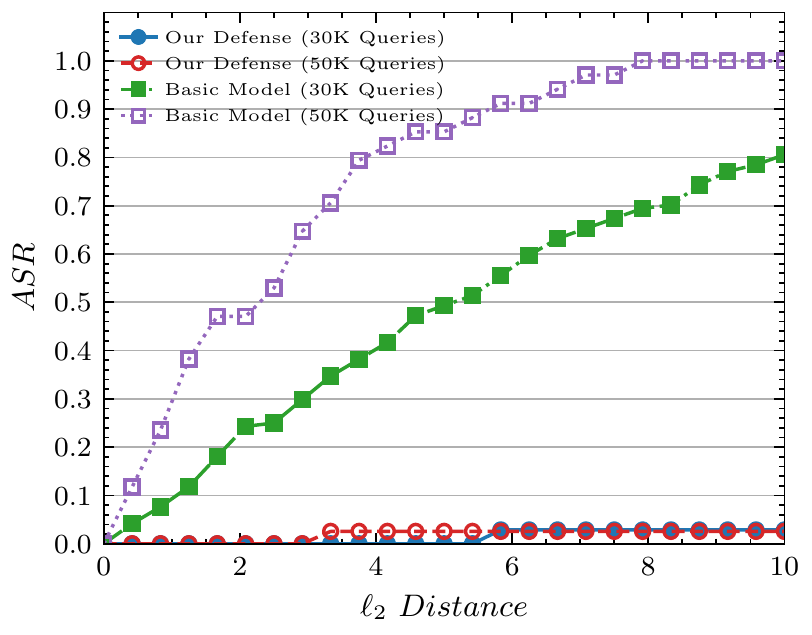}}
\caption{ ASR of SFA and HSJA attacks across various settings with different thresholds for perturbation magnitude, against the target models for CIFAR-10 and ImageNet, with and without \name{}.}
\label{fig:asr}
\end{figure*}

\begin{figure}[!t]
\centering

\subfigure[Basic Image\label{fig:basic}]{\includegraphics[width=0.32\columnwidth]{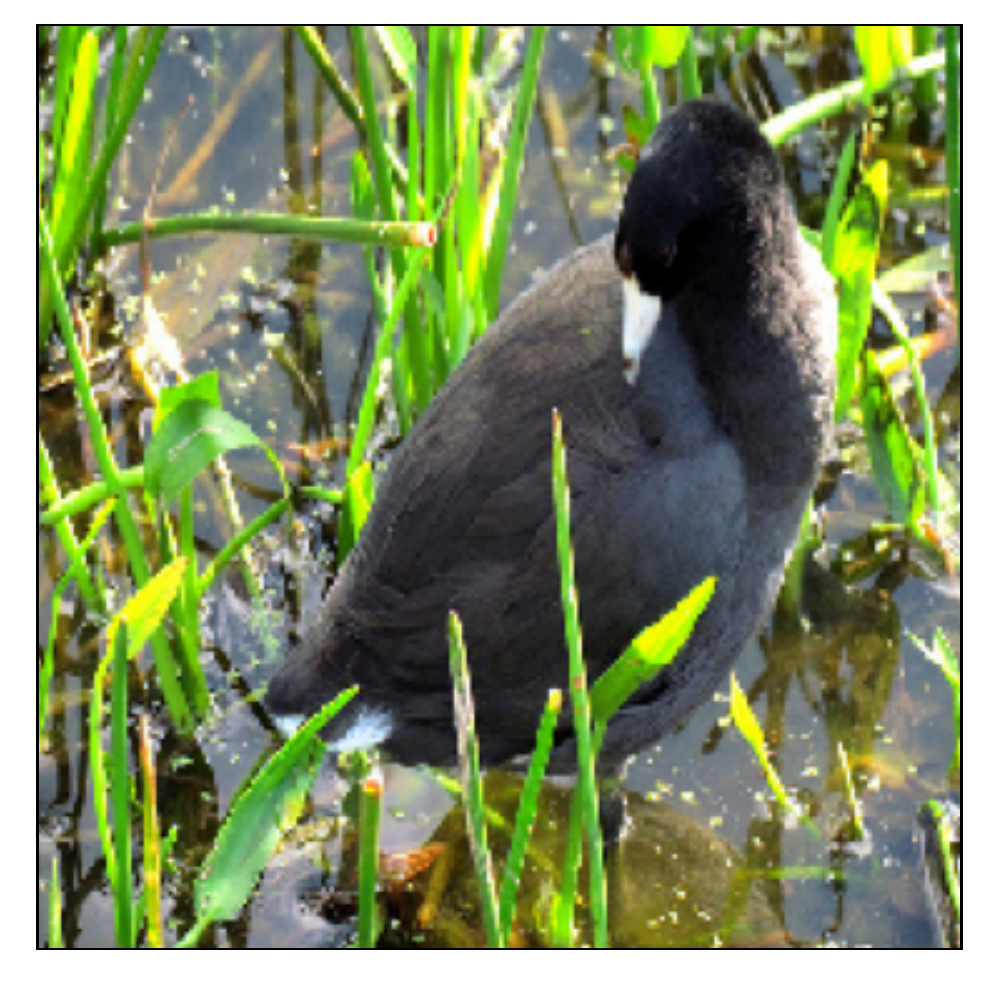}}
\subfigure[Adversarial image without applying \name{} Approach\label{fig:basic}]{\includegraphics[width=0.32\columnwidth]{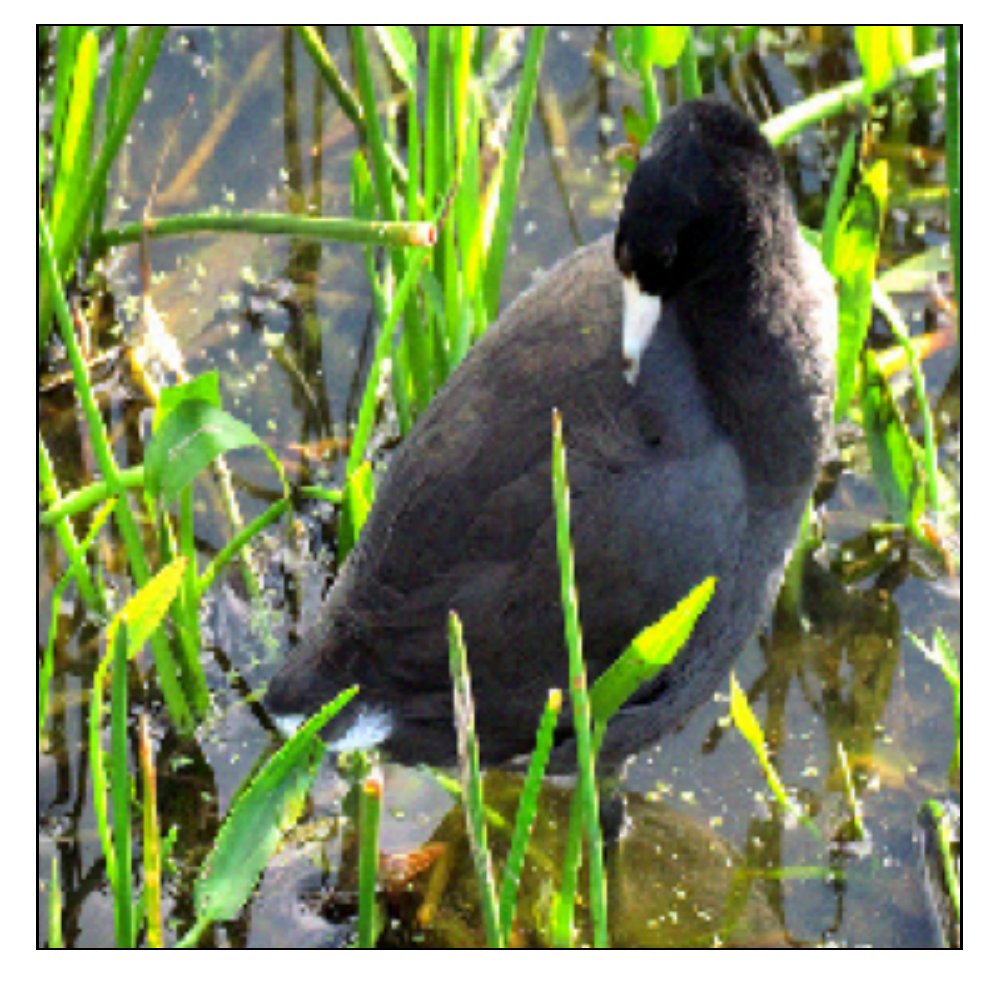}}
\subfigure[Adversarial image with applying \name{}\label{fig:basic}]{\includegraphics[width=0.32\columnwidth]{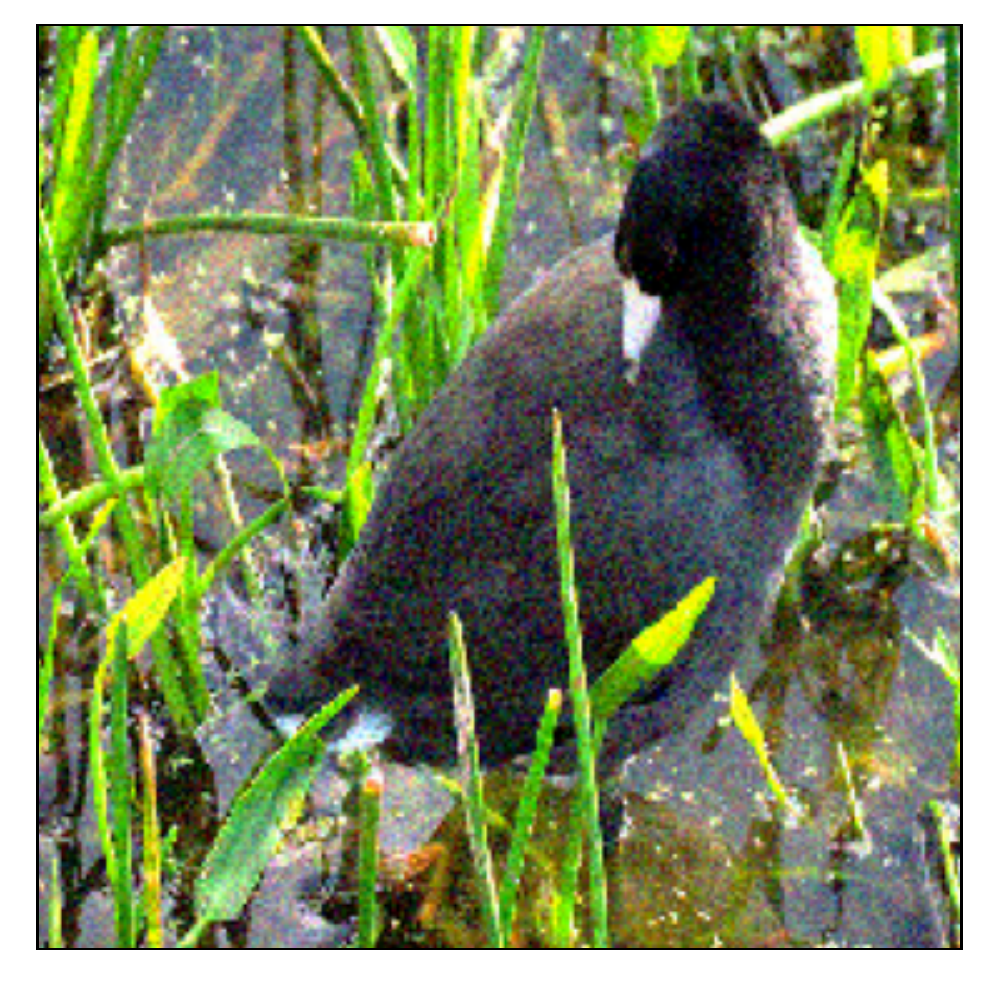}}

 \caption{Illustrating an example original sample in ImageNet and the corresponding crafted adversarial samples by SFA in the $\ell_\infty$ settings with and without applying \name{}. $\ell_\infty$ between the original image and the generated images is: (b) 0.004, (c) 0.2031. }
\label{fig:visual}

\end{figure}

\begin{table}[!t]
\centering
\begin{tabular}{|c|c|}
\hline
Task     & $\bigtriangleup_{acc} (\%)$ \\ \hline\hline
MNIST    & 0.17               \\ \hline
CIFAR-10 & 1.16               \\ \hline
GTSRB    & 0.36               \\ \hline
ImageNet & 6.31               \\ \hline
\end{tabular}

\caption{Accuracy Loss incurred due to \name{} in each task on the test dataset.}
\label{table:accuracy_loss}
\end{table}

\noindent\textbf{Attack Configurations.}
We implement QBHL methods, following the prior works~\cite{ba,DBLP:journals/corr/abs-1909-10773,sfa,hsja}.
Notably, the Sign-OPT attack, whose design is specialized for the  $\ell_{2}$ setting, is ineffective in the $\ell_{\infty}$ setting (success rate $\leq 21\%$ and $\norm{\epsilon}_{\infty} \leq$ 0.031 for CIFAR-10, ImageNet tasks). Similar observations are reported by other works~\cite{sfa}, we thus do not evaluate \name{} against Sign-OPT in the $\ell_{\infty}$ setting.



We evaluate our  approach with 1,000 correctly classified test inputs for each attack, sampled randomly from the original dataset and evenly distributed across different classes. We set the query budgets as 30K, and 50K for each attack since previous works ~\cite{hsja,sfa,DBLP:journals/corr/abs-1909-10773,ba} show that 20K queries are sufficient for each attack.

\noindent\textbf{Evaluation Metrics.}
To evaluate the robustness and practicality of \name{} against various QBHL, we adopt these three metrics: 
\begin{itemize}
    \item \textbf{Median $\ell_p$ Distance.} Consistent with prior work (e.g., $C\&W$ attack~\cite{cw}, HSJA~\cite{hsja}), we use the median $\ell_{p}$ norm distance between $x^*$ and its corresponding perturbed image $x_t$ as a metric for evaluating robustness. Under the same conditions, a larger median $\ell_{p}$ value implies higher robustness.     
    
    \item \textbf{Accuracy Success Rate (ASR).} Aligned with prior work~\cite{hsja}, we compute the attack success rate (ASR) as:
    \begin{equation}
        ASR:= \frac{\text{\# of misclassified ~samples}}{\text{\# of ~test~samples}},
    \end{equation}
    the attack success rate directly measures a model's robustness under certain query budgets and perturbation magnitudes. 
    

    \item \textbf{Conventional Accuracy Loss ($\bigtriangleup_{acc}$).}
    Conventional Accuracy Loss is computed as the change in accuracy (on the validation set) between the target model with and without the defensive method. It is used to check whether a defensive method can preserve the target model's overall accuracy. 
    
    \item \textbf{Inference Time Ratio.}
    Inference Time Ratio is the ratio between the inference time for the target model with and without a defensive method. Inference time ratio is used to evaluate the computational impact of the defensive method with various target models. It is also used to evaluate the defensive method's scalability by comparing this metric on target models with various computational complexity.

\end{itemize}

\subsection{Identify $\gamma$}
\label{sec:trade-off}

As discussed in Sec~\ref{sec:build}, we need a threshold $\gamma$ in $\mathcal{F_Q}$ to identify $x_t$. Thus, before evaluating our approach, we first investigate how $\gamma$  affects \name{}'s robustness and $\bigtriangleup_{acc}$.  We evaluate \name{} on the CIFAR-10 dataset with varying $\gamma$ in the $\ell_{2}$ settings using 100 randomly selected inputs. The results are shown in Fig.~\ref{fig:trade_off}; a very small $\gamma$ makes $\mathcal{F_Q}$ identify almost every input $x$ as $x_t$, resulting in low accuracy ($\bigtriangleup_{acc} \geq 10\%$) and high vulnerability  (small Median $\ell_2$ distance $\leq 0.16$). Interestingly, $\bigtriangleup_{acc}$ rapidly drops ($\leq 2\%$), and the delivered robustness significantly increases when $\gamma$ increases to a specific value ($\approx 0.09$). However, when $\gamma$ becomes large enough ($\geq 0.3$), the delivered robustness will go down since the FN rates of $\mathcal{F_Q}$ would increase as $\mathcal{F_Q}$ will miss more $x_t$. Therefore, we use a binary search to get an appropriate value for $\gamma$ efficiently, as shown in Algorithm~\ref{alg:binary_search}.

\subsection{Overall Performance}
This section investigates the overall performance of \name{} against four QBHL attacks for four tasks. Table~\ref{table:median_distance} summarizes the overall median $\ell_p$ distances across various models, settings, and attack methods under 30K and 50K query budgets. We find that \name{} can increase the median perturbation significantly in both $\ell_{2}$  (1.7X to 18X) and $ \ell_{\infty}$ (2.2X to 18X) settings across various models. Note that, for the MNIST task, \name{} is less robust (implied by a relatively smaller increase) compared to other tasks. This might be due to the inputs for MNIST being much simpler gray-scale images. Thus, DNNs for MNIST are inherently robust against adversarial samples (implied by larger magnitude perturbations by each attack), which is consistent with observations from previous works~\cite{hsja,DBLP:journals/corr/abs-1909-10773}.  
Due to space constraints, Fig. \ref{fig:asr} only illustrates the ASR for the two most recent attacks (SFA and HSJA) on MNIST, CIFAR-10, and ImageNet; the remaining ASR results are included in the Appendix. 

As seen in Fig. \ref{fig:asr}, even with an excessive amount of queries ($\geq$ 30K), \name{} can still provide significant robustness, which implies that \name{} can prevent the attacker from crafting  visually-indistinguishable adversarial samples in a black-box setting. An interesting observation is that, unlike other attack methods, HSJA is more sensitive to query budgets (ASR increases as the number of queries increases); this is due to HSJA requiring more samples for its gradient estimation step. Hence a larger query budget (more gradient estimation steps) allows HSJA to approach closer to its optimization point. 
Another interesting observation is that  HSJA is the most robust attack method (inferred by its low average increase in the distance from base to \name{}) against \name{} compared to other attack methods. Additionally, SFA outperforms HSJA the $\ell_{\infty}$ settings but severely under-performs in the $\ell_2$ settings.

Fig. \ref{fig:visual} shows a visual comparison of the original image and the generated adversarial samples by SFA attack in the $\ell_\infty$ setting, with and without \name{}. Due to space restrictions, we include the detailed demonstration for adversarial images across various settings in Fig.~\ref{fig:visual_appendix} in the Appendix. We observe that \name{} enhances the visual differences between the original image $x^*$ and the crafted adversarial sample $x_t$ for various attacks under both $\ell_2$ and $\ell_\infty$ settings, especially in the $\ell_\infty$ settings. Furthermore, in Table~\ref{table:accuracy_loss} we evaluate $\bigtriangleup_{acc}$ of \name{} for each task; \name{} has an acceptable impact on the model's accuracy with  at most $\bigtriangleup_{acc} (\leq 6.31\%$) for all tasks, which demonstrates the practicality of \name{}.

\subsection{Comparison with Existing Methods}


We compare \name{} with existing state-of-art defense mechanisms designed for white-box attacks. We choose four defense mechanisms for comparison: defensive distillation~\cite{distill}, region-based classification~\cite{region-based}, TRADES~\cite{trades}, BitDepth~\cite{deepbit}, and adversarial training (ADV-TRAINING)~\cite{adv_train}. For these defense mechanisms, we test against the best performing attacks, HSJA and SFA, which are robust against several defensive methods~\cite{hsja,sfa}.  Since most state-of-art defensive methods support the CIFAR-10 dataset only, we conduct this study on CIFAR-10. Additionally, the unscalable nature of state-of-art robust optimization-based approaches (ADV-TRAINING, TRADES) makes them impractical to complicated tasks (e.g., ImageNet). Moreover, previous work ~\cite{sfa} shows that existing defensive methods that apply to ImageNet (e.g., BitDepth) are more vulnerable than the defensive methods for CIFAR-10.

We implement each defense mechanism following the configurations of previous works~\cite{sfa,hsja} and  use publicly available models proposed by \cite{adv_model, trade_model} to evaluate TRADES~\cite{trades}, ADV-TRAINING~\cite{adv_train}.

\begin{figure}[!t]
    \centering
    \subfigure[CIFAR-10(HSJA)\label{fig:cifar_hsja}]{\includegraphics[width=0.22\textwidth]{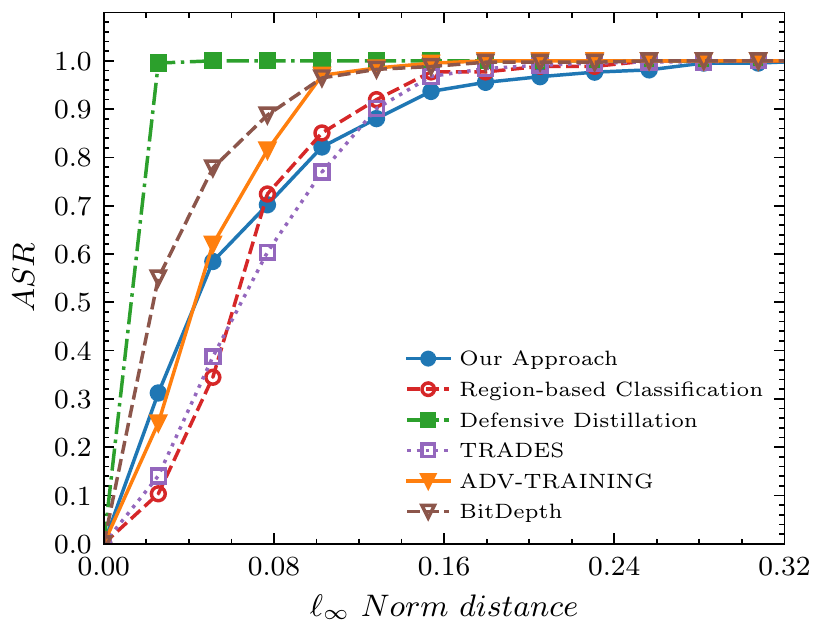}}
    \subfigure[CIFAR-10(SFA)\label{fig:cifar_sfa}]{\includegraphics[width=0.22\textwidth]{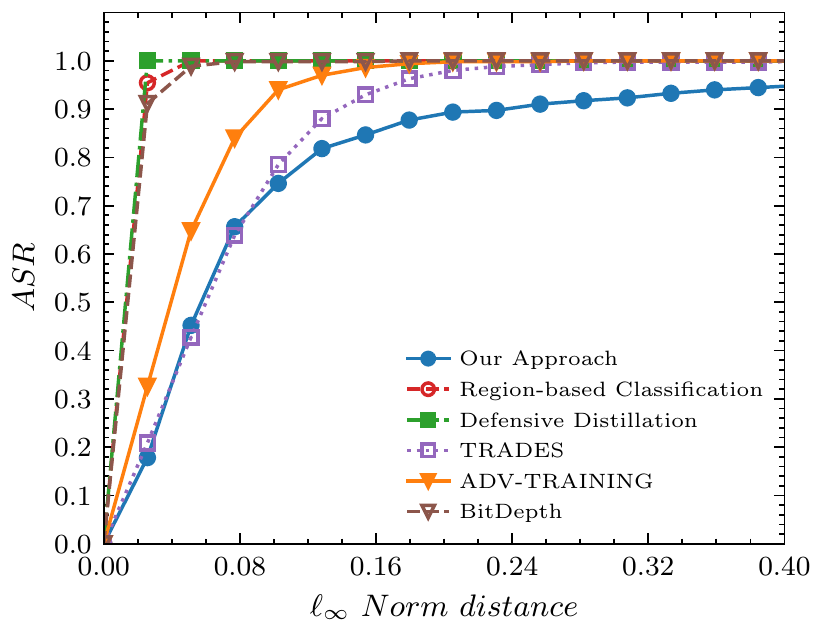}}
    \subfigure[CIFAR-10(HSJA)\label{fig:cifar_hsja_l2}]{\includegraphics[width=0.22\textwidth]{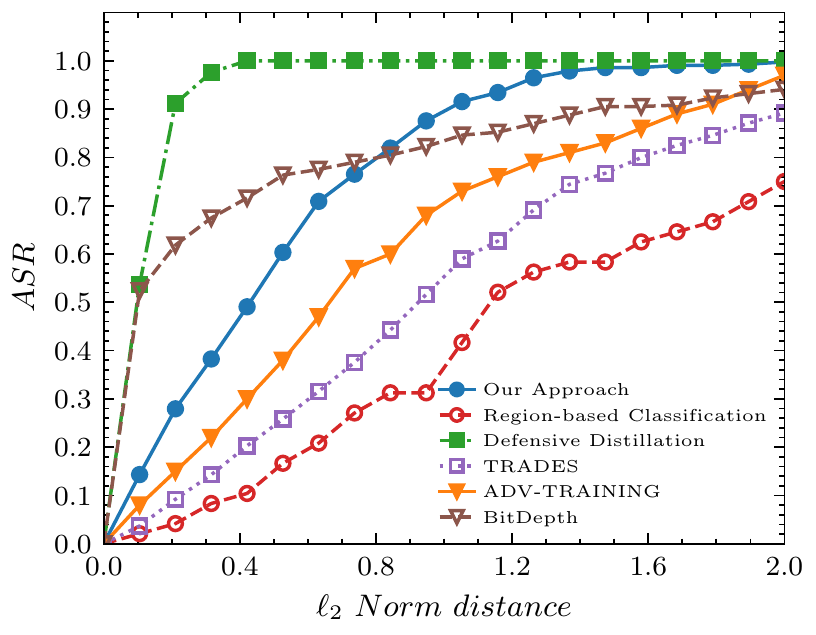}}
    \subfigure[CIFAR-10(SFA)\label{fig:cifar_sfa_l2}]{\includegraphics[width=0.22\textwidth]{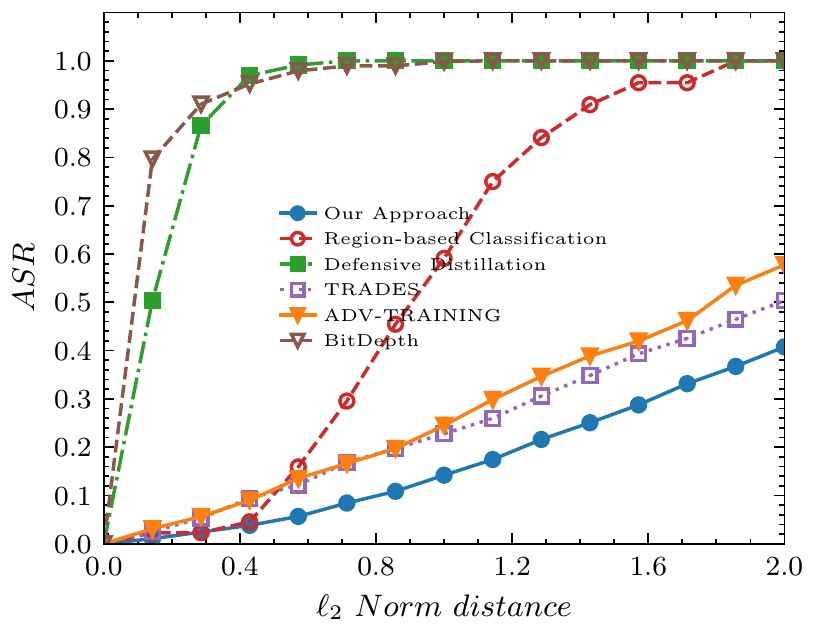}}

    \vspace{1mm}
    \caption{ASR of \name{} and state-of-the-art defense methods against  QBHL (SFA and HSJA) attacks, across different settings and tasks.}
      \vspace{2mm}
    \label{fig:compare_performance}
\end{figure}

\noindent\textbf{Defense Performance.}
Fig. \ref{fig:compare_performance} compares the performance of \name{} and existing defense mechanisms in the $\ell_\infty$ and $\ell_2$ settings. In the $\ell_\infty$ setting, against SFA, \name{} achieves similar $ASR$ as the best state-of-art white-box defensive method TRADES while $\ell_\infty \leq 0.1$; while $\ell_\infty \geq 0.1$, \name{} outperforms TRADES. We also observe that the inference-phase defensive methods (BitDepth, Region-Based Classification) fail against the SFA attack. Only the robust optimization-based approaches (i.e., TRADES, ADV-TRAIN) hinder the SFA attack. As for HSJA, consistent with previous work~\cite{sfa} observations, we find that HSJA is much more sensitive to white-box defensive mechanisms. Hence, \name{} performs a little worse than the white-box defensive methods when $\ell_\infty$ is smaller. As $\ell_\infty$ increases, \name{} starts outperforms other state-of-art methods.

In the $\ell_2$ settings, for HSJA attack, \name{} has a significant impact on $ASR$ but performs worse than TRADES, ADV-TRAINING, and Region-Based Classification methods. In the case of SFA, \name{} outperforms all existing defensive methods. Interestingly, we find that BitDepth can enhance SFA in the $\ell_2$ settings and makes the target model more vulnerable, implying a potential new vulnerability introduced by BitDepth. Moreover, similar to the $\ell_\infty$ settings, SFA is more robust to the set of inference-phase defensive methods (i.e., Region-based Classification, BitDepth).

While TRADES performs much better in the $\ell_2$ setting against HSJA and is comparable to our approach in other settings, \name{} is the first successful, scalable defensive method. As discussed earlier (also see evaluation on scalability later in this section), robust optimization-based approaches (like TRADES) suffer from unnecessary and rather expensive training costs, making them inapplicable to more complex tasks such as ImageNet. Moreover, \name{} can be in conjunction with the TRADES, enhancing their effectiveness as shown in Sec. \ref{sec:enchance_trades}.

\noindent\textbf{Inference Efficacy.} We also evaluate the inference time ratio for each defense mechanism on CIFAR-10 and ImageNet datasets. We feed each defensive model and its corresponding target model a batch of 256 inputs and calculate the ratio between their inference times. Results are shown in Table. \ref{table:inference_time_cost}, we can see that \name{} only suffers a 1.21x and 1.08x slow down for CIFAR-10 and ImageNet datasets, which is comparable to robust optimization-based approaches (ADV-TRAINING, TRADES) and surpasses other inference-phase defensive mechanisms (Region-Based Classification and BitDepth). Such low inference cost property demonstrates the practicality and suitability of \name{} to time-sensitive situations (embedding systems).

\begin{table}[!t]
\centering
\scalebox{0.8}{
\begin{tabular}{|c|c|c|}
\hline
Task                      & Approach                    &  Inference Time Ratio \\ [0.5ex] 
 \hline\hline
\multirow{6}{*}{CIFAR-10} & \textbf{\name{}(Ours)}       & \textbf{1.219}          \\ \cline{2-3} 
                          & Region-based Classification & 28.978                  \\ \cline{2-3} 
                          & Defensive Distillation      & 1.000                   \\ \cline{2-3} 
                          & TRADES                      & 1.000                   \\ \cline{2-3} 
                          & ADV-TRAINING                & 1.000                   \\ \cline{2-3} 
                          & BitDepth                    & 1.544                   \\ \hline
\multirow{4}{*}{ImageNet} & \textbf{\name{}(Ours)}       & \textbf{1.085}          \\ \cline{2-3} 
                          & Region-based Classification & 446.538                 \\ \cline{2-3} 
                        
                          & BitDepth                    & 1.508                   \\ \hline
\end{tabular}}
 \caption{Inference Time Ratio of various defense mechanisms for Image Classification task.}
\label{table:inference_time_cost}
\end{table}

\begin{table}[!t]
\centering
\scalebox{0.8}{
\begin{tabular}{|c|c|c|}
\hline
Task                      & Approach                    & $\bigtriangleup_{acc} (\%)$ \\ [0.5ex] 
 \hline\hline
\multirow{6}{*}{CIFAR-10} & \textbf{\name{}(Ours)}       & \textbf{1.16}          \\ \cline{2-3} 
                          & Region-based Classification & $0$                \\ \cline{2-3} 
                          & Defensive Distillation      & $1.27$                 \\ \cline{2-3} 
                          & TRADES                      & $9.37$                   \\ \cline{2-3} 
                          & ADV-TRAINING                & $7.75$                   \\ \cline{2-3} 
                          & BitDepth                    & $5.3$                \\ \hline
\multirow{4}{*}{ImageNet} & \textbf{\name{}(Ours)}       & \textbf{6.31}          \\ \cline{2-3} 
                          & Region-based Classification & 0.3                \\ \cline{2-3} 
                        
                          & BitDepth                    & 7.30                  \\ \hline
\end{tabular}}
 \caption{$\bigtriangleup_{acc} (\%)$ for each defensive methods under CIFAR-10 and ImageNet}
\label{table:acc_com}
\end{table}

\noindent\textbf{Accuracy Loss.}
Table \ref{table:acc_com} evaluates each defensive mechanism's overall accuracy on regular test inputs in CIFAR-10 and ImageNet datasets. We calculate $\bigtriangleup_{acc} (\%)$ based on the models incorporated with defensive methods and its corresponding target model. We can see that compared to the robust optimization-based approaches (i.e., TRADES, ADV-TRAINING), \name{} has a negligible effect ($1.16\%$) on $\bigtriangleup_{acc} (\%)$ for the CIFAR-10 dataset. As for ImageNet, \name{} has a smaller $\bigtriangleup_{acc} (\%)$ than BitDepth but underperforms to region-based classification.

\begin{table}[!t]\centering
\scalebox{0.98}{
\begin{tabular}{|c|c|c|c|}
\hline
Attack                & Model         & $\ell_\infty$ Distance & $\ell_2$ Distance  \\ \hline
\multirow{6}{*}{SFA}  & DenseNet121   &  \textbf{0.126}(0.003)                       &     \textbf{27.891}(2.103)                               \\ \cline{2-4} 
                      & ResNet.V2.50  & \textbf{0.131}(0.003)                        &     \textbf{30.037}(2.145)                                    \\ \cline{2-4} 
                      & ResNet.V2.101 &      \textbf{0.130}(0.003)                  &     \textbf{31.103}(2.209)                                    \\ \cline{2-4} 
                      & VGG16         &       \textbf{0.143}(0.003)                  &     \textbf{26.743}(1.716)                                   \\ \cline{2-4} 
                      & Inception.V3  &          \textbf{0.125}(0.003)              &   \textbf{30.026}(2.276)                                      \\ \cline{2-4} 
                      & Xception      &           \textbf{0.125}(0.003)             &    \textbf{31.015}(2.743)                                      \\ \hline
\multirow{6}{*}{HSJA} & DenseNet121   &    \textbf{0.133}(0.006)  &   \textbf{9.48}(0.89)                       \\ \cline{2-4} 
                      & ResNet.V2.152  &      \textbf{0.127}(0.006)  &     \textbf{8.82}(0.83)                         \\ \cline{2-4} 
                      & ResNet.V2.101 &        \textbf{0.134}(0.007)  &      \textbf{8.72}(0.83)                            \\ \cline{2-4} 
                      & VGG16         &     \textbf{0.126}(0.006)  &      \textbf{7.48}(0.82)                        \\ \cline{2-4} 
                      & Inception.V3  &           \textbf{0.158}(0.062)  &     \textbf{13.86}(1.145)                              \\ \cline{2-4} 
                      & Xception      &      \textbf{0.171}(0.067)  &      \textbf{12.13}(1.103)                         \\ \hline
\end{tabular}}
  \vspace{1mm}
\caption{The Summary Of median $\ell_p$ of \name{} Under Various Publicly available models for SFA and HSJA with the query budget as 50K. The values out of the bracket represent the median perturbation under the model incorporated with our approach, while the values inside the bracket represent that under the basic model.}
\label{table：sum}
\end{table}

\begin{table}[!t]
\centering\scalebox{0.9}{
\begin{tabular}{|c|c|c|}
\hline
Model         & Accuracy Loss(\%) & Inference Time Ratio \\ \hline
DenseNet121    & 6.37              & 1.034                   \\ \hline
ResNet.V2.50  & 6.31              & 1.087                   \\ \hline
ResNet.V2.101 &  6.33              & 1.034                   \\ \hline
VGG16         &  6.27              & 1.056                   \\ \hline
Xception      &  6.38              & 1.037                   \\ \hline
InceptionV3   &  6.35              & 1.036                   \\ \hline
\end{tabular}}
\caption{Summary Of Accuracy Loss and Inference Time Ratio of \name{} on a set of  publicly available DNN models.}
\label{table：sum_b}
\end{table}

\noindent\textbf{Scalability.} Unfortunately, most existing defensive methods may be impractical due to their low scalability. For instance, the most potent defensive methods, robust optimization-based approaches (e.g., TRADES, ADV-TRAINING), typically require the defender to build the model from scratch using sufficient training data with data augmentation. Such training procedures require at least dozens of hours, even on environments with multiple GPUs for each model. Hence, robust optimization-based approaches do not scale to complex datasets (e.g., ImageNet).
Moreover, in addition to the computational cost, optimization-based approaches cannot directly be built upon existing models, which significantly constrains its application under many real-world scenarios. For instance, considering a scenario where the defender can only obtain the model $\mathcal{C}$ from a third party, they may not have enough computation resources or adequate training data for robust training. Under such a scenario, a company may replace their whole model with new architectures and parameters; it is impractical to conduct costly robust optimization for each update.  

In contrast, \name{} can be built upon existing publicly shared models with a training time of under 30 minutes using a single GPU. To further illustrate our approach's  scalability and generalizability, we evaluate \name{} on 6 publicly available state-of-art pre-trained models for ImageNet provided by Keras Team~\cite{Keras_model}. The results, as shown in Tables \ref{table：sum}, \ref{table：sum_b}\footnote{For models with different pre-processing procedures, we intentionally normalise the perturbation into [0,1].} indicate \name{} can provide significant robustness against state-of-art QBHL attacks while having a low impact on their accuracy and inference times. 


\begin{figure}[!t]
    \centering
    \subfigure[CIFAR-10(HSJA)\label{fig:cifar_hsja}]{\includegraphics[width=0.22\textwidth]{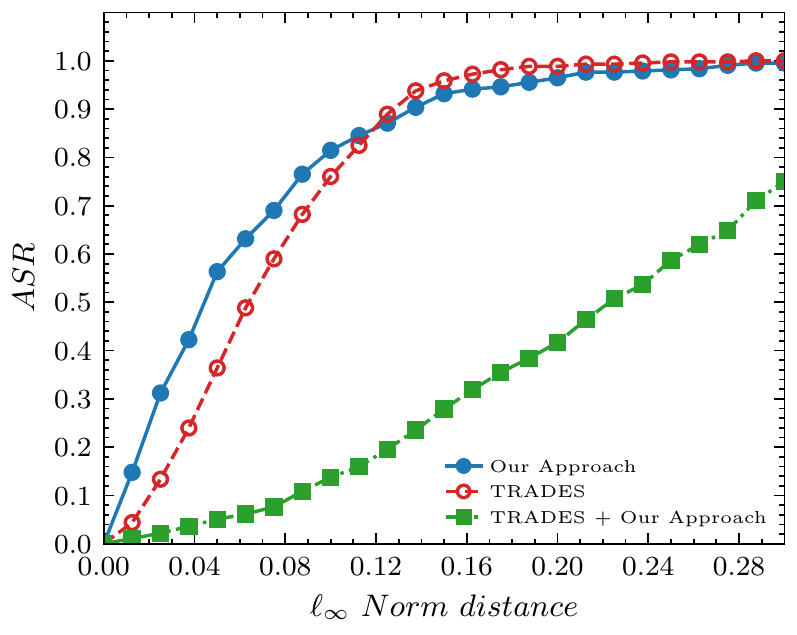}}
    \subfigure[CIFAR-10(SFA)\label{fig:cifar_sfa}]{\includegraphics[width=0.22\textwidth]{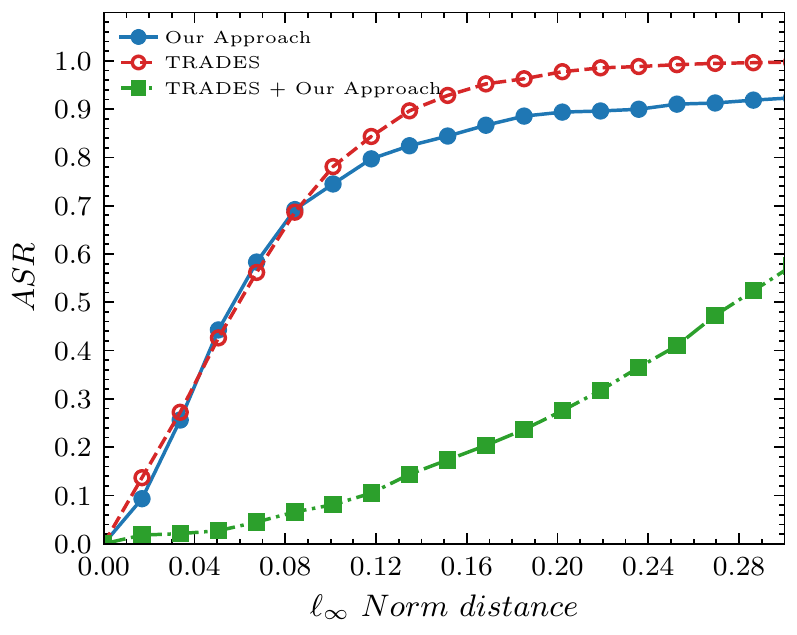}}
    \subfigure[CIFAR-10(HSJA)\label{fig:cifar_hsja_l2}]{\includegraphics[width=0.22\textwidth]{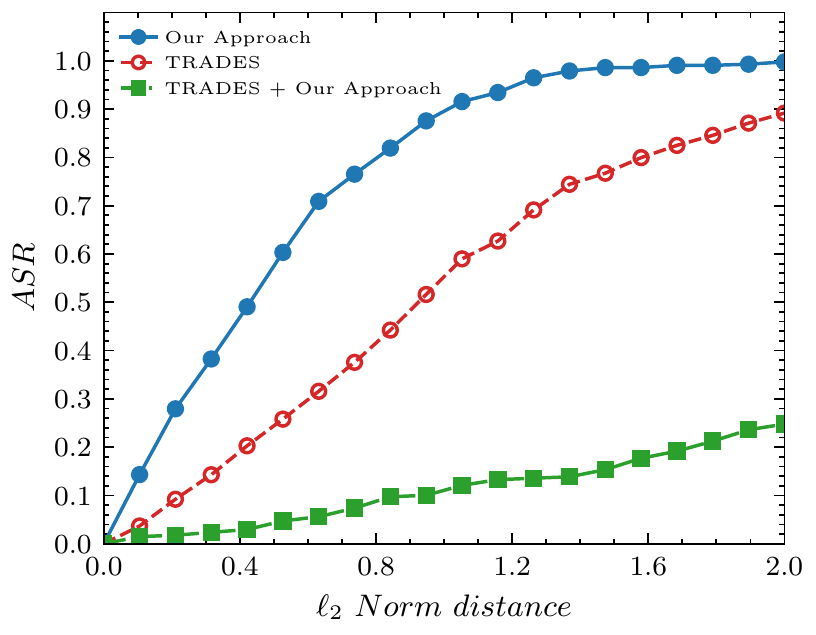}}
    \subfigure[CIFAR-10(SFA)\label{fig:cifar_sfa_l2}]{\includegraphics[width=0.22\textwidth]{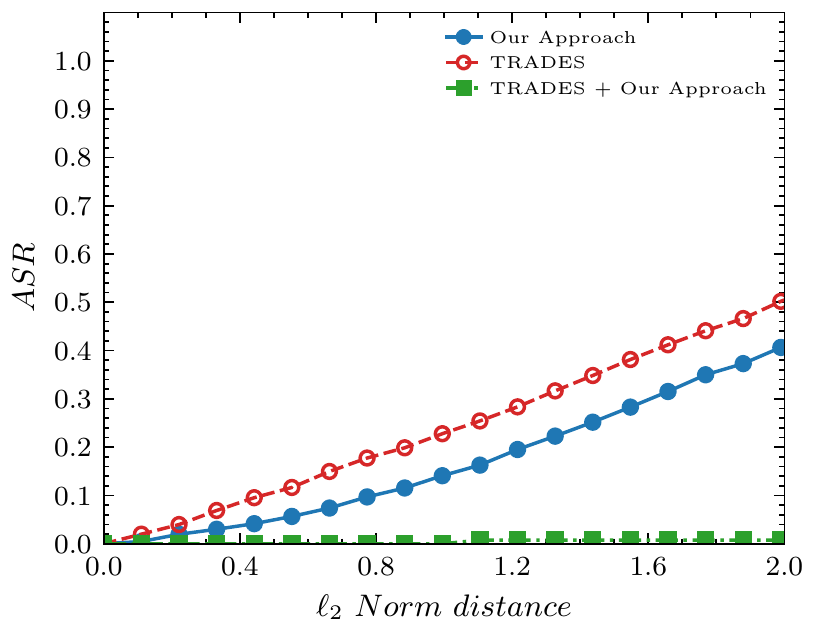}}

    \caption{ ASR comparison between TRADES, our approach, the combination of TRADES and our approach.}
    \label{fig:boost}
\end{figure}

\subsection{Enhancing Existing White-box Defenses} 
\label{sec:enchance_trades}
Since \name{} is an inference phase defensive method, we investigate whether our approach could be used in conjunction with methods like TRADES~\cite{trade_model}, which conduct robust optimization during the target model's training phase. We conduct our experiments on the CIFAR-10 dataset, whose results are shown in Fig. \ref{fig:boost}. We observe that \name{} can significantly enhance TRADES under HSJA and SFA in both $\ell_2$ and $\ell_\infty$. Moreover, the accuracy loss compared  to the adversary trained model produced by TRADES is merely $3.93\%$, and the inference time ratio is 1.219, both comparable to the original \name{} performance. Hence, \name{} can be used in conjunction with state-of-art white-box defensive methods to achieve a significantly more robust model.

\section{Robustness Against Defense-Aware Attack}
This section investigates whether \name{}  is robust to adaptive attacks. We evaluate the robustness of \name{} on the CIFAR-10 dataset under the worst-case scenario where the attacker is fully aware of the details of \name{} and the parameters of $\mathcal{F_Q}$. Since \name{} has two main steps: \textit{(i)} Detection of $x_t$ by $\mathcal{F_Q}$ \textit{(ii)} Non-deterministic mechanism, we modify HSJA and SFA each to produce two types of adaptive attacks that bypass each step: \textit{(i)} Bypassing the detection of $\mathcal{F_Q}$ \textit{(ii)} Conducting Uncertainty-Aware Attack.


\subsection{Bypassing the Detection of $\mathcal{F_Q}$}

\label{adaptive_attack1}
The effectiveness of \name{} critically depends on the accuracy of $\mathcal{F_Q}$. While computing the gradient estimation, an attacker could bypass $\mathcal{F_Q}$ by producing queries that are adversarial to $\mathcal{F_Q}$. Given some boundary point $x_t$ and $\{ u_b \}_{b=1}^{B}$, to evaluate the robustness of \name{} against such an adaptive attack, we formulate an optimization problem to generate a gradient estimation queries $x_t + \delta u_b$ as follows: 

$$ \text{With} \; (y^1_Q, y^2_Q) = \mathcal{F_Q}(\overline{\mathcal{F}}(x_t + \delta u_b ; \theta )$$
\begin{equation}
     \delta_b = \min \big\{ \; \delta \; \big| \; y^1_Q < \gamma \; \text{and} \;  x_t + \delta u_b \in [0,1]^d  \big\} 
    \label{eq:optimization}
\end{equation}




\begin{figure}[!t]
    \centering
    \subfigure[CIFAR-10(SFA)\label{fig:cifar_sfa}]{\includegraphics[width=0.22\textwidth]{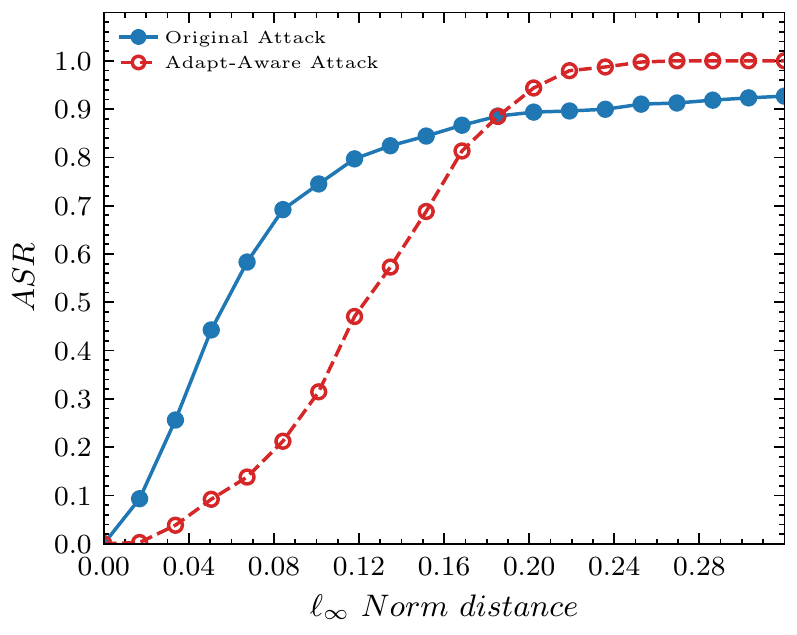}}
    \subfigure[CIFAR-10(SFA)\label{fig:cifar_sfa_l2}]{\includegraphics[width=0.22\textwidth]{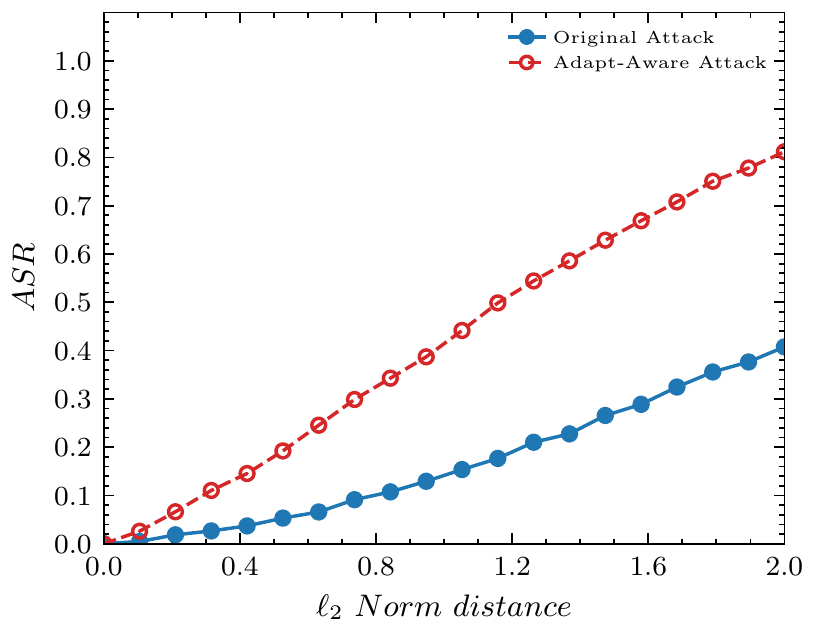}}
    \subfigure[CIFAR-10(HSJA)\label{fig:cifar_hsja}]{\includegraphics[width=0.22\textwidth]{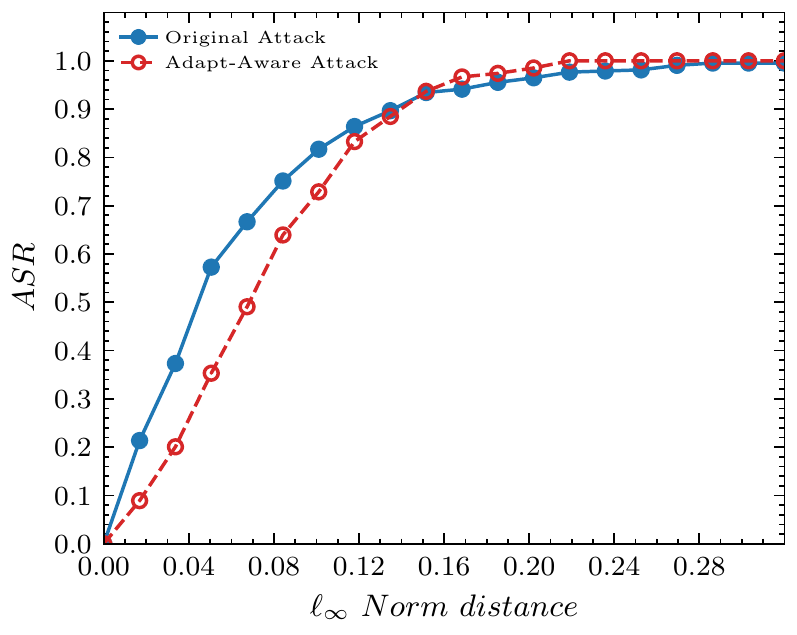}}
    \subfigure[CIFAR-10(HSJA)\label{fig:cifar_hsja_l2}]{\includegraphics[width=0.22\textwidth]{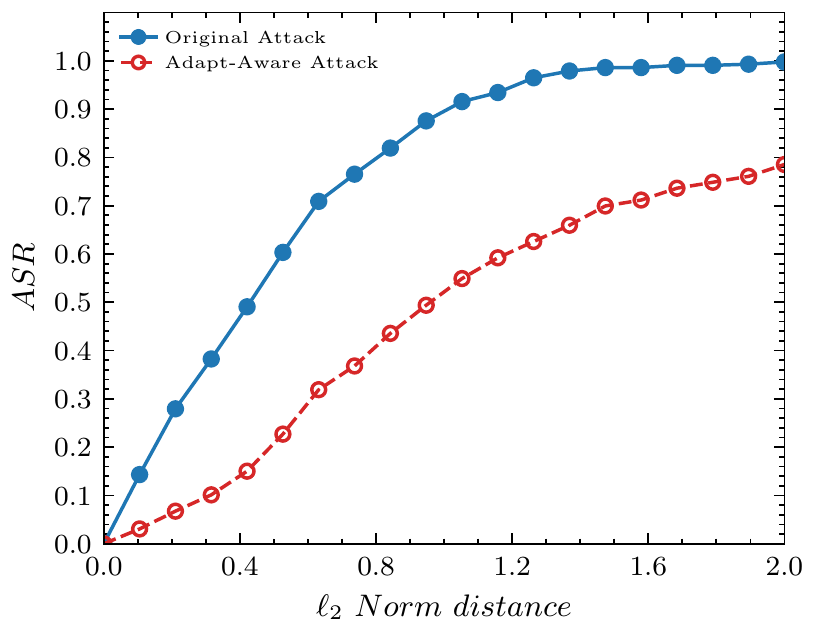}}
    \caption{ASR of \name{} under adaptive attacks for Bypassing the Detection of $\mathcal{F_Q}$.}
    \label{fig:adapt}
\end{figure}
\noindent recall that $y^1_Q \geq \gamma$ is the condition for detection in the proposed $\mathcal{F_Q}$. We test 1,000 random boundary points $x_t$ computed using the CIFAR-10 dataset. Interestingly, since $\mathcal{F_Q}$ takes the output of $\mathcal{C}$ ($\overline{\mathcal{F}}(x_t;\theta)$) as input, for the adaptive attack to succeed,  the gradient estimation queries should bypass both $\mathcal{C}$ and $\mathcal{F_Q}$ simultaneously, which from our experiments is very difficult. On average, for a boundary point, $x_t$ a valid $\delta_b$ (i.e., a query that bypasses $\mathcal{F_Q}$) is found for only 6\% of the samples ($x_t+\delta_b u_b$).

The adaptive attack performs much worse than the original attack in most cases except for SFA ($\ell_2$), as shown in Fig. \ref{fig:adapt}. 
Such observation could be attributed to two reasons: first, there are very few $x_{t}+ \delta_b u_b$  (around 6\%) which can bypass $\mathcal{F_Q}$, limiting the number of distinct queries sent for gradient estimation leading to  more flawed estimates; second, from Eq. \ref{eq:to_break}, accurate gradient estimates require smaller $\delta$ but, a large $\delta$ is required for bypassing $\mathcal{F_Q}$ again ensuring inaccurate gradient estimates. 

\begin{figure}[!t]
    \centering
    \subfigure[CIFAR-10(SFA)\label{fig:cifar_sfa}]{\includegraphics[width=0.22\textwidth]{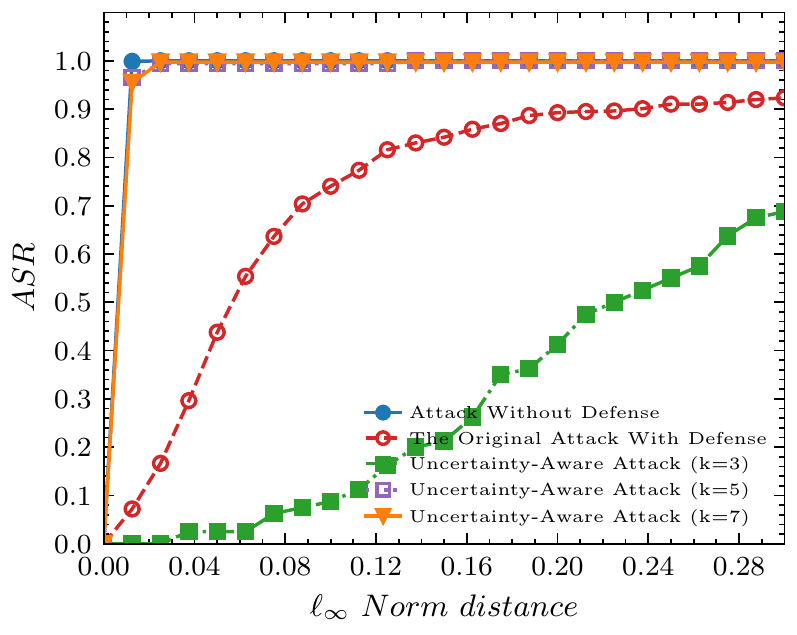}}
    \subfigure[CIFAR-10(SFA)\label{fig:cifar_sfa}]{\includegraphics[width=0.22\textwidth]{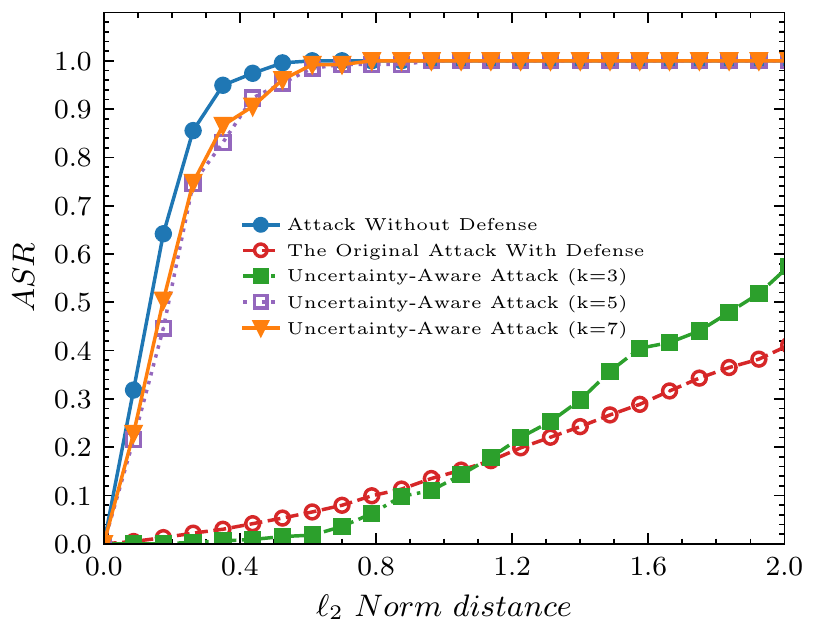}}
    \subfigure[CIFAR-10(HSJA)\label{fig:cifar_hsja}]{\includegraphics[width=0.22\textwidth]{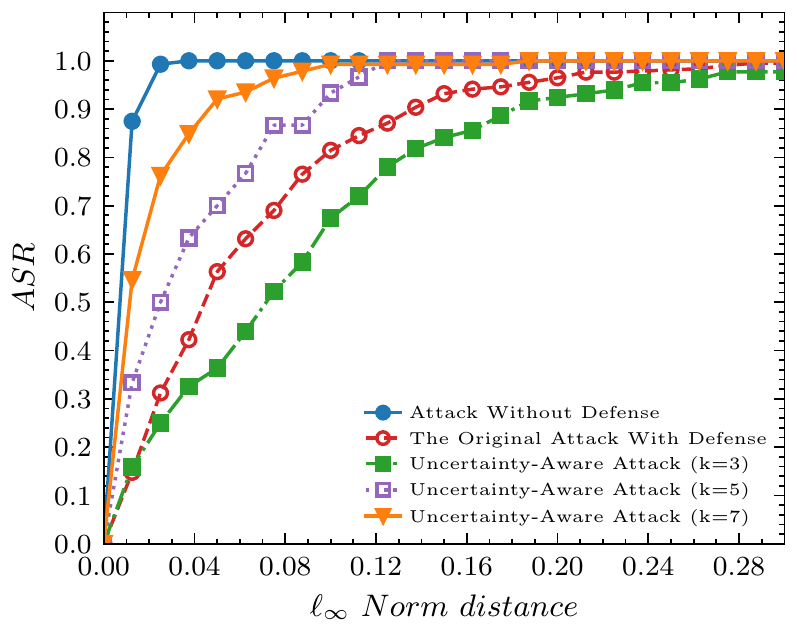}}
    \subfigure[CIFAR-10(HSJA)\label{fig:cifar_hsja_l2}]{\includegraphics[width=0.22\textwidth]{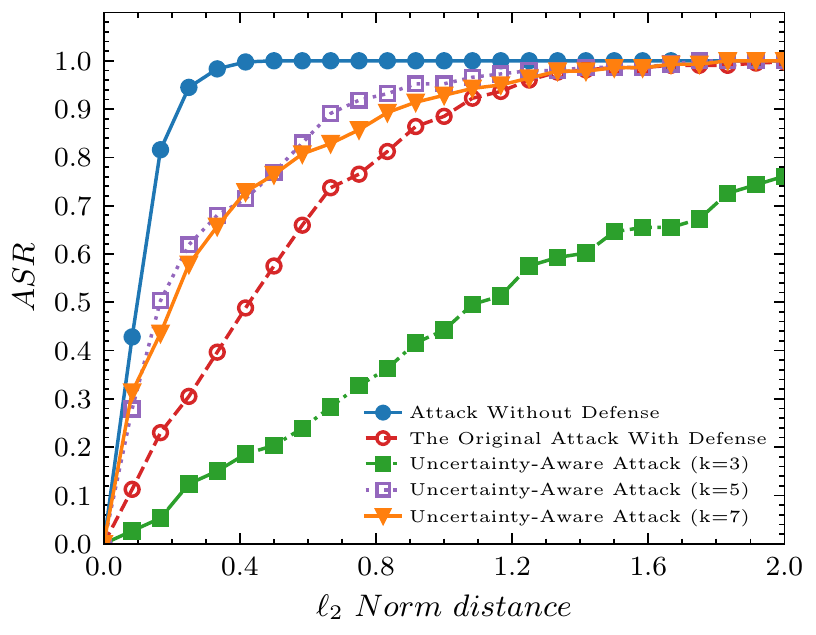}}
    \caption{ASR of \name{} against Uncertainty-Aware Attacks with various values of $k$. }
    \label{fig:uncertainty}
    \vspace{2mm}
\end{figure}

\begin{figure}[!t]
    \centering
    \subfigure[CIFAR-10(SFA)\label{fig:cifar_sfa}]{\includegraphics[width=0.22\textwidth]{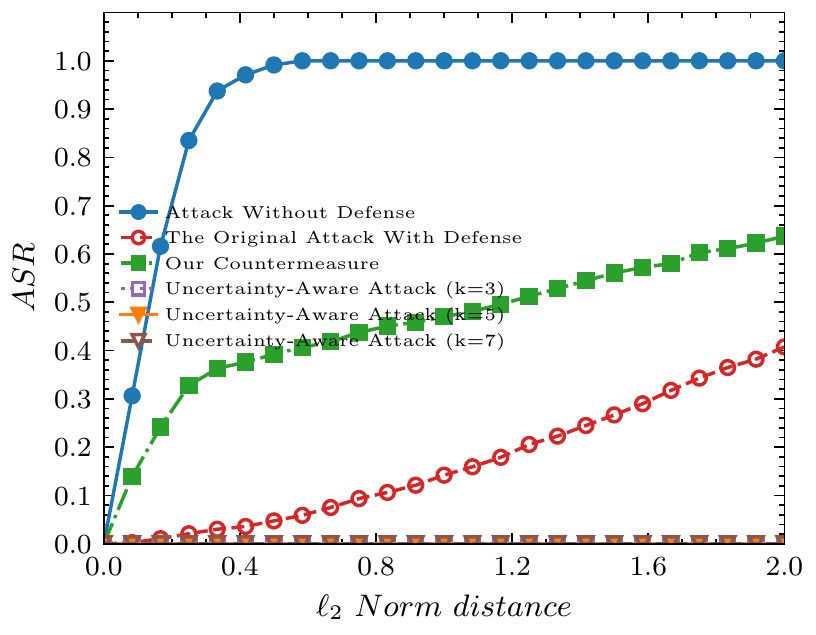}}
    \subfigure[CIFAR-10(SFA)\label{fig:cifar_sfa_li}]{\includegraphics[width=0.22\textwidth]{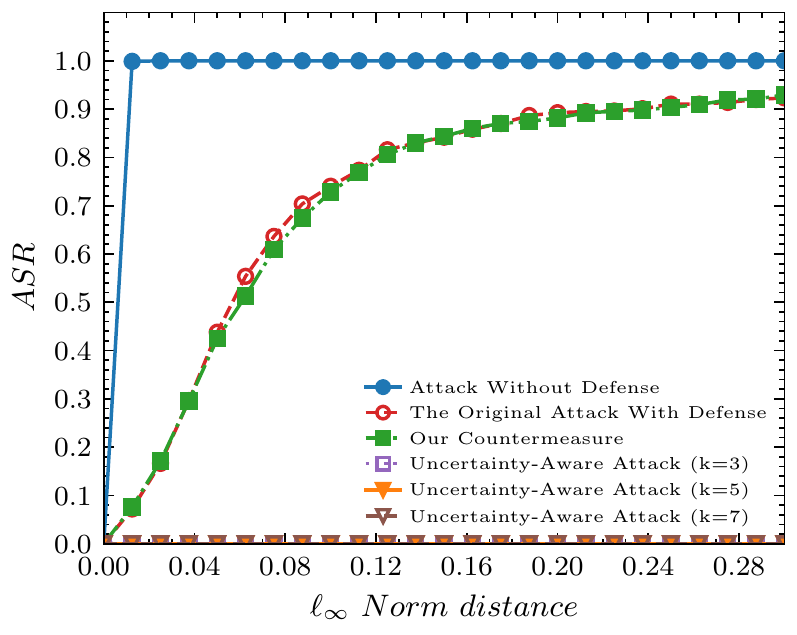}}
    \subfigure[CIFAR-10(HSJA)\label{fig:cifar_hsja}]{\includegraphics[width=0.22\textwidth]{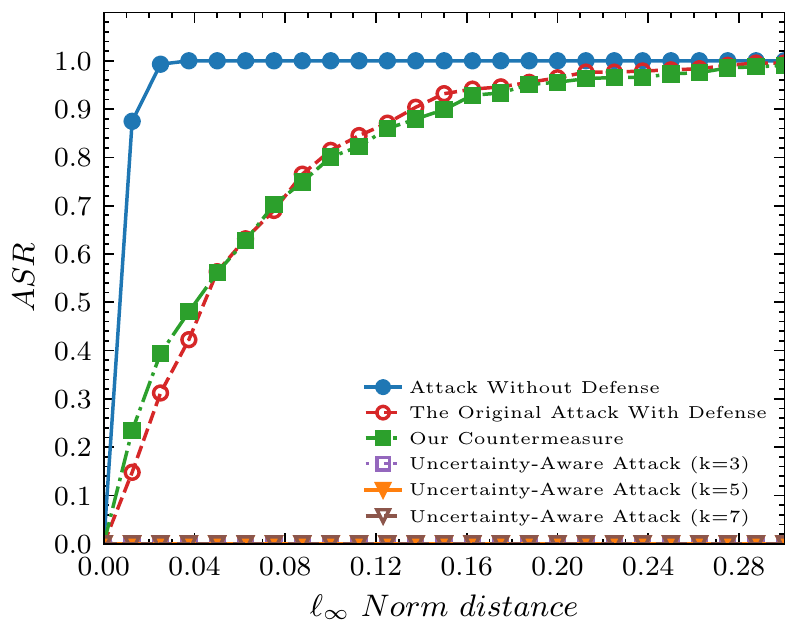}}
    \subfigure[CIFAR-10(HSJA)\label{fig:cifar_hsja_l2}]{\includegraphics[width=0.22\textwidth]{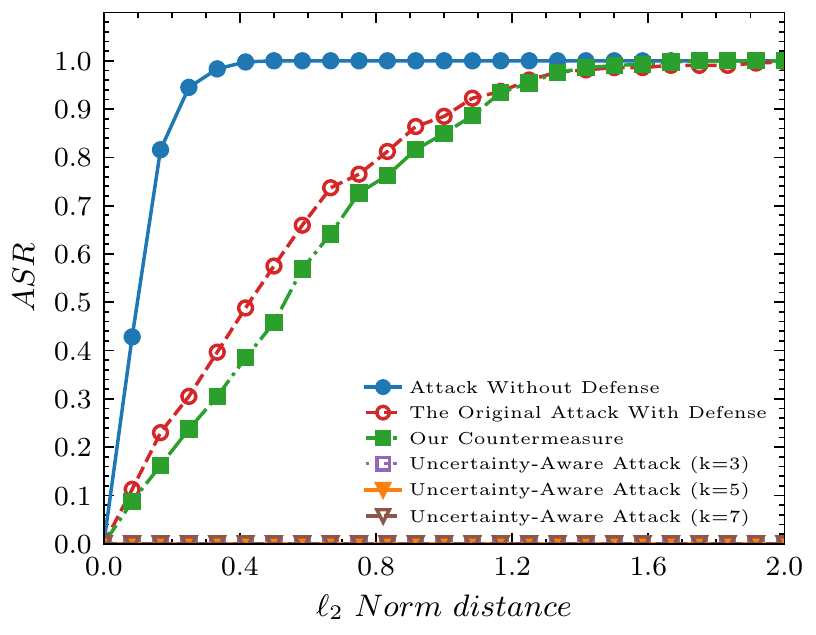}}
    \caption{ASR of our countermeasure and the original \name{} against original and Uncertainty-Aware attacks with various $k$.}
    \label{fig:uncertainty_counter}
\end{figure}

\begin{figure}[!t]
    \centering
    \subfigure[CIFAR-10(SFA)\label{fig:cifar_sfa}]{\includegraphics[width=0.22\textwidth]{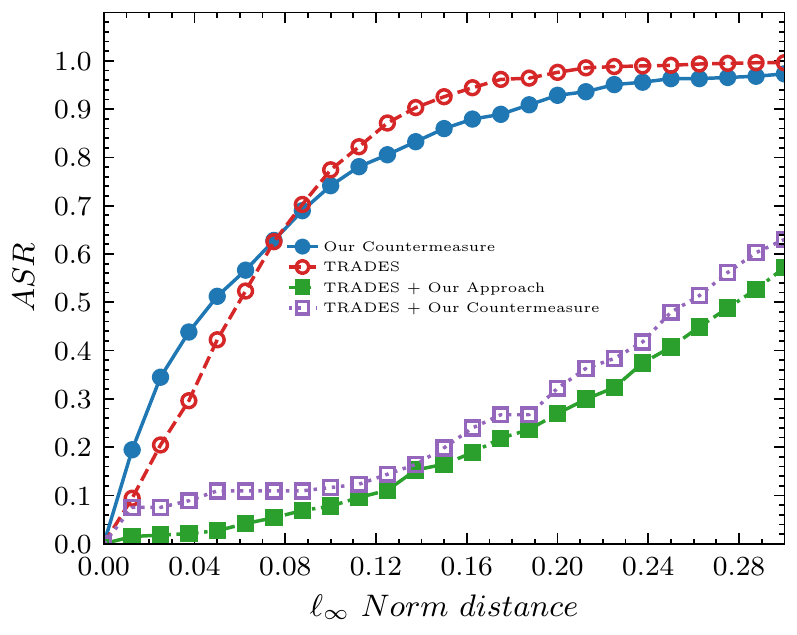}}
    \subfigure[CIFAR-10(SFA)\label{fig:cifar_sfa_li}]{\includegraphics[width=0.22\textwidth]{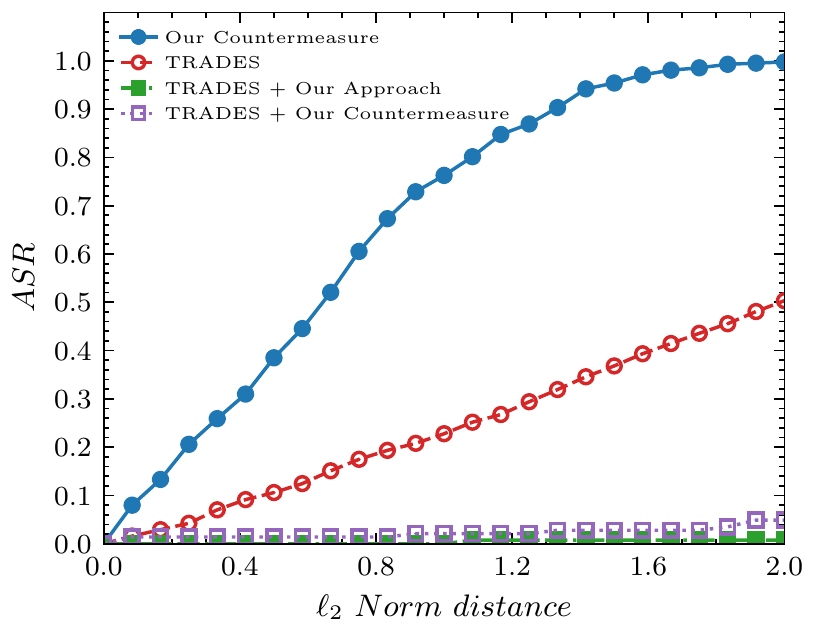}}
    \subfigure[CIFAR-10(HSJA)\label{fig:cifar_hsja}]{\includegraphics[width=0.22\textwidth]{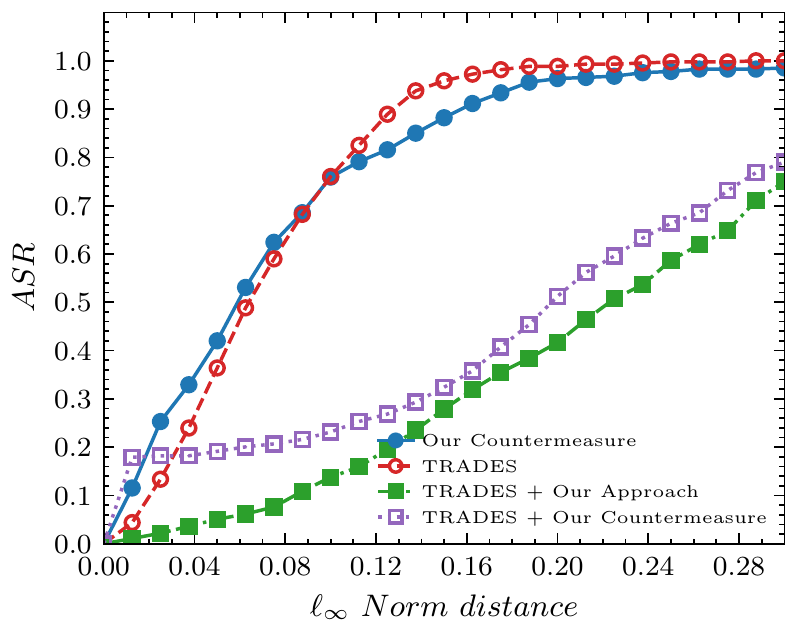}}
    \subfigure[CIFAR-10(HSJA)\label{fig:cifar_hsja_l2}]{\includegraphics[width=0.22\textwidth]{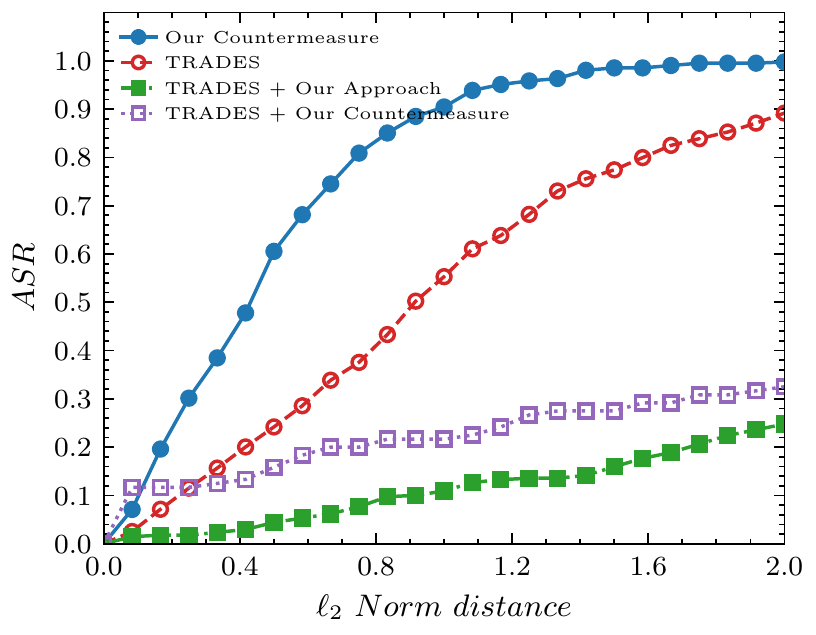}}
    \setlength{\belowcaptionskip}{-10pt}
    \caption{ASR comparison between TRADES, \name{}, countermeasure, combination of TRADES and \name{} and combination of TRADES and countermeasure.}
    \label{fig:uncertainty_counter_com}
    
\end{figure}

\subsection{Uncertainty-Aware Attack}
\label{sec:uncertainity_aware}
Another option for the attacker is to exploit \name{}'s non-deterministic nature. Specifically, the attacker can send $k$ ($k \geq$ 2) identical samples in the gradient estimation step, making $\overline{\mathcal{F}}(x_t;\theta)$ approach $\mathcal{F_Q}$'s decision boundary, which causes $\mathcal{C}$ to misclassify $x_t$. The algorithm  is detailed as Algorithm \ref{alg:uncertainty_aware} in the Appendix.                                                                                                       

We evaluate the uncertainty-aware attack with various values of $k$, as shown in Fig. \ref{fig:uncertainty}. We observe that as  $k$ increases,  \name{} gradually becomes invalid. This is because each inference's uncertainty is associated with $(\frac{1}{2})^{k-1}$ 
under our approach; thus, when $k \geq 5$, \name{} can not cause enough uncertainty to disrupt the gradient estimation step. However, such adaptive attacks have two main issues: first, multiple passes of the same input simultaneously can easily be detected. Second, the adaptive attack's adversarial samples have a $50\%$ success rate.

We propose a countermeasure to bypass such uncertainty-aware attack. 
Instead of adding a probabilistic nature to $\mathcal{\phi}_{x^*}(x_{t})$,  we could add an input induced behavior to ensure $\mathbb{E}[ |\beta_{1}|v_1] \to 0$, securing the validity of Eq. \ref{eq:fi}. This removes the uncertainty in \name{}, thwarting these uncertainty-aware attacks. Specifically, for each detected $x_t$, we leverage the last digit $f$ of the sum of $Conv(x_t)$, where $Conv(x_t)$ is the output of the first convolution layer of the base DNN. 
If $f$ is an even (odd) number, we change $\phi_{x^*}(x_{t})$; otherwise, not change it. Intuitively, the high-complexity and non-linearity of DNNs, 
ensures $f$ would be even or odd with a probability close to $50\%$. 
Furthermore, empirical observations show that the countermeasure causes similar accuracy loss as the original approach ($\approx 1.7\%$ for CIFAR-10 and $6.71\%$ for ImageNet).

Fig. \ref{fig:uncertainty_counter} shows our countermeasure results; we can see that the performance is comparable to our approach against HSJA ($\ell_2$, $\ell_\infty$) and SFA ($\ell_\infty$).
Like the adaptive attack in Sec. \ref{adaptive_attack1}, the uncertainty-aware attack performs well with SFA in the $\ell_2$ setting, indicating that the gradient estimation step in SFA ($\ell_2$) is more effective than other settings. 
Also, the adaptive attacks fail for all $k$ values against our countermeasure. The downside to our countermeasure is the additional cost in computing the sum of $Conv(x_t)$. We can easily combine our original approach and countermeasure to provide additional robustness against QBHL. Specifically, we can arbitrarily choose to apply the countermeasure for every set number of inputs.

Similar to Sec. \ref{sec:enchance_trades}, we investigate whether our countermeasure is also viable to use in conjunction with robust optimization-based approaches. As seen in Fig. \ref{fig:uncertainty_counter_com}, our countermeasure does enhance the performance of TRADES significantly. 

\section{Conclusion}

This paper presents \name{}, the first practical and generalizable framework to provide robustness for DNN models against hard-label query-based attacks. Inspired by the theoretical analysis of gradient estimation, the most critical procedure within most hard-label query-based attacks, \name{} can defend against hard-label query-based attacks by breaking the underlying condition for gradient estimation. \name{} leverages a second DNN model to distinguish between  normal and adversary queries. \name{} then introduces uncertainty into the prediction of the adversary queries. \name{} provides significant robustness against state-of-the-art hard-label query-based attacks under practical scenarios, with minimal accuracy loss and computation cost. \name{} can enhance state-of-art white-box defensive methods against query-based hard-label attacks. Finally, \name{} is still robust against a set of adaptive attacks that know its internal mechanisms.

\newpage
\bibliographystyle{plain}
\bibliography{ref.bib}

\begin{thebibliography}{10}

\bibitem{adv_model}
adv-training model.
\newblock {adv-training model}.
\newblock \url{https://github.com/MadryLab/cifar10_challenge}.

\bibitem{bojarski2017explaining}
Mariusz Bojarski, Philip Yeres, Anna Choromanska, Krzysztof Choromanski,
  Bernhard Firner, Lawrence Jackel, and Urs Muller.
\newblock Explaining how a deep neural network trained with end-to-end learning
  steers a car.
\newblock {\em arXiv preprint arXiv:1704.07911}, 2017.

\bibitem{ba}
Wieland Brendel, Jonas Rauber, and Matthias Bethge.
\newblock Decision-based adversarial attacks: Reliable attacks against
  black-box machine learning models, 2018.

\bibitem{bug_3}
Amar Budhiraja, Kartik Dutta, Raghu Reddy, and Manish Shrivastava.
\newblock Dwen: deep word embedding network for duplicate bug report detection
  in software repositories.
\newblock In {\em Proceedings of the 40th International Conference on software
  engineering: companion proceeedings}, pages 193--194, 2018.

\bibitem{pgd}
Paul~H Calamai and Jorge~J Mor{\'e}.
\newblock Projected gradient methods for linearly constrained problems.
\newblock {\em Mathematical programming}, 39(1):93--116, 1987.

\bibitem{region-based}
Xiaoyu Cao and Neil~Zhenqiang Gong.
\newblock Mitigating evasion attacks to deep neural networks via region-based
  classification.
\newblock In {\em Proceedings of the 33rd Annual Computer Security Applications
  Conference}, pages 278--287, 2017.

\bibitem{carlini2017adversarial}
Nicholas Carlini and David Wagner.
\newblock Adversarial examples are not easily detected: Bypassing ten detection
  methods.
\newblock In {\em Proceedings of the 10th ACM Workshop on Artificial
  Intelligence and Security}, pages 3--14, 2017.

\bibitem{cw}
Nicholas Carlini and David Wagner.
\newblock Towards evaluating the robustness of neural networks.
\newblock In {\em 2017 ieee symposium on security and privacy (sp)}, pages
  39--57. IEEE, 2017.

\bibitem{hsja}
Jianbo Chen, Michael~I Jordan, and Martin~J Wainwright.
\newblock Hopskipjumpattack: A query-efficient decision-based attack.
\newblock In {\em 2020 ieee symposium on security and privacy (sp)}, pages
  1277--1294. IEEE, 2020.

\bibitem{fgsm}
Jinyin Chen, Yangyang Wu, Xuanheng Xu, Yixian Chen, Haibin Zheng, and Qi~Xuan.
\newblock Fast gradient attack on network embedding.
\newblock {\em arXiv preprint arXiv:1809.02797}, 2018.

\bibitem{stateful}
Steven Chen, Nicholas Carlini, and David Wagner.
\newblock Stateful detection of black-box adversarial attacks.
\newblock In {\em Proceedings of the 1st ACM Workshop on Security and Privacy
  on Artificial Intelligence}, pages 30--39, 2020.

\bibitem{sfa}
Weilun Chen, Zhaoxiang Zhang, Xiaolin Hu, and Baoyuan Wu.
\newblock Boosting decision-based black-box adversarial attacks with random
  sign flip.
\newblock In {\em European Conference on Computer Vision}, pages 276--293.
  Springer, 2020.

\bibitem{DBLP:journals/corr/abs-1909-10773}
Minhao Cheng, Simranjit Singh, Patrick~H. Chen, Pin{-}Yu Chen, Sijia Liu, and
  Cho{-}Jui Hsieh.
\newblock Sign-opt: {A} query-efficient hard-label adversarial attack.
\newblock {\em CoRR}, abs/1909.10773, 2019.

\bibitem{nlp_3}
Ronan Collobert and Jason Weston.
\newblock A unified architecture for natural language processing: Deep neural
  networks with multitask learning.
\newblock In {\em Proceedings of the 25th international conference on Machine
  learning}, pages 160--167, 2008.

\bibitem{evolutionary}
Kalyanmoy Deb, Ashish Anand, and Dhiraj Joshi.
\newblock A computationally efficient evolutionary algorithm for real-parameter
  optimization.
\newblock {\em Evolutionary computation}, 10(4):371--395, 2002.

\bibitem{deng2012mnist}
Li~Deng.
\newblock The mnist database of handwritten digit images for machine learning
  research [best of the web].
\newblock {\em IEEE Signal Processing Magazine}, 29(6):141--142, 2012.

\bibitem{nlp_2}
Yoav Goldberg.
\newblock A primer on neural network models for natural language processing.
\newblock {\em Journal of Artificial Intelligence Research}, 57:345--420, 2016.

\bibitem{nlp_4}
Yoav Goldberg.
\newblock Neural network methods for natural language processing.
\newblock {\em Synthesis lectures on human language technologies},
  10(1):1--309, 2017.

\bibitem{gong2017adversarial}
Zhitao Gong, Wenlu Wang, and Wei-Shinn Ku.
\newblock Adversarial and clean data are not twins, 2017.

\bibitem{adv}
Ian~J Goodfellow, Jonathon Shlens, and Christian Szegedy.
\newblock Explaining and harnessing adversarial examples.
\newblock {\em arXiv preprint arXiv:1412.6572}, 2014.

\bibitem{google_cloud}
Google.
\newblock {Google Cloud}.
\newblock \url{https://cloud.google.com/}.

\bibitem{grosse2016adversarial}
Kathrin Grosse, Nicolas Papernot, Praveen Manoharan, Michael Backes, and
  Patrick McDaniel.
\newblock Adversarial perturbations against deep neural networks for malware
  classification, 2016.

\bibitem{guo2020practical}
Junfeng Guo and Cong Liu.
\newblock Practical poisoning attacks on neural networks.
\newblock In {\em European Conference on Computer Vision}, pages 142--158.
  Springer, 2020.

\bibitem{resnet}
Kaiming He, Xiangyu Zhang, Shaoqing Ren, and Jian Sun.
\newblock Deep residual learning for image recognition.
\newblock In {\em Proceedings of the IEEE conference on computer vision and
  pattern recognition}, pages 770--778, 2016.

\bibitem{ql}
Andrew Ilyas, Logan Engstrom, Anish Athalye, and Jessy Lin.
\newblock Black-box adversarial attacks with limited queries and information.
\newblock In {\em International Conference on Machine Learning}, pages
  2137--2146. PMLR, 2018.

\bibitem{robot_2}
Alaa~Abdulhady Jaber and Robert Bicker.
\newblock Fault diagnosis of industrial robot gears based on discrete wavelet
  transform and artificial neural network.
\newblock {\em Insight-Non-Destructive Testing and Condition Monitoring},
  58(4):179--186, 2016.

\bibitem{adv_2}
Robin Jia and Percy Liang.
\newblock Adversarial examples for evaluating reading comprehension systems.
\newblock {\em arXiv preprint arXiv:1707.07328}, 2017.

\bibitem{Keras_model}
Keras.
\newblock {Keras model}.
\newblock \url{https://keras.io/api/applications/}.

\bibitem{phsGAN}
Zelun Kong, Junfeng Guo, Ang Li, and Cong Liu.
\newblock Physgan: Generating physical-world-resilient adversarial examples for
  autonomous driving.
\newblock In {\em Proceedings of the IEEE/CVF Conference on Computer Vision and
  Pattern Recognition}, pages 14254--14263, 2020.

\bibitem{blacklight}
Huiying Li, Shawn Shan, Emily Wenger, Jiayun Zhang, Haitao Zheng, and Ben~Y
  Zhao.
\newblock Blacklight: Defending black-box adversarial attacks on deep neural
  networks.
\newblock {\em arXiv preprint arXiv:2006.14042}, 2020.

\bibitem{nattack}
Yandong Li, Lijun Li, Liqiang Wang, Tong Zhang, and Boqing Gong.
\newblock Nattack: Learning the distributions of adversarial examples for an
  improved black-box attack on deep neural networks.
\newblock In {\em International Conference on Machine Learning}, pages
  3866--3876. PMLR, 2019.

\bibitem{bug_2}
Yi~Li, Shaohua Wang, Tien~N Nguyen, and Son Van~Nguyen.
\newblock Improving bug detection via context-based code representation
  learning and attention-based neural networks.
\newblock {\em Proceedings of the ACM on Programming Languages},
  3(OOPSLA):1--30, 2019.

\bibitem{binary_4}
Bingchang Liu, Wei Huo, Chao Zhang, Wenchao Li, Feng Li, Aihua Piao, and Wei
  Zou.
\newblock $\alpha$diff: cross-version binary code similarity detection with
  dnn.
\newblock In {\em Proceedings of the 33rd ACM/IEEE International Conference on
  Automated Software Engineering}, pages 667--678, 2018.

\bibitem{chen}
Yanpei Liu, Xinyun Chen, Chang Liu, and Dawn Song.
\newblock Delving into transferable adversarial examples and black-box attacks.
\newblock {\em arXiv preprint arXiv:1611.02770}, 2016.

\bibitem{trojan}
Yingqi Liu, Shiqing Ma, Yousra Aafer, Wen-Chuan Lee, Juan Zhai, Weihang Wang,
  and Xiangyu Zhang.
\newblock Trojaning attack on neural networks.
\newblock 2017.

\bibitem{adv_train}
Aleksander Madry, Aleksandar Makelov, Ludwig Schmidt, Dimitris Tsipras, and
  Adrian Vladu.
\newblock Towards deep learning models resistant to adversarial attacks.
\newblock {\em arXiv preprint arXiv:1706.06083}, 2017.

\bibitem{deep_id}
Wanli Ouyang, Xiaogang Wang, Xingyu Zeng, Shi Qiu, Ping Luo, Yonglong Tian,
  Hongsheng Li, Shuo Yang, Zhe Wang, Chen-Change Loy, et~al.
\newblock Deepid-net: Deformable deep convolutional neural networks for object
  detection.
\newblock In {\em Proceedings of the IEEE conference on computer vision and
  pattern recognition}, pages 2403--2412, 2015.

\bibitem{pa}
Nicolas Papernot, Patrick McDaniel, Ian Goodfellow, Somesh Jha, Z~Berkay Celik,
  and Ananthram Swami.
\newblock Practical black-box attacks against machine learning.
\newblock In {\em Proceedings of the 2017 ACM on Asia conference on computer
  and communications security}, pages 506--519, 2017.

\bibitem{distill}
Nicolas Papernot, Patrick McDaniel, Xi~Wu, Somesh Jha, and Ananthram Swami.
\newblock Distillation as a defense to adversarial perturbations against deep
  neural networks.
\newblock In {\em 2016 IEEE symposium on security and privacy (SP)}, pages
  582--597. IEEE, 2016.

\bibitem{bug}
Michael Pradel and Koushik Sen.
\newblock Deepbugs: A learning approach to name-based bug detection.
\newblock {\em Proceedings of the ACM on Programming Languages},
  2(OOPSLA):1--25, 2018.

\bibitem{auto_deep}
Ahmad~EL Sallab, Mohammed Abdou, Etienne Perot, and Senthil Yogamani.
\newblock Deep reinforcement learning framework for autonomous driving.
\newblock {\em Electronic Imaging}, 2017(19):70--76, 2017.

\bibitem{face_adv}
Mahmood Sharif, Sruti Bhagavatula, Lujo Bauer, and Michael~K Reiter.
\newblock Accessorize to a crime: Real and stealthy attacks on state-of-the-art
  face recognition.
\newblock In {\em Proceedings of the 2016 acm sigsac conference on computer and
  communications security}, pages 1528--1540, 2016.

\bibitem{binary}
Eui Chul~Richard Shin, Dawn Song, and Reza Moazzezi.
\newblock Recognizing functions in binaries with neural networks.
\newblock In {\em 24th $\{$USENIX$\}$ Security Symposium ($\{$USENIX$\}$
  Security 15)}, pages 611--626, 2015.

\bibitem{shokri2017membership}
Reza Shokri, Marco Stronati, Congzheng Song, and Vitaly Shmatikov.
\newblock Membership inference attacks against machine learning models, 2017.

\bibitem{deep_face}
Yi~Sun, Xiaogang Wang, and Xiaoou Tang.
\newblock Deep learning face representation from predicting 10,000 classes.
\newblock In {\em Proceedings of the IEEE conference on computer vision and
  pattern recognition}, pages 1891--1898, 2014.

\bibitem{tencent}
Tencent.
\newblock {Tencent Image Classification API}.
\newblock \url{https://ai.qq.com/hr/youtu.shtml}.

\bibitem{deep_auto2}
Yuchi Tian, Kexin Pei, Suman Jana, and Baishakhi Ray.
\newblock Deeptest: Automated testing of deep-neural-network-driven autonomous
  cars.
\newblock In {\em Proceedings of the 40th international conference on software
  engineering}, pages 303--314, 2018.

\bibitem{deeptest}
Yuchi Tian, Kexin Pei, Suman Jana, and Baishakhi Ray.
\newblock Deeptest: Automated testing of deep-neural-network-driven autonomous
  cars.
\newblock In {\em Proceedings of the 40th international conference on software
  engineering}, pages 303--314, 2018.

\bibitem{trade_model}
TRADES.
\newblock {TRADES model}.
\newblock \url{https://github.com/yaodongyu/TRADES}.

\bibitem{e_adv_train}
Florian Tram{\`e}r, Alexey Kurakin, Nicolas Papernot, Ian Goodfellow, Dan
  Boneh, and Patrick McDaniel.
\newblock Ensemble adversarial training: Attacks and defenses.
\newblock {\em arXiv preprint arXiv:1705.07204}, 2017.

\bibitem{robot}
Arun~T Vemuri and Marios~M Polycarpou.
\newblock Neural-network-based robust fault diagnosis in robotic systems.
\newblock {\em IEEE Transactions on neural networks}, 8(6):1410--1420, 1997.

\bibitem{adv_gan}
Chaowei Xiao, Bo~Li, Jun-Yan Zhu, Warren He, Mingyan Liu, and Dawn Song.
\newblock Generating adversarial examples with adversarial networks.
\newblock {\em arXiv preprint arXiv:1801.02610}, 2018.

\bibitem{bug_4}
Yan Xiao, Jacky Keung, Qing Mi, and Kwabena~E Bennin.
\newblock Bug localization with semantic and structural features using
  convolutional neural network and cascade forest.
\newblock In {\em Proceedings of the 22nd International Conference on
  Evaluation and Assessment in Software Engineering 2018}, pages 101--111,
  2018.

\bibitem{deepbit}
Weilin Xu, David Evans, and Yanjun Qi.
\newblock Feature squeezing: Detecting adversarial examples in deep neural
  networks.
\newblock {\em arXiv preprint arXiv:1704.01155}, 2017.

\bibitem{binary_3}
Xiaojun Xu, Chang Liu, Qian Feng, Heng Yin, Le~Song, and Dawn Song.
\newblock Neural network-based graph embedding for cross-platform binary code
  similarity detection.
\newblock In {\em Proceedings of the 2017 ACM SIGSAC Conference on Computer and
  Communications Security}, pages 363--376, 2017.

\bibitem{nlp}
Wenpeng Yin, Katharina Kann, Mo~Yu, and Hinrich Sch{\"u}tze.
\newblock Comparative study of cnn and rnn for natural language processing.
\newblock {\em arXiv preprint arXiv:1702.01923}, 2017.

\bibitem{trades}
Hongyang Zhang, Yaodong Yu, Jiantao Jiao, Eric Xing, Laurent El~Ghaoui, and
  Michael Jordan.
\newblock Theoretically principled trade-off between robustness and accuracy.
\newblock In {\em International Conference on Machine Learning}, pages
  7472--7482. PMLR, 2019.

\bibitem{deepbillboard}
Husheng Zhou, Wei Li, Zelun Kong, Junfeng Guo, Yuqun Zhang, Bei Yu, Lingming
  Zhang, and Cong Liu.
\newblock Deepbillboard: Systematic physical-world testing of autonomous
  driving systems.
\newblock In {\em 2020 IEEE/ACM 42nd International Conference on Software
  Engineering (ICSE)}, pages 347--358. IEEE, 2020.

\bibitem{binary_2}
Fei Zuo, Xiaopeng Li, Patrick Young, Lannan Luo, Qiang Zeng, and Zhexin Zhang.
\newblock Neural machine translation inspired binary code similarity comparison
  beyond function pairs.
\newblock {\em arXiv preprint arXiv:1808.04706}, 2018.

\end{thebibliography}
\newpage
\appendix

\section{Proof for gradient estimation in HSJA}
\label{sec:theorem2}
\begin{theorem}
\label{theorem:1}
According to HSJA~\cite{hsja}, given an orthogonal base of $\mathbb{R}^{d}:v_1=\frac{\nabla\mathcal{S}_{x^*}(x_{t})}{\norm{\nabla\mathcal{S}_{x^*}{x_{t}}}_{2}},v_2,v_3,...,v_d$, each $u_b$ can be represented as $u_b=\sum_{i=1}^{d}\beta_{i}v_i$ with $|\beta_i|\in(0,1)$.
\end{theorem}

\noindent With sufficient samples for Monte Carlo estimation, the effectiveness of the gradient estimation for HSJA can be guaranteed as   \[ \lim_{\delta \to 0} cos\angle(\widetilde{\nabla S}_{x^*}(x_{t}) ,\nabla \mathcal{S}_{x^*}(x_{t}))=1 \].

\begin{proof} 
By Taylor's expansion, we have that $\mathcal{S}_{x^*}(x_t+\delta u)=\mathcal{S}_{x^*}(x_t)+\delta \nabla \mathcal{S}_{x^*}(x_{t})^Tu+\frac{1}{2}\delta^2 u^{T}\nabla^{2}\mathcal{S}_{x^{*}}(x_t) u$, where $u \in (0,1)$ is a random vector and $x_t$ lies between $x$ and $x+\delta u$. Through intentionally making  $\mathcal{S}_{x^*}(x_t)=0$,
 we have:
 \begin{equation}\label{eq:1}
     \mathcal{S}_{x^*}(x_t+\delta u)=\delta \nabla \mathcal{S}_{x^*}(x_{t})^Tu+\frac{1}{2}\delta^2 u^{T}\nabla^{2}\mathcal{S}_{x^*}(x_t) u
 \end{equation}
 
\noindent Since $\mathcal{S}$ is Lipschitsz continuous, the second-order can be bounded as:
\begin{equation}\label{eq:2}
     |\frac{1}{2}\delta^2 u^{T}\nabla^{2}\mathcal{S}_{x^{*}}(x_t) u| \leq \frac{1}{2}L\delta^2
\end{equation}

\noindent Therefore, we have that:

\begin{equation}\label{eq:condition}
    \phi_{x^{*}}(x_{t}+\delta u_b)=\left\{
             \begin{array}{lr}
                1  &\text{if} \; \nabla\mathcal{S}_{x^*}(x_t)^{T}u > \frac{1}{2}L\delta, \\
            
                 -1 &\text{if} \; \nabla{S}_{x^*}(x_t)^{T}u < - \frac{1}{2}L\delta
             
             \end{array}
\right.
\end{equation}

Given an orthogonal base of~$\mathbb{R}^{d}$:$v_1=\frac{\nabla\mathcal{S}_{x^*}(x_t)}{||\nabla\mathcal{S}_{x^*}(x_t)||_2}$,$v_2$,$\dots$,$v_d$. 
Thus each $u_b$ can be represented as $u_b=\sum_{i=1}^{d} \beta_{i}v_{i}$, with $\beta_{i} \in (0,1)$. Let $q$ be the probability of event $|\nabla{S}_{x^*}(x_t)^{T}u|\leq\frac{1}{2}L\delta$. Therefore, we can bound the difference between $\widetilde{\nabla S}_{x^*}(x_{t})=\mathbb{E}[\phi_{x^*}(x_t+u_b)u_b]$ and $\mathbb{E}[|\beta_1|v_1]$ as:

\begin{equation}\label{eq:bounce1}
    ||\widetilde{\nabla S}_{x^*}(x_{t})-\mathbb{E}[|\beta_1|v_1]||\leq 3q
\end{equation}

\noindent  Basically, $\mathbb{E} [|\beta_1|v_1]$ and $\nabla \mathcal{S}_{x^*}(x_{t})$ can be easily bridged as: 

\begin{equation} \label{eq:bridge}
    \mathbb{E} [|\beta_1|v_1]=\sum_{b=1}^{B} |\beta_1|\frac{\nabla \mathcal{S}_{x^*}(x_{t})}{||\nabla \mathcal{S}_{x^*}(x_{t})||}=\frac{\nabla \mathcal{S}_{x^*}(x_{t})}{||\nabla \mathcal{S}_{x^*}(x_{t})||}\mathbb{E}[|\beta_1|],
\end{equation}

Therefore, combining Eq. \ref{eq:bounce1} and Eq. \ref{eq:bridge}, we have:
\begin{equation}\label{eq:final}
    cos\angle(\widetilde{\nabla S}_{x^*}(x_{t}) ,\nabla \mathcal{S}_{x^*}(x_{t})) \geq 1-\frac{1}{2} \bigg(\frac{3q}{\mathbb{E} [|\beta_1|]}\bigg)^2,
\end{equation}

By observing that $<\frac{\nabla\mathcal{S}_{x^*}(x_t)}{||\nabla\mathcal{S}_{x^*}(x_t)||_2},u>$ is a Beta distribution $\mathcal{B}(\frac{1}{2},\frac{d-1}{2})$, we can bound $q$ as:

\begin{equation}\label{eq:p}
q\leq \frac{L\delta}{\mathcal{B}(\frac{1}{2},\frac{d-1}{2})||\nabla\mathcal{S}_{x^*}(x_t)||_2}, 
\end{equation}

Plugging Eq. \ref{eq:p} into Eq. \ref{eq:final}, we get:

\begin{equation}
cos\angle(\widetilde{\nabla S}_{x^*}(x_{t}) ,\nabla \mathcal{S}_{x^*}(x_{t})) \geq 1-\frac{9L^2\delta^2(d-1)^2}{8||\nabla\mathcal{S}_{x^*}(x_t)||_2^{2}},
\end{equation}

As a consequence, we established: \[ \lim_{\delta \to 0} cos\angle(\widetilde{\nabla S}_{x^*}(x_{t}) ,\nabla \mathcal{S}_{x^*}(x_{t}))=1 \]

\end{proof}

\section{Target model descriptions}
 Tables~\ref{table:MNIST1}-\ref{table:training_configuration} show the structure and parameters of target models used in our evaluation for various datasets as discussed in Sec. \ref{sec:exp}.
The structures of models for MNIST, 
 GTSRB datasets are shown in Tables~\ref{table:MNIST1} and ~\ref{table:MNIST2}. The training configuration for these models is shown in Table ~\ref{table:training_configuration}. 
The ResNet.V1.50 model for CIFAR-10 and ImageNet are implemented using prior work~\cite{resnet} structures and training configurations.

\begin{table}[H]
\centering
\begin{tabular}{ c c c  } 
\hline
Layer Type & Known Model\\
\hline
Convolution + ReLU&3$\times$3$\times$32\\
Convolution + ReLU&3$\times$3$\times$32\\
Max Pooling&2$\times$2\\
Convolution + ReLU&3$\times$3$\times$32\\
Max Pooling&2$\times$2\\
Fully Connected + ReLU&256\\
Fully Connected + ReLU&256\\
Softmax&10\\
\hline

\end{tabular}
\caption{The Structure of MNIST Model}
\label{table:MNIST1}
\end{table}

\begin{table}[H]
\centering
\begin{tabular}{ c c c  } 
\hline
Layer Type & Known Model\\
\hline
Convolution + ReLU&3$\times$3$\times$32\\
Convolution + ReLU&3$\times$3$\times$32\\
Max Pooling&2$\times$2\\
Convolution + ReLU&3$\times$3$\times$128\\
Convolution + ReLU&3$\times$3$\times$64\\
Convolution + ReLU&3$\times$3$\times$64\\
Max Pooling&2$\times$2\\
Fully Connected + ReLU&256\\
Fully Connected + ReLU&256\\
Fully Connected + ReLU&128\\
Softmax&43\\
\hline
\end{tabular}
\caption{The Structure of GTSRB Model}
\label{table:MNIST2}

\end{table}

\begin{table}[H]
\centering
\begin{tabular}{ c c} 
\hline
Parameter & Models(MNIST,GTSRB)\\
\hline
Learning Rate&0.01\\
Momentum&0.9\\
Dropout&0.5\\
Batch Size&128\\
Epochs&50\\
\hline
\end{tabular}
\caption{Training Configuration}
\label{table:training_configuration}
\end{table}

\section{Uncertainty-Aware Attack Algorithm}
For each QBHL attack , we incorporate Algorithm~\ref{alg:uncertainty_aware} into the gradient estimation procedure to conduct Uncertainty-Aware Attack.

\begin{algorithm}[h]
    \DontPrintSemicolon
    \LinesNumbered
    \SetKwInOut{Input}{input}
    \SetKwInOut{Output}{output}
    \SetKwComment{Comment}{\#}{}

    \Input{Target Model $\mathcal{C}$}

    \Input{The values of $k$}
   
    \Input{The basic input $x^*$}

    Set \textbf{$x_t$}  approaches the boundary of $\mathcal{C}$~($\phi_{x^*}(x_t)=1$);
    
    Set $i$=0;
    
    Set A=0
    
    Given each $x_t+\delta u$

    \While{$i \leq k$}{

        A=A+$\phi_{x^*}(x_t+\delta u)$}
    \Return{$A$}
    
    \If{$\frac{A}{k}=0$}
    {We set $\phi_{x^*}(x_t+\delta u)=-1$ }
    \Else{We set $\phi_{x^*}(x_t+\delta u)=1$}

    \caption{\textbf{Uncertainty-Aware Attack}}
    \label{alg:uncertainty_aware}
\end{algorithm}

\begin{figure*}[!t]
\centering
\subfigure[MNIST(BA)\label{fig:mnist_ba}]{\includegraphics[width=0.24\textwidth]{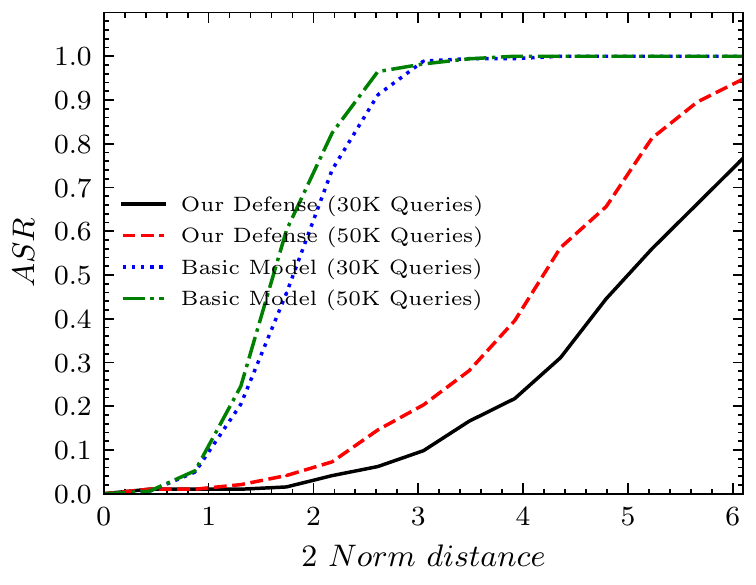}}
\subfigure[MNIST(BA)\label{fig:mnist_ba_li}]{\includegraphics[width=0.24\textwidth]{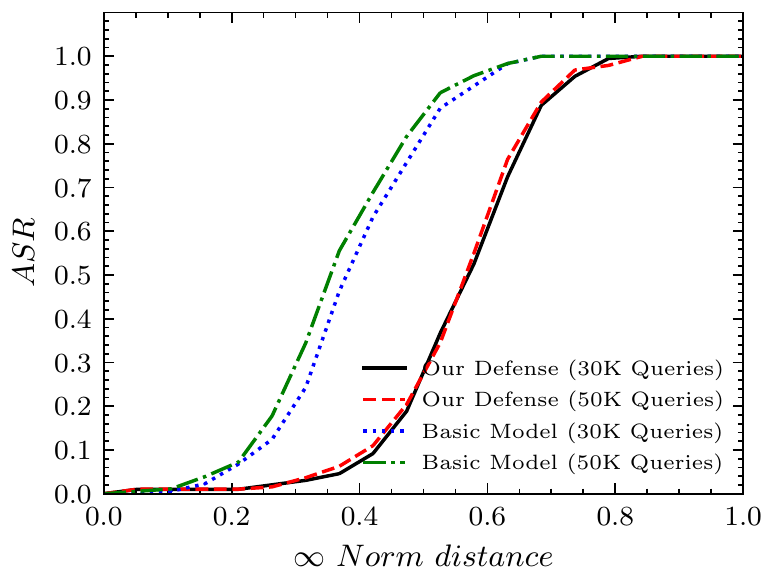}}
\subfigure[CIFAR(BA)\label{fig:cifar_ba}]{\includegraphics[width=0.24\textwidth]{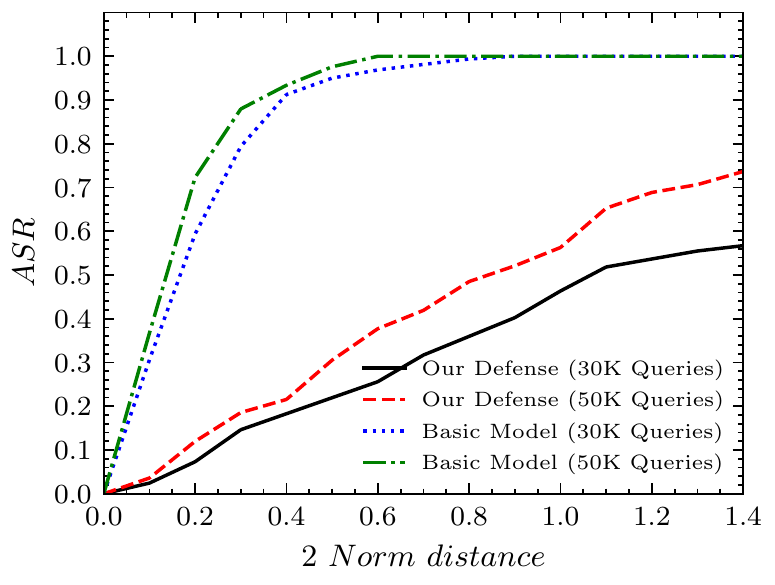}}
\subfigure[CIFAR(BA)\label{fig:cifar_ba_li}]{\includegraphics[width=0.24\textwidth]{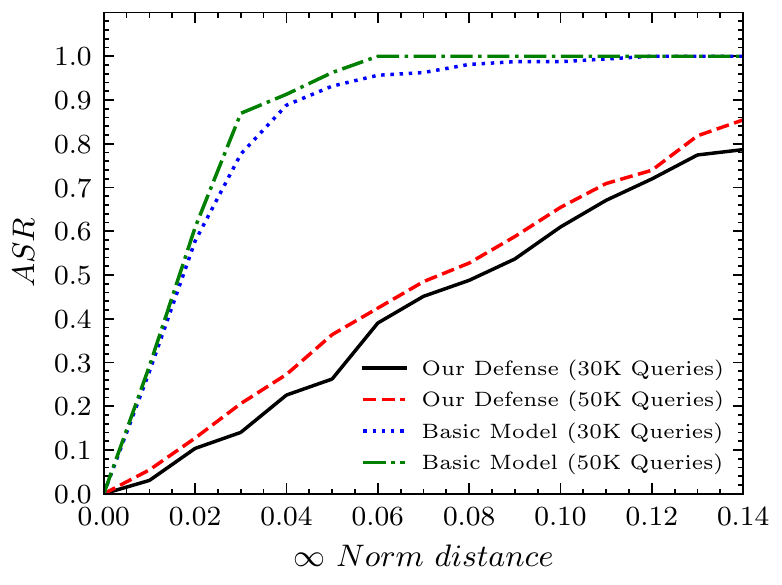}}
\subfigure[GTSRB(BA)\label{fig:gtsrb_ba}]{\includegraphics[width=0.24\textwidth]{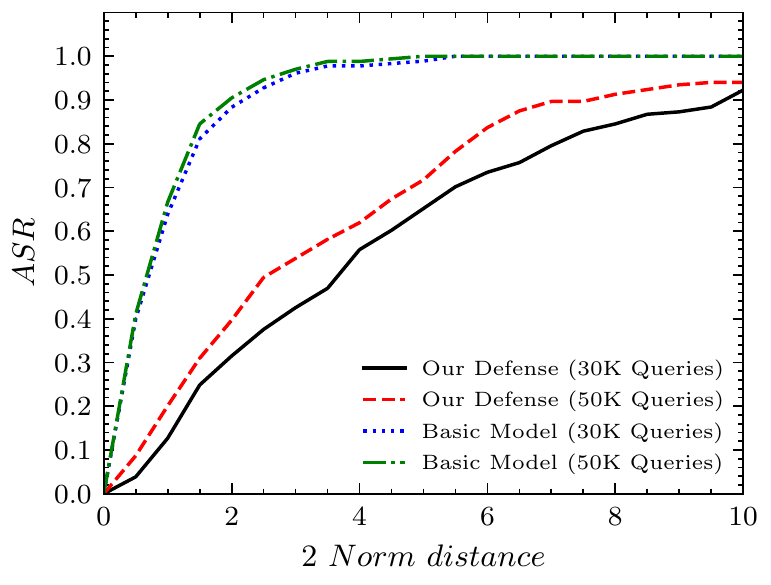}}
\subfigure[GTSRB(BA)\label{fig:gtsrb_ba_li}]{\includegraphics[width=0.24\textwidth]{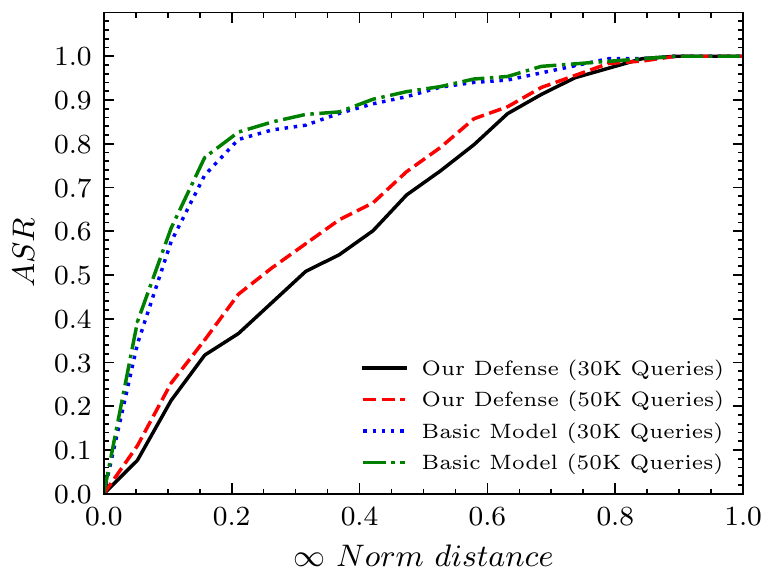}}
\subfigure[ImageNet(BA)\label{fig:imagenet_ba_l2}]{\includegraphics[width=0.24\textwidth]{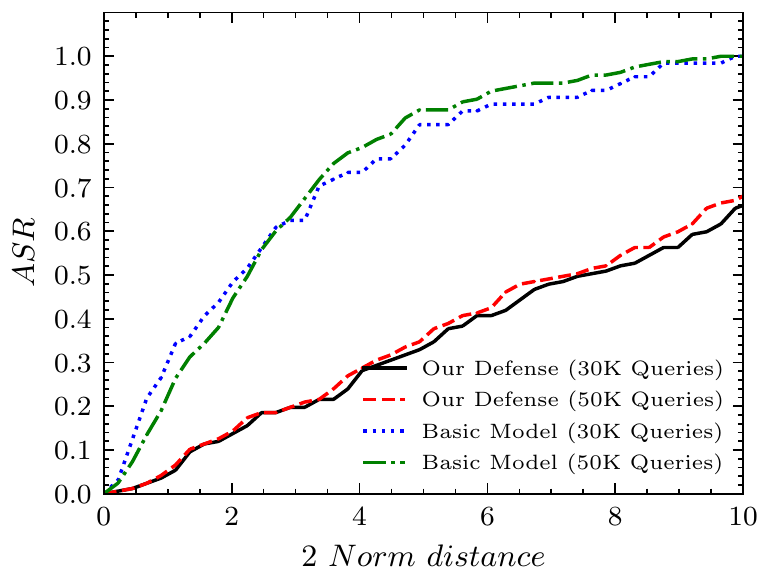}}
\subfigure[ImageNet(BA)\label{fig:imagenet_ba_li}]{\includegraphics[width=0.24\textwidth]{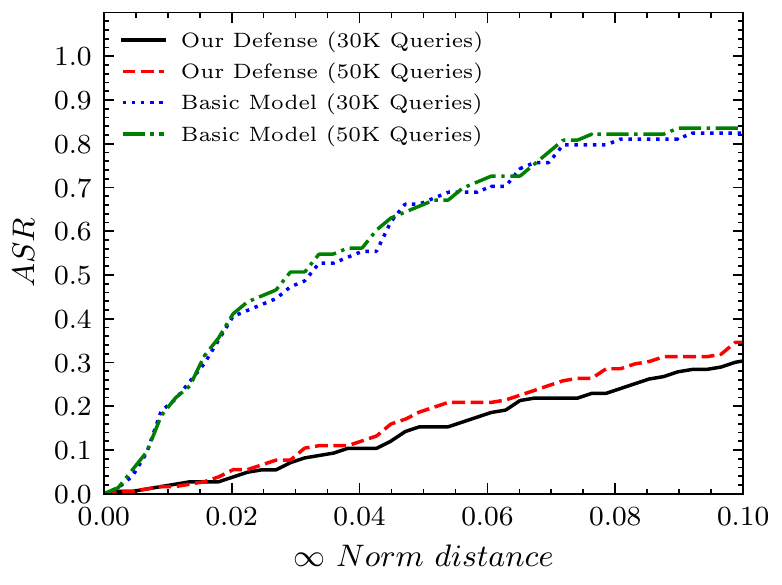}}
\subfigure[CIFAR-10(Sign-OPT)\label{fig:cifar_sopt}]{\includegraphics[width=0.24\textwidth]{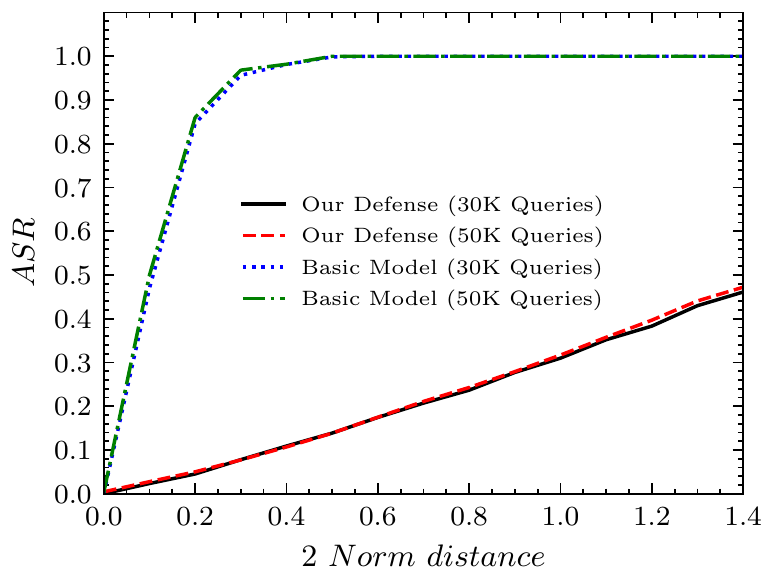}}
\subfigure[ImageNet(Sign-OPT)\label{fig:cifar_sopt}]{\includegraphics[width=0.24\textwidth]{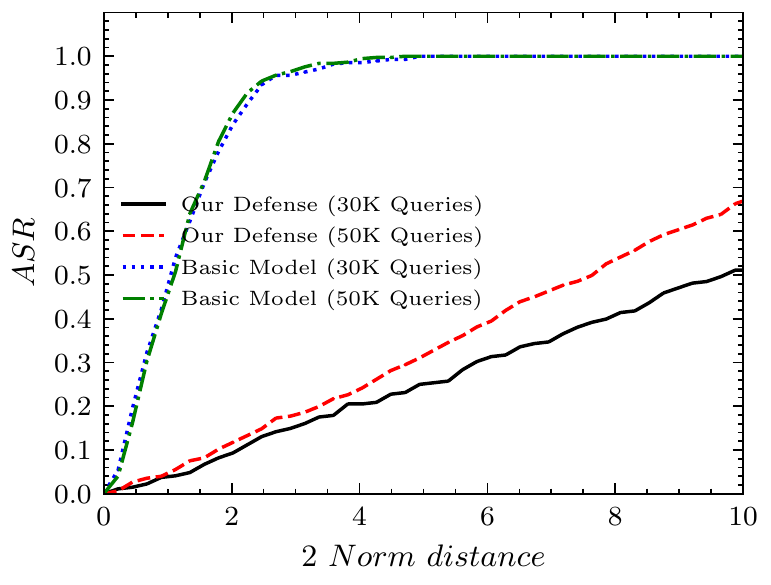}}
\subfigure[GTSRB(HSJA)\label{fig:gtsrb_hsja_li}]{\includegraphics[width=0.24\textwidth]{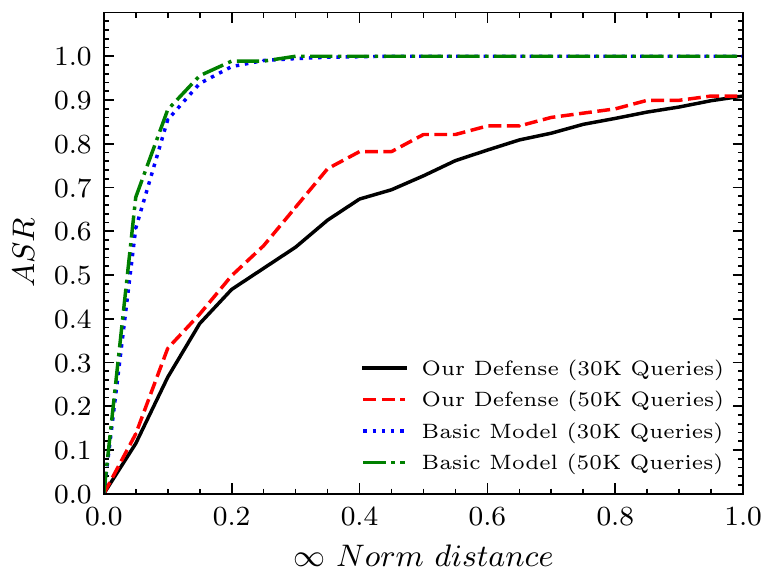}}
\subfigure[GTSRB(HSJA)\label{fig:gtsrb_hsja}]{\includegraphics[width=0.24\textwidth]{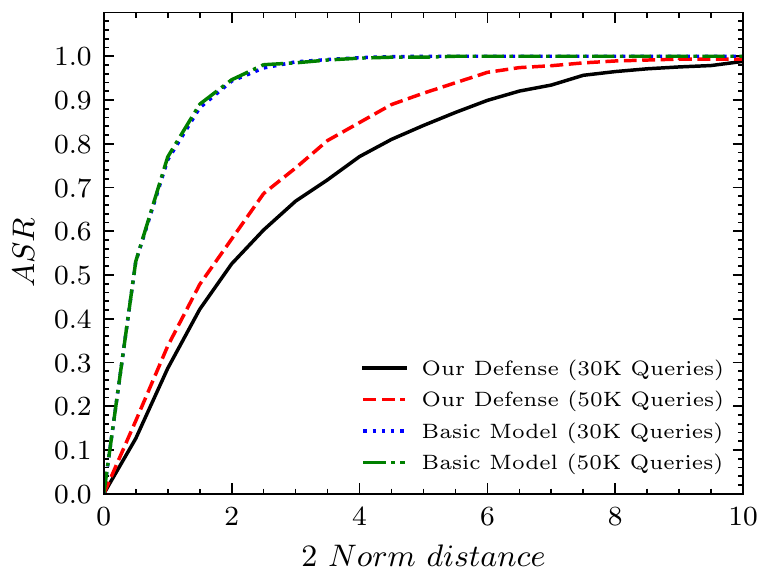}}
\subfigure[GTSRB(SFA)\label{fig:gtsrb_sfa_l2}]{\includegraphics[width=0.24\textwidth]{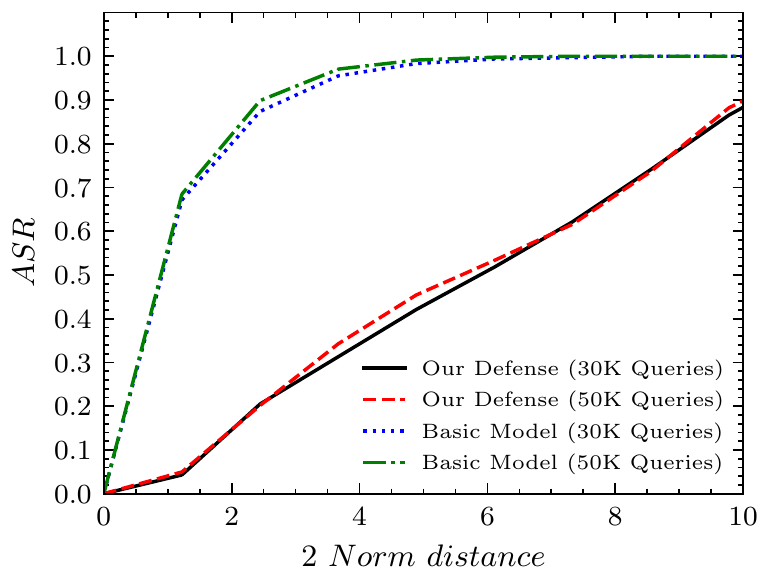}}
\subfigure[GTSRB(SFA)\label{fig:gtsrb_sfa}]{\includegraphics[width=0.24\textwidth]{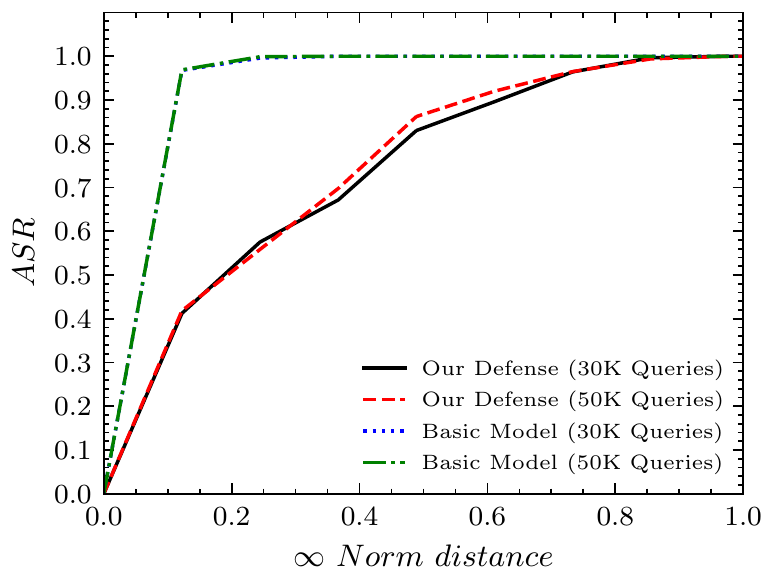}}
\caption{ASR under different thresholds for perturbation under SFA and HSJA, across various tasks and settings.}
\label{fig:asr_2}
\end{figure*}

\begin{figure*}[!t]
\centering
\subfigure[Basic Image\label{fig:basic}]{\includegraphics[width=0.2\textwidth]{Analasis/basic_img.pdf}}
\subfigure[BA\label{fig:hsja}]{\includegraphics[width=0.42\textwidth]{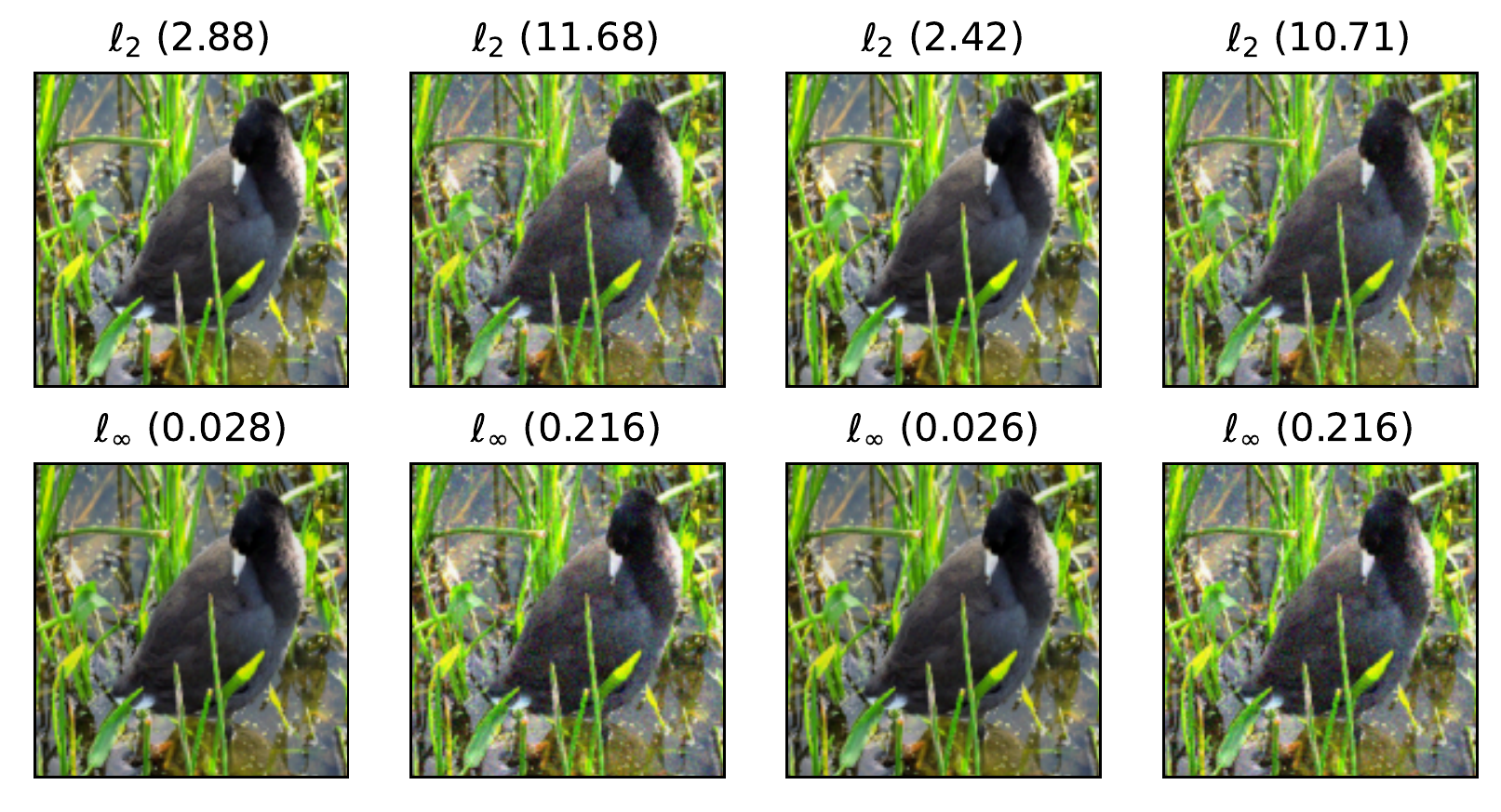}}
\subfigure[Sign-OPT\label{fig:sign-opt}]{\includegraphics[width=0.21\textwidth]{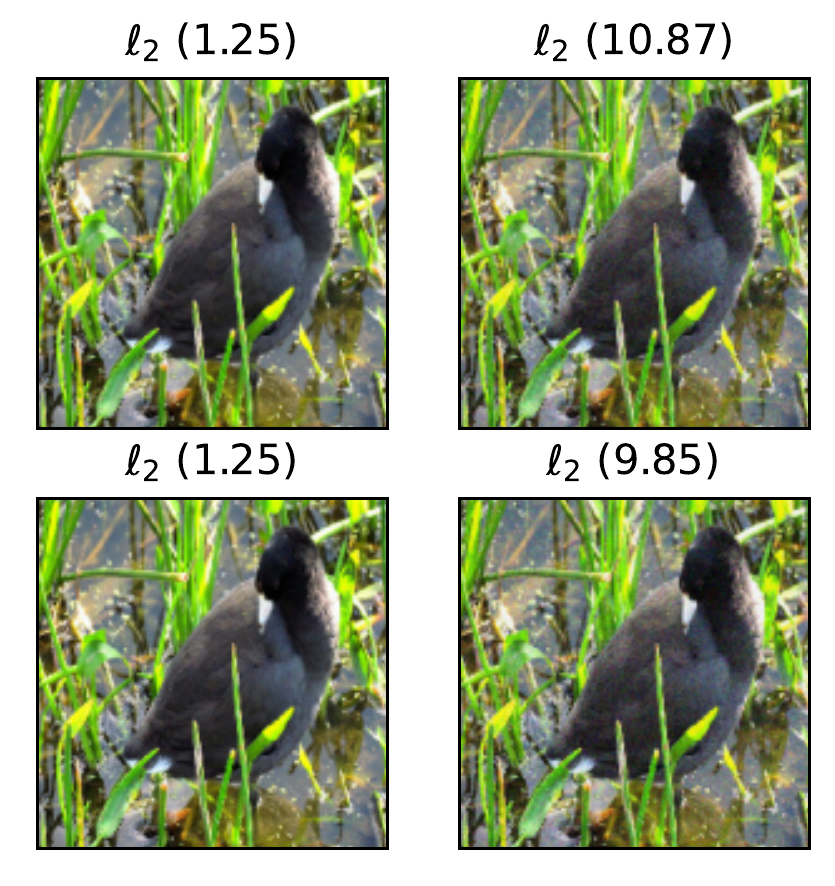}}
\subfigure[SFA\label{fig:sfa}]{\includegraphics[width=0.42\textwidth]{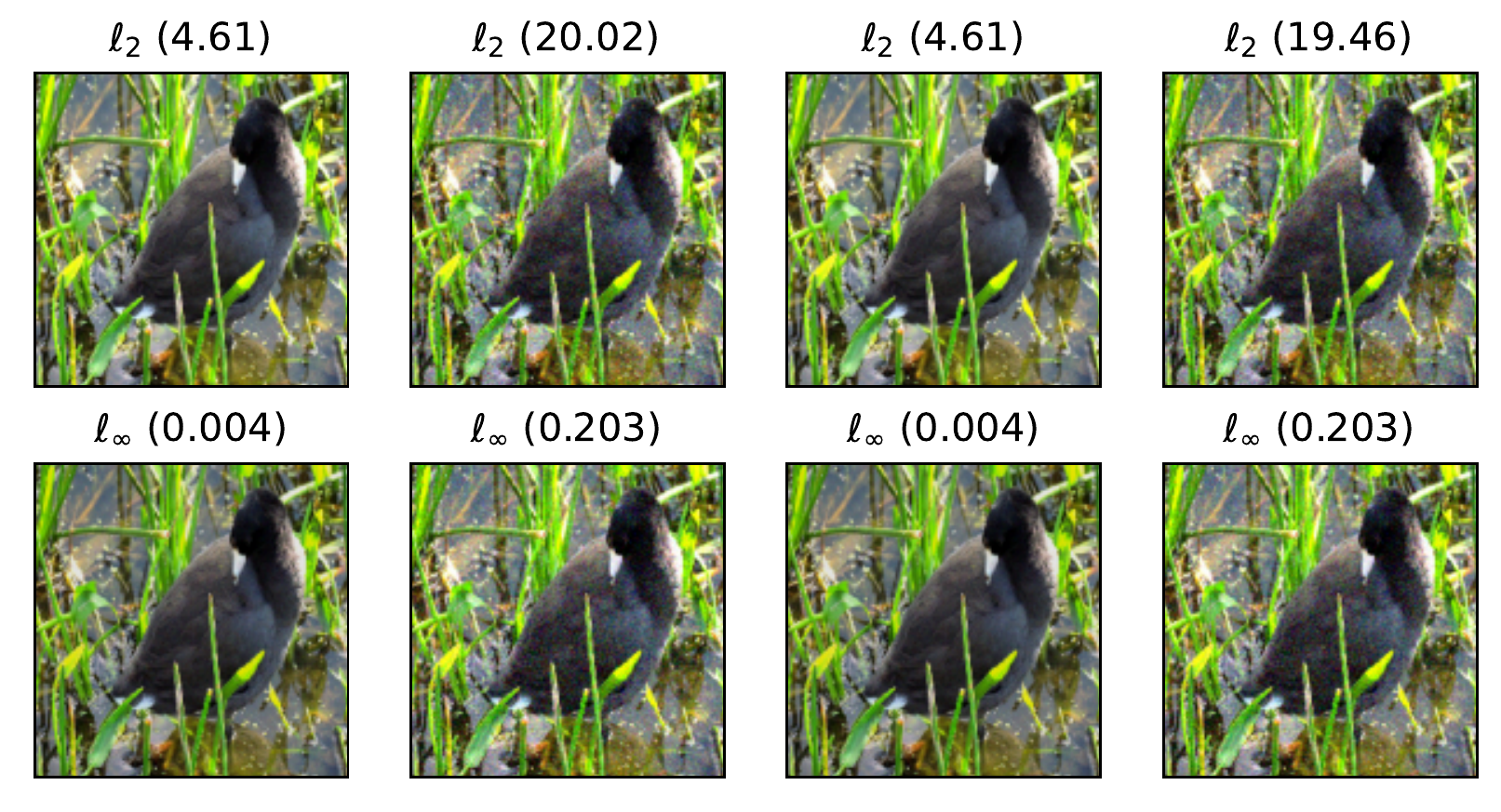}}
\subfigure[HSJA\label{fig:hsja}]{\includegraphics[width=0.42\textwidth]{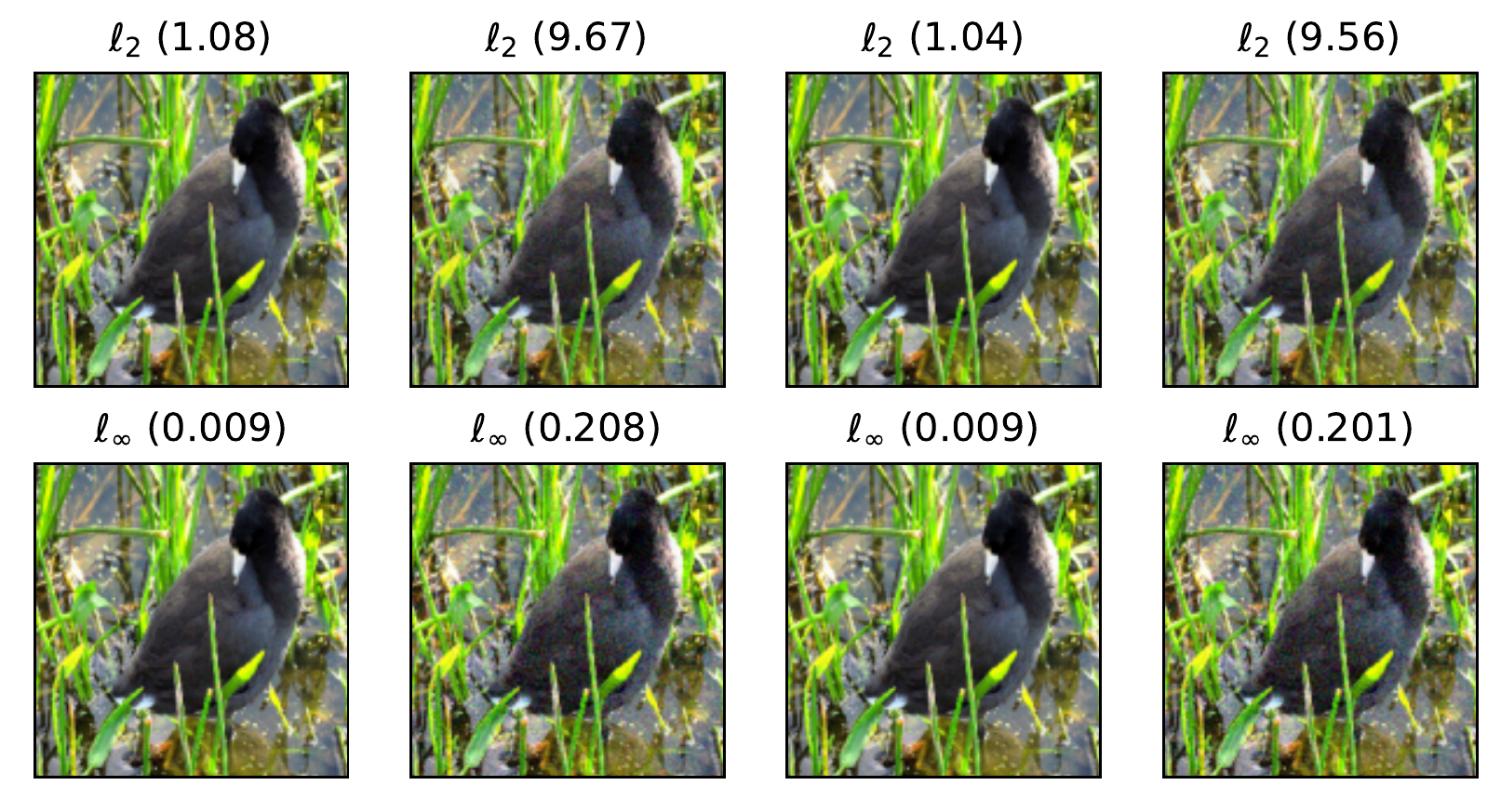}}

 \caption{The visualized demonstration of the crafted adversarial sample with or without our approach. For BA, HSJA and SFA attacks, in the first row, from left to right,  the first two images represent the crafted sample with and without \name{}under 30K query budget and the last two images represent the crafted sample with and without \name{}under 50K query budget in the $\ell_2$ settings , respectively; the bottom row represents the crafted images in the $\ell_\infty$ settings. As for Sign-opt, in the first two, from left to right, the images represent the crafted input with or without defense under 30K query budget in the $\ell_2$ settings, while the bottom row represents the same crafted inputs under 50K query budgets. And the value below each adversarial sample represents the corresponding perturbation distance under the specific norm settings (shown in the bottom of the figure). }
\label{fig:visual_appendix}
\end{figure*}

\end{document}